# Dimensionality Reduction and State Space Systems: Forecasting the US Treasury Yields Using Frequentist and Bayesian VARs[1]

Sudiksha Joshi
Pennsylvania State University, University Park

## Abstract

Using a state-space system, I forecasted the US Treasury yields by employing frequentist and Bayesian methods after first decomposing the yields of varying maturities into its unobserved term structure factors. Then, I exploited the structure of the state space model to forecast the Treasury yields, and compared the forecast performance of each model using the metric − mean squared forecast error, as the loss function. Among the frequentist methods, I applied the two-step Diebold-Li, two-step principal components, and one-step Kalman filter approaches. Likewise, I imposed the five different priors in Bayesian $VARs$ − Diffuse, Minnesota, natural conjugate, independent normal inverse − Wishart, and the stochastic search variable selection priors. After forecasting the Treasury yields for 9 different forecast horizons, I found that the $BVAR$ with Minnesota prior generally minimizes the loss function. I augmented the above $BVARs$ by including macroeconomic variables and constructed impulse response functions with a recursive ordering identification scheme. Finally, I fitted a sign-restricted $BVAR$ with dummy observations.

---

[1] I gratefully acknowledge the helpful advice and suggestions from Ross Doppelt. All errors are mine only.

# 1. Introduction

In this research paper, I have forecasted the yield curve using the dynamic Nelson Siegel model, also known as the Diebold-Li model using Bayesian and frequentist methods. The yield curve is a graph that plots the yields of various bonds such as Treasury securities against their term to maturity. Depicting a snapshot of the current yield levels in the market, it has both a cross-sectional and a temporal dimension. Investigating how the yield curve dynamically evolves is crucial not only in financial markets in pricing assets, hedging risks, and allocating portfolios, but also essential while conducting monetary policy, restructuring fiscal debt, and in general gauging the overall economic health. Often economists at the Federal Reserve use the yield curve to signal the expected ex-ante economic state. Stock and Watson (1993), Estrella and Adrian (2009) consider an inverted yield curve as a harbinger of a recession.

The structure of the research paper is as follows. Section 2 reviews the literature on yield curve modeling; section 3 describes the data; section 4 specifies the state-space system and the model's assumptions; section 5 presents the analysis – elaborates on the methods applied to decompose the dataset into its three latent factors using principal component analysis; the yield curve modeling using the two-step Diebold-Li, two-step principal components, and the one-step Kalman filter approaches; and discusses the findings. Section 5.2 presents the caveats of the methods and elaborates the Bayesian methods with different priors on the latent factors obtained from the first step of the two-step Diebold-Li method. Furthermore, it discusses an augmented Bayesian Vector Autoregression ($BVAR$) model by adding macroeconomic variables to the existing $BVARs$ of latent factors; and finally depicts the impulse response functions and predictive distributions from each of the priors. Also, I fitted a complete Bayesian sign-restricted $VAR$ on the Treasury securities and macro variables. Lastly, section 6 summarizes the results and concludes.

# 2. Literature Review

In this section, I have described the relevant research on term structure modeling, wherein I explained the seminal works of the Nelson-Siegel model proposed by Nelson and Siegel (1987), the affine no-arbitrage models made largely in the finance literature, which are now a workhorse model used in central banks, and finally the Bayesian methods. Each of these methods attempts to improve on the deficiencies of the previous works.

Typically in the finance literature, researchers have represented the yield curve via the no-arbitrage factor approach. Although such factor models fit the cross-sectional component of the yields very well at a particular time, these are unable to capture the temporal dynamics manifested in the yield curve as stressed by Brousseasu (2002) and Duffee (2002). However, to understand the evolution of the yield curve to changes in macroeconomic variables, fitting a model that captures the dynamic trend is crucial. Gürkaynak et al. (2007) estimated the daily values of the US Treasury yield curve from 1961 by applying a smoothing method that fitted the data very well. Moreover, they generated estimates of greater than thirty-five years of 10-year yields. They constructed the Nelson-Siegel-Svensson functional form, which is an extension of the method by Svensson (1994).

Diebold, Rudebusch, Aruoba (2006) developed a dynamic Nelson-Siegel model, based on Nelson and Siegel (1987), that summarizes the yield curve using unobserved or latent factors – level, slope, and curvature. Their three-factor term structure model suggests strong evidence of the impact of macroeconomic variables in forecasting the ex-ante yield curve. They estimated the "yields only" model using a state-space specification by fitting the yield curve at each time and simultaneously estimated the latent factors. The results from the one-step state-space model yield more accurate forecasts than those from the two-step Diebold-Li approach. In another similar model, they also incorporated the observed macroeconomic variables – inflation rate, capacity utilization (as a measure of real activity), and federal funds rate. They estimated a VAR with the unobserved factors or principal components obtained from a larger dataset of yields. Whilst this methodology confines the factors to be orthogonal, no such restriction is imposed on the factor loadings.

Diebold and Li (2006) reinterpreted the Nelson-Siegel yield curve as a dynamic model by reducing the dimension to three factors. They fitted AR models on the latent factors and forecasted the term structure at both long and short term horizons. Besides the autoregressive model, they estimated a VAR separately on yield changes and levels, random walk, and error correction models. Particularly, their long-term (12-months ahead) forecasts outperform the short-term (6-month ahead) counterparts for all maturities by wide margins.

Although the dynamic Nelson-Siegel model performs empirically well and is theoretically appealing, it lacks the requisite restrictions to eliminate riskless arbitrage opportunities as conveyed by Filipović (2000). Therefore, Christensen et al. (2011) introduced the affine no-arbitrage yield cure model that maintains the factor loading structure present in the DNS model. Known as the arbitrage-free Nelson-Siegel (AFNS) yield curve models, the parsimonious models enhance the process of estimating the parameters and consequently generate more precise out-of-sample predictions, not to mention the high forecasting accuracy for medium to long-term yields and forecast horizons.

The AFNS and the Diebold-Li models are variants of the same underlying dynamic yield curve forecasting and modeling variant, but with slightly different mechanisms. As bonds of longer horizons are riskier, risk-averse investors require risk-premium to hold these long-term bonds, paving the way for arbitrage opportunities. Nonetheless, markets can eradicate such arbitrage opportunities if the long-term yields are the risk-adjusted expectations of the mean ex-ante short-term interest rates. To depict no-arbitrage in models, and capture the comovements in the cross-section of the yields, the standard class of affine models imposes cross-equation restrictions in the VAR of the yields. Such restrictive assumptions yield tractable solutions for bond yields, eschewing the need to use Monte Carlo or other computationally expensive methods to estimate the parameters. Therefore, Vasicek (1977) and Cox et al. (1985) obtained closed-form solutions. Amongst the first to develop no-arbitrage models, the models of Vasicek (1977) and Cox et al. (1985) described the dynamic nature of risk premia and the risk-neutral evolution of yield curve factors.

Similarly, Ang and Piazzesi (2003) built a three-factor affine term structure model comprising macroeconomic variables. Introducing two macro factors into an affine model of yield curve similar to that generated by Dai and Singleton (2000), they extracted the first principal component from measures of real economic activity. Likewise, the second principal component is extracted from various indices measuring price. From their results, the macroeconomic factors explain 85 percent of the variation in the shorter-term yields and enhance the forecasts, although have smaller effects on longer yields.

Following the strategy and novel findings by Ang and Piazzesi (2003), numerous economists, including Hordahl et al. (2006) and Dewachter and Lyrio (2006) have adopted similar methodologies to model the term structure. Similarly, Bernanke et al. (2005) amalgamated factor analysis and structural VAR analysis by fitting a joint VAR of interest rates and latent factors derived from the cross-section of macroeconomic time series. Calling it Factor Augmented VAR (FAVAR), they applied this model to scrutinize the behavior of short-term interest rates and the impact of monetary policy on a wide array of macroeconomic variables.

Taking it a step further, Moench (2008) applied the FAVAR model to examine the affine term structure in the arbitrage-free model (with restrictions on the parameters), given short-term rates and four macro variables that fit the yield curve. With the state equation being the FAVAR, Moench (2008) called his model the No-Arbitrage Factor-Augmented VAR. Relative to other benchmark models in the literature, their model generated superior recursive out-of-sample forecasts. For instance, the model significantly outperformed the standard three-factor affine model that Bernanke (2005) proposed and the Diebold-Li model by Diebold and Li (2006) for all except for very long maturities.

Unlike the promising results exhibited by Ang and Piazzesi (2003), Duffee (2002) showed that the standard class of three-factor affine models generate poor in-sample and out-of-sample forecasts of the ex-ante Treasury yields, in contrast to the forecasts from a random walk. He attributed the poor forecasting ability to the fact that risk premium (extra compensation to buying a risky bond) is correlated with the volatility inherent in the interest rates. To overcome this shortcoming, he defined an extensive class of tractable affine models called the "essentially affine" models which enable risk premiums to be independent of interest rate volatility. These models retain the affine cross-sectional and time series characteristics of the bond prices. These alterations enhance the forecasting precision of the future yields. However, he pointed out that there is a tradeoff between flexibility in fitting the variance in interest rates and flexibility in calibrating ex-ante yields.

Whilst most of the researchers have explored how the latent factors drive the term structure, fewer economists have studied how the observable macroeconomic factors drive yields. Evans and Marshall (2006) concluded that macro shocks (impulses) explain most of the variation present in the nominal Treasury yields, including parallel shifts in the level factor of the yield curve. Specifically, the macro factors explain 85 percent of the variance in the 5-year Treasury yield. For instance, shocks to the preferences on current consumption directly influence the real interest rates and expected inflation. These changes in the monetary policy transmission mechanism indirectly affect the yields. Their analysis displayed minuscule evidence on the role of fiscal policy in propelling variation in interest rates. Consistent with the findings of Ang and Piazzesi (2003) that the macro factors explain significant variability in the medium and short-term yields; however; their model uses interest rate smoothing, wherein the interest rates hinge upon their own lags. Finally, they deduced that omission of interest rate smoothing attenuates the significance of macroeconomic shocks on interest rates.

Finally, in the spectrum of Bayesian methods, Carriero (2011) constructed an arbitrage-free, Gaussian affine term structure model (ATSM) that considers possible misspecification. Instead of dogmatically imposing restrictions, he used the restrictions on a VAR implied by ATSM as prior information. In enforcing the ATSM prior, he followed the technique of Del Negro and Schorfheide (2004). Setting ATSM as prior boosts the forecasting accuracy. In fact, incorporating the ATSM prior to the VAR outperforms the exact ATSM at all horizons while forecasting the one, three, twelve, and thirty-six month yields with statistically significant gains. Furthermore, the forecasts were not only superior relative to those when he constructed a VAR with Minnesota prior, but also the VAR with both priors outperformed those from an unrestricted VAR. This corroborated the argument that the magnitude and direction of shrinkage increase the forecasting performance. Finally, he demonstrated that the ATSM prior yielded better results, in contrast to those from the Nelson-Siegel models and linear models.

Hautsch and Yang (2012) applied Bayesian inference based on Markov Chain Monte Carlo (MCMC) algorithm to a Stochastic Volatility Nelson–Siegel (SVNS) model introduced by Hautsch and Ou (2008). Extending the DNS to incorporate stochastic volatility in the latent factors, this framework flexibly yet parsimoniously captures the variance in the yield curve. After applying the SVNS model to the monthly data on US zero-coupon yields, they found significant evidence of temporal volatility embedded in the latent factors. Including stochastic volatility diminishes the forecasting and parameter uncertainties, especially in less volatile periods.

Pooter et al. (2007) examined the uncertainty in the parameters and models that forecast the yield curve. They used the time series of US zero-coupon bonds to analyze the forecasting performance of each model. Applying the frequentist maximum likelihood and Bayesian methods, they examined the effects of incorporating parameter uncertainty in the models. Amongst the frequentist methods they apply are the two-step and one-step approaches, while the Bayesian technique they implement is one-step. Moreover, they scrutinized the impact of adding macroeconomic variables by fitting each model with and without the macro variables. While the accuracy from individual models varies over time, the results showed that by combining forecasts, the uncertainty in the models reduces and forecasts are more precise, especially for longer maturities. In particular, the forecasts are more precise when they combine each method by weighing each forecasting method based on historical performance. Lastly, the out-of-sample forecasts substantially improve after adding macroeconomic factors.

## 3. Data

I summarized the yield curve through the latent factors – level, slope, and curvature, by fitting the Diebold-Li model of yield curve to a time series of monthly values of yield curves derived from the US Treasury yields. Diebold and Li (2006) converted the prices of the government bonds to unsmoothed Fama-Bliss $US$ Treasury zero-coupon yields to develop the Dynamic Nelson Siegel ($DNS$) model to approximate the dynamic nature of the structure of factors. However, I constructed the $DNS$ by directly using the Treasury yields from the Federal Reserve Economic Data repository. The monthly data on 10 Treasury yields are: 3, 6, 12, 24, 36, 60, 84, 120, 240, and 360 months from January 1990 to January 2021, totalling 374 time series observations. Furthermore, I collected data on 21 other monetary variables that affect the yield curve – 7 measures of yield spreads, such as the $TED$ spread, term premium on zero-coupon bonds of maturities ranging from 1 to 10 years, effective federal funds rate, U-3 unemployment rate, inflation rate measured via the personal consumption expenditures, and the year to year change in total capacity utilization ($TCU$). It is the percent of total capacity used to produce finished goods in the spectrum of manufacturing, mining, gas and electric utilities in the United States. Then, I divided the whole dataset into a train and test set, where the train set comprises the first 85 percent of the observations ranging from January 1990 to June 2016. The test set has the remaining 25 percent of the observations, from July 2016 to January 2021. This is the out-of-sample data on which I calculated the forecasts of the latent variables and the Treasury yields; and compared the magnitude of the deviations of those forecasts with the actual yields.

The descriptive statistics of the Treasury yields in table 1 indicates that on average short-term yields (such as 3, 6 and 12 months yields) are less than long-term yields (such as 120, 240, and 360 months yields). This is because investors require extra compensation or risk premium as an incentive to hold securities for longer duration as longer term yields have more inflation risk than the short term yields. Moreover, short-term yields fluctuate more as they have higher standard deviations than their long-term counterparts. We can assumably attribute this to the expectation hypothesis of the term structure of interest rates which encompasses that the long-term yields are averages of the expected future short-term yields. Furthermore, results from the Augmented Dickey Fuller ($ADF$) test of stationarity provide evidence that all Treasury yields have unit roots.

| Time to Maturity (months) | Minimum | Median | Mean | Maximum | Standard Deviation | $ADF$ Test Statistic ($p-$value) |
|---|---|---|---|---|---|---|
| 3 | 0 | 2.205 | 2.6775 | 8.07 | 2.2877 | $-2.0361$ (0.271) |
| 6 | 0.03 | 2.355 | 2.8033 | 8.44 | 2.3228 | $-2.7249$ (0.0698) |
| 12 | 0.08 | 2.45 | 2.9281 | 8.58 | 2.3289 | $-2.6002$ (0.0929) |
| 24 | 0.11 | 2.85 | 3.2131 | 8.96 | 2.3588 | $-2.135$ (0.2307) |
| 36 | 0.11 | 3.095 | 3.4237 | 9.05 | 2.318 | $-1.8785$ (0.3422) |
| 60 | 0.21 | 3.685 | 3.8132 | 9.04 | 2.2113 | − 1.7541 (0.4035) |
| 84 | 0.39 | 3.99 | 4.1174 | 9.06 | 2.1235 | − 1.7091 (0.4264) |
| 120 | 0.55 | 4.315 | 4.353 | 9.04 | 2.0206 | − 1.89 (0.3368) |
| 240 | 0.98 | 4.845 | 4.83 | 9.02 | 1.9278 | − 1.631 (0.467) |
| 360 | 1.2 | 4.865 | 4.9012 | 9 | 1.8321 | − 0.7138 (0.424) |

*Table 1. Descriptive statistics of the Treasury yields.*

## 4. State-Space System

Firstly, I represented the Diebold-Li model in a parametric state-space form. The Dynamic Nelson Siegel ($DNS$) model is:

$$y_t(\tau_m) = L_t + S_t\left(\frac{1-e^{-\lambda \tau_m}}{\lambda \tau_m}\right) + C_t\left(\frac{1-e^{-\lambda \tau_m}}{\lambda \tau_m} - e^{-\lambda \tau_m}\right) + \varepsilon_t(\tau_m), \text{ where} \qquad (1)$$

$t = 1, 2, ..., T$, $m = 1, 2, ..., M$ represent the yields at time $t$ of maturity $m$. Therefore, the data is:

$Y = \{y_{11}, y_{12}, ..., y_{TM}; \tau_1, ..., \tau_M\}$

$L_t$, $S_t$, $C_t$ are level, slope, and curvature factors, respectively, that measure the effect of long, short, and medium-term Treasury yields, respectively. The idiosyncratic stochastic errors $\varepsilon_t(\tau_m)$ are specific to maturities. Let the vector of factors be $f_t = [L_t, S_t, C_t]'$. $\lambda$ is the exponential decay term. Since the latent factors adjusted for the means follow $VAR(1)$, I represented it as a state-transition equation in the state-space system. The state transition equation (2) is:

$$\begin{pmatrix} L_t - \mu_L \\ S_t - \mu_S \\ C_t - \mu_C \end{pmatrix} = \begin{pmatrix} a_{11} & a_{12} & a_{13} \\ a_{21} & a_{22} & a_{23} \\ a_{31} & a_{32} & a_{33} \end{pmatrix} \begin{pmatrix} L_{t-1} - \mu_L \\ S_{t-1} - \mu_S \\ C_{t-1} - \mu_C \end{pmatrix} + \begin{pmatrix} \eta_t(L) \\ \eta_t(S) \\ \eta_t(C) \end{pmatrix}$$

It expresses the dynamics of the demeaned latent state (unobserved) variables. Correspondingly, the measurement equation (3), which connects the observed yields with the unobserved states, is:

$$\begin{pmatrix} y_t(\tau_1) \\ y_t(\tau_2) \\ \vdots \\ y_t(\tau_M) \end{pmatrix} = \begin{pmatrix} 1 & \frac{1-e^{-\lambda\tau_1}}{\lambda\tau_1} & \frac{1-e^{-\lambda\tau_1}}{\lambda\tau_1} - e^{-\lambda\tau_1} \\ 1 & \frac{1-e^{-\lambda\tau_2}}{\lambda\tau_2} & \frac{1-e^{-\lambda\tau_2}}{\lambda\tau_2} - e^{-\lambda\tau_2} \\ & \vdots & \\ 1 & \frac{1-e^{-\lambda\tau_M}}{\lambda\tau_M} & \frac{1-e^{-\lambda\tau_M}}{\lambda\tau_M} - e^{-\lambda\tau_M} \end{pmatrix} \begin{pmatrix} L_t \\ S_t \\ C_t \end{pmatrix} + \begin{pmatrix} \epsilon_t(\tau_1) \\ \epsilon_t(\tau_2) \\ \vdots \\ \epsilon_t(\tau_M) \end{pmatrix}$$

In matrix notation, we can rewrite the observed state-space system given the observed yields $y_t$, and demeaned factors as the latent states: $X_t = f_t - \mu$:

$$f_t - \mu = A(f_{t-1} - \mu) + \eta_t \tag{4}$$
$$\Rightarrow X_t = AX_{t-1} + \eta_t \tag{5}$$

The measurement equation is:

$$Y_t = \Lambda f_t + \varepsilon_t \tag{6}$$

Here, $\varepsilon_t$, and $\eta_t$ are Gaussian and orthogonal white noise $(WN)$ processes such that:

$$\begin{bmatrix} \eta_t \\ \epsilon_t \end{bmatrix} \sim WN\left( \begin{bmatrix} 0 \\ 0 \end{bmatrix}, \begin{bmatrix} \Sigma_\eta & 0 \\ 0 & \Sigma_\epsilon \end{bmatrix} \right)$$

Equivalently, $\eta_t \sim WN_3(0, \Sigma_\eta)$, and $\varepsilon_t \sim WN_{10}(0, \Sigma_\varepsilon)$ i.e. the models are homoskedastic. Since the dataset consists of yields of 10 different maturities, I assumed that the $(10 \times 10)$ matrix $\Sigma_\varepsilon$ is diagonal as when the yields of different maturities deviate from the yield curve, they are uncorrelated. This induces computations to be tractable, particularly when the number of maturities of yields is very large. Alternatively, the $(3 \times 3)$ matrix $\Sigma_\eta$ is non-diagonal i.e. the shocks to the three term structure or latent factors are correlated with each other. $\Sigma_\eta$ is the variance-covariance matrix of the residuals from $VAR(1)$.

## 5. Methods and Analysis

In the frequentist methods section 5.1, I estimated the unobserved factors that define the yield curve using the three methods – the two-step Diebold-Li, two-step principal components, and the one-step Kalman filter approaches. These require us to estimate the parameters using the reduced form restricted (frequentist) vector autoregressions. Following the Kalman filter approach described in Diebold, Rudebusch, Aruoba (2006), section 5.1.4 compares the results, including the forecasts from the three methods. In section 5.2, I estimated Bayesian $VARs$ by applying the diffuse, Minnesota, natural conjugate, independent normal inverse – Wishart, and stochastic search variable selection $(SSVS)$ priors. Through probability distributions, I formed prior beliefs about the estimated parameters, and updated those beliefs based on data on the latent factors i.e. by using likelihood functions. Using Bayes law, I combined the likelihood function and the prior distributions to derive the posterior distributions of the parameters. While sections 5.2.1 − 5.2.4 outline the structure of each of the priors and their respective posterior distributions, section $A.2.$ in the Appendix elucidates the likelihood functions and the functional forms of the posterior means and variances.

## 5.1 Frequentist Methods

### 5.1.1 Two-Step Principal Component Analysis

Principal component analysis (PCA) decomposes the high-dimensional set of observed yields and other 19 macroeconomic variables − consisting of monetary, labor, inflation, and output variables − into a parsimonious lower-dimensional set of three factors that describe the characteristics of the yield curve. The first three $PCs$ or the latent factors − $PC1, PC2, PC3,$ are also referred to as the level, slope, and curvature factors, respectively. Table 2 shows that the level and slope factors explain 95.493 and 4.155 percent of the variances, respectively.

| $PC$ | Variance Explained by Each $PC$ |
|---|---|
| 1 | 0.95493 |
| 2 | 0.041556 |
| 3 | 0.0025064 |
| 4 | 0.00046683 |
| 5 | 0.00028291 |
| 6 | 0.00014685 |

*Table 2. Proportion of variance explained by each principal component or the eigenvalues estimated on the train set*

### 5.1.2 Two-Step Diebold-Li Model with Fixed λ

In contrast with $PCA$ where I estimated both the latent factors and the factor loadings, I imposed a specific structure of the factor loadings in the dynamic Nelson-Siegal model. This not only enables us to interpret the latent factors as level, slope, and curvature, but also obtain very accurate estimates of the factors. It is founded upon the underlying notion that because the yield curve varies across time, the parameters are time-varying as well. Finding the ex-ante value of the yield curve is tantamount to finding the ex-ante values of the term structure factors as the yield curve is a function of the factors. Then, I constructed the Diebold-Li model in the train set with the explanatory variables as factor forecasts.

Firstly, fixing $\lambda$ to a constant, I estimated the static Nelson-Siegal model $\forall t = 1, 2, ..., T$ by $OLS$. This results in a three-dimensional or a $(10 \times 3)$ matrix of the estimated latent factor loadings $[1, (\frac{1-e^{-\lambda \tau_m}}{\lambda \tau_m}), (\frac{1-e^{-\lambda \tau_m}}{\lambda \tau_m} - e^{-\lambda \tau_m})]$ associated with the latent factors $\{L_t, S_t, C_t\}_{t=1}^{T}$. Keeping $\lambda$ fixed simplifies the estimation method from non-linear least squares to $OLS$. Secondly, I fitted a $VAR(1)$ to the time series of estimated factors derived in the first step.

In the dynamic Nelson-Siegel model given by equation (1), $\lambda$ determines the time to maturity at which the loadings on the curvature or the medium-term factor are maximized. Diebold and Li (2005) set $\lambda_t = 0.0609 \ \forall \ t$ as it maximizes the curvature loadings maturing in 30 months (average of 24 and 36 months). However, their dataset ranges from January 1985 to December 2000, while this analysis is based on time series ranging from January 1990 to June 2016. Equating $\lambda$ to 0.0609 maximizes the curvature loadings in exactly 28.58 months, instead of 30 months. Therefore, I have set $\lambda$ to 0.0598. Figure 1 shows the factor loadings on $L_t$, $S_t$, and $C_t$ that are functions of maturity.

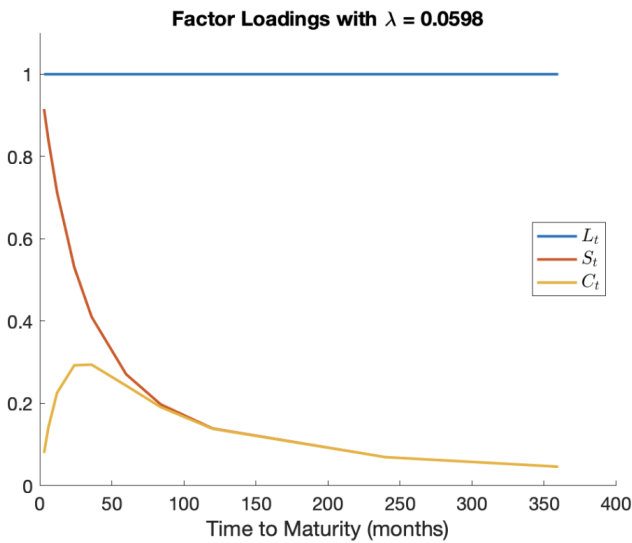

*Fig 1. Factor loadings associated with the term structure factors*

The loading on $L_t$ in equation (1) is a constant as the level factor is a flat line parallel to the maturity axis. It does not decay to 0 and affects the long-term yields; hence it is also known as a "long-term factor." The loading on $S_t$ is a function $\frac{1-e^{-\lambda \tau_m}}{\lambda \tau_m}$, which starts at approximately 1 but monotonically decays to 0; therefore, it is a convex-shaped curve. Also known as a "short-term factor", it affects short-term yields. Finally, the loading on $C_t$ is a function $\frac{1-e^{-\lambda \tau_m}}{\lambda \tau_m} - e^{-\lambda \tau_m}$ that starts at 0 and gradually converges to 0. As a "medium-term factor", it affects the medium-term yields. $\lambda$ governs the pace of exponential decay wherein small values of $\lambda$ slows the decay and better fits the curve at longer-term maturities. This contrasts with large values of $\lambda$ which fastens the rate of decay and better fits shorter-term maturities. Figure 2 compares the empirical path of the level, slope, and curvature with their corresponding factors estimated from the first step ($OLS$ regression) in the Diebold-Li model, and the first three principal components. The 10-year yield is an example of empirical measure of the level. The slope or the yield spread is the difference between the 3-month and 10-year

Treasury yield. Lastly, the curvature is measured as $2 \times 2y - 10y - 3mo$. The first $PC$ fluctuates considerably relative to the trajectories of the 10-year yield and is also evident from its largest standard deviation in the summary statistics.

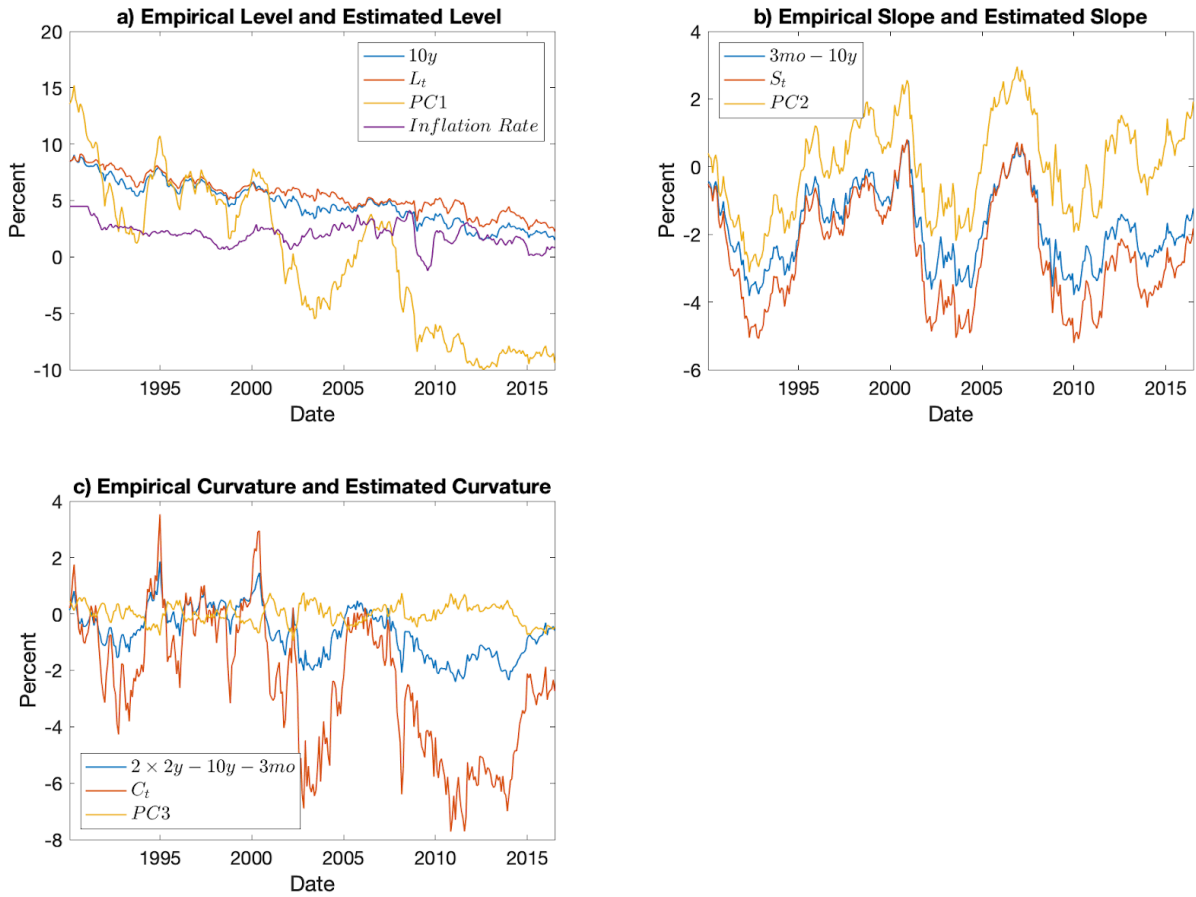

Fig 2a) Long-term factor – $\hat{L}_t$ , PC1, a empirical measures of level: 10-year Treasury yield, and inflation rate

b) Short-term factor – $\hat{S}_t$, PC2, and an empirical measure of slope: difference between 3-month and 10-year Treasury yields

c) Medium-term factor – $\hat{C}_t$, PC3, PC3 and an empirical measure of curvature : $2 \times 2y - 10y - 3mo$

The summary statistics in table 2 indicates that $PC1$ and $PC3$ are the most ($\sigma = 6.6389$) and least ($\sigma = 0.3401$) volatile. Evidently, the means of $PCs$ are 0 as they are demeaned values.

| Latent Factors | Minimum | Median | Mean | Maximum | Standard Deviation ($\sigma$) |
|---|---|---|---|---|---|
| $\hat{L}_t$ | 2.256 | 5.4608 | 5.5365 | 9.1356 | 1.6091 |
| $\hat{S}_t$ | − 5.2034 | − 2.7796 | − 2.5536 | 0.8066 | 1.5677 |
| $\hat{C}_t$ | − 7.7157 | − 2.2653 | − 2.4449 | 3.5335 | 2.4514 |
| $PC1$ | − 9.973 | 1.2499 | $6.144 \times 10^{-15}$ | 15.207 | 6.6389 |
| PC2 | −3.1217 | 0.0203 | −3.809 × 10⁻¹⁵ | 2.9579 | 1.3849 |
| PC3 | −0.9508 | 0.0022 | 1.254 × 10⁻¹⁵ | 0.757 | 0.3401 |

Table 2. Summary statistics of the latent factors

The empirical level, slope, and curvature factors, alongside their corresponding factor loadings, closely follow each other, as also reflected by the high positive correlation with each other :

$$\rho\left(10y, \hat{L}_t\right) = 0.9714, \; \rho\left(3mo - 10y, \hat{S}_t\right) = 0.9934, \; \rho\left(2 \times 2y - 10y, \hat{C}_t\right) = 0.9693, \; \rho\left(\%\Delta pce, \hat{L}_t\right) = 0.5088$$

$\%\Delta pce$ refers to the inflation rate that I calibrated as year-to-year percent change in personal consumption expenditures. It is positively correlated with both the level factor and $PC1$, and similar paths of the unobserved factors and inflation rate in figure 2 a) bolster the claim that $\hat{L}_t$, and $\%\Delta pce$ are closely tied. Likewise, the 10-year yield, and the yield spread have strong positive correlations with the first two $PCs$, respectively.

$$\rho(10y, PC\;1) = 0.966, \; \rho(3mo - 10y, PC\;2) = 0.897, \; \rho(2 \times 2y - 3mo, PC\;3) = -0.489, \; \rho(\%\Delta pce, PC\;1) = 0.507$$

The $ACF$ plots in figure 3 examine the autocorrelation present in the three latent factors from OLS and three $PCs$. Both display similar characteristics. All variables except $PC3$ are heavily autocorrelated even after 50 lags, thereby showing high level of persistence, although the serial correlation subsides for $PC2$ and $S_t$ over time. Whilst the yield dynamics depicted by $L_t$ are vastly persistent, the slope dynamics, $S_t$ − closely linked to the difference between long and short term yields or yield spread − are less persistent.

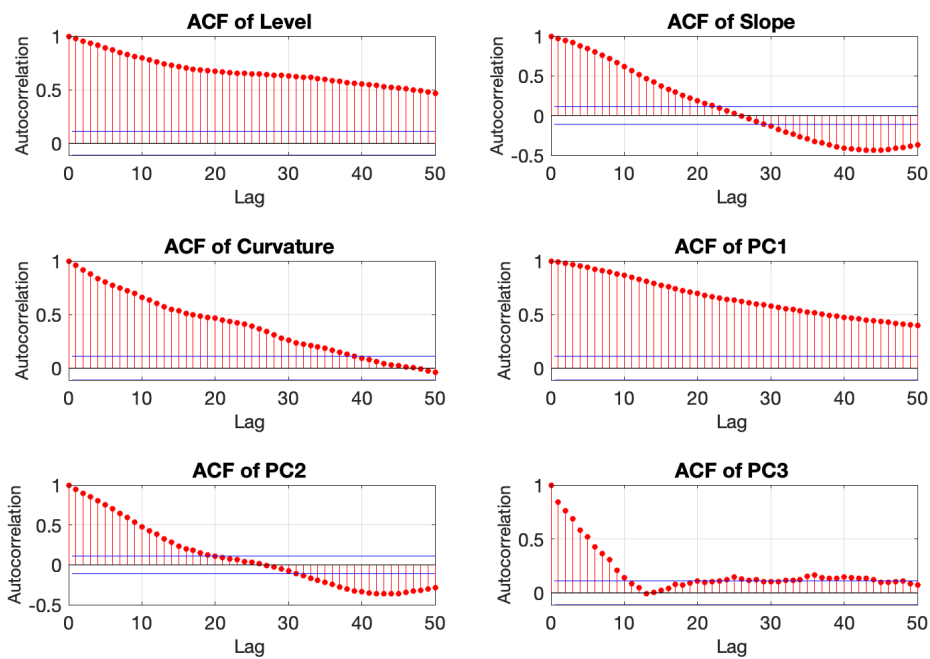

*Fig 3. Autocorrelation function plots of the latent factors show high levels of persistence.*

After fitting the three-dimensional data of factors, I fitted a $VAR(1)$ model on this estimated time series of factors and $PCs$ separately. Whilst the conventional procedure of $VAR$ entails no more than six to eight variables in a $VAR$ to conserve degrees of freedom, this information set does not capture the plethora of macroeconomic and financial time series indicators that central bankers and financial markets use in forecasting macro variables. Since the orthogonal principal components effectively summarize the macroeconomic variables, I constructed a $VAR(1)$ model on the principal components.

Fitting $VARs$ on the latent factors (obtained from $OLS$, and $PCA$) is the second step of the two-step of the Diebold-Li, and the principal components approach, respectively. As the state-space equation works with factors adjusted for mean, I incorporated an additive constant into the model. The $VAR(1)$ is

$$f_t = a_0 + A_1 f_{t-1} + \eta_t. \tag{7}$$

Using the multivariate least squares ($MLS$) method, which is a multivariate analog to the $OLS$ performed equation-by-equation, I estimated the coefficient matrix $A_1$.[2]

### 5.1.3 One-Step Kalman Filter Approach

So far, I have estimated the two-step Diebold-Li and $PC$ models by keeping $\lambda$ fixed. Now, using the one-step Kalman filter approach, I modeled the state-space equation without the constant $\lambda$, and instead, estimated it alongside other variables. The Kalman filter in the state-space model helps to output maximum likelihood estimates, optimal smoothed, and filtered estimates of the latent factors. I initialized the estimation by inputting the results from the two-step Diebold-Li method to acquire the initial transition matrix $A$ in equation (5). However, before applying the Kalman filter, I adjusted the state-space system, keeping in mind the demeaned factors in the state equation (4). From the state-space system, I wrote the yields adjusted for intercepts as $Y_{1,t} = Y_t - \Lambda\mu$

Upon rearranging, $Y_{1,t} = (\Lambda f_t + \varepsilon_t) - \Lambda\mu = \Lambda(f_t - \mu) + \varepsilon_t \Rightarrow Y_{1,t} = \Lambda X_t + \varepsilon_t$

Thus, another way of writing the Diebold-Li state-space system is the previously mentioned equation (5) and the measurement equation (8):

$$X_t = AX_{t-1} + \eta_t \tag{5}$$
$$Y_{1,t} = \Lambda X_t + \varepsilon_t \tag{8}$$

I decomposed the positive-definite symmetric matrix $\Sigma_\eta$ into a product of two matrices: $Z$ and its transpose such that $\Sigma_\eta = ZZ'$. Here, $Z$ is the lower Cholesky factor of $\Sigma_\eta$, i.e. $Z$ is a lower triangular matrix with positive diagonal entries. To allow the off-diagonal elements of the matrix $\Sigma_\eta$ to be non-0 so that the elements are correlated with each other, I allocated the six elements of the lower Cholesky factor $Z$ as a column vector of initial parameter values. Then, I initialized the elements of the initial parameters' vector by taking a square root of the variance of the estimated innovations from the $VAR(1)$ model. Since the Diebold-Li model also assumes that the variance-covariance matrix $\Sigma_\varepsilon$ is diagonal, I again decomposed it into a product of two matrices: $W$ and its transpose such that $\Sigma_\varepsilon = WW'$. I initialized $W$ to be the square root of the diagonal elements of the $(10 \times 10)$ covariance matrix $\Sigma_\varepsilon$.[3]

---

[2] Section 5.1.4 shows the parameters of the VAR(1) coefficient matrix and its forecasts after discussing the one-step Kalman filter method.

After initializing $\lambda$ to 0.0598, I set the sample average of the three $OLS$ regression coefficients (in the first step) as the vector of initial latent factor parameters $\mu$. Therefore, in total, the numerical optimization estimates 29 free parameters. These are as follows:

- 9 parameters from the $(3 \times 3)$ matrix $A$;
- 3 parameters from the $(3 \times 1)$ vector of means $\mu$;
- 1 parameter from the measurement matrix $\Lambda$, which is the exponential decay rate $\lambda$;
- 6 parameters from the transition disturbance matrix $\Sigma_\eta$ : for each of the three latent factors, I computed the error variance $\sigma_l^2$, $\sigma_s^2$, $\sigma_c^2$ ; and three covariance terms $\sigma_{lc}$, $\sigma_{ls}$, $\sigma_{sc}$. But to obtain $\Sigma_\eta$, I first estimated $Z$ from the optimization technique;
- 10 parameters from the measurement disturbance covariance matrix $\Sigma_\varepsilon$ : for each of the ten yields, I computed the error variances as above. Similarly, I directly optimized the estimates of $W$ to calculate $\Sigma_\varepsilon$.

### 5.1.4 Results from the Three Frequentist Methods

After calibrating the initial values, I established the optimization parameters and estimated the model via the Kalman filter. Below I contrasted the results gathered from the two-step and the one-step methods to analyze the extent of similarities in the results from the aforementioned techniques. First, I compared the $(3 \times 3)$ coefficient matrix $A_1^{2-step}$, and $A_1^{2-stepPC}$ from the two-step Diebold-Li, and two-step $PC$ approaches with the state transition matrix $A^{1-step}$ from the one-step Kalman filter approach. They are as follows:

$$A_1^{2-step} = \begin{bmatrix} 0.9955 & 0.0200 & -0.0079 \\ -0.0587 & 0.9095 & 0.0734 \\ 0.0555 & 0.0426 & 0.9185 \end{bmatrix}, A_1^{2-stepPC} = \begin{bmatrix} 0.9914 & 0.0086 & -0.1574 \\ 0.0002 & 0.9708 & -0.2610 \\ 0.0015 & 0.0030 & 0.8620 \end{bmatrix}$$

$$A^{1-step} = \begin{bmatrix} 0.9862 & 0.0144 & 0.0030 \\ -0.0470 & 0.9279 & 0.0657 \\ 0.0445 & 0.0452 & 0.9284 \end{bmatrix}$$

---

[3] Section A.1.2 in the Appendix explains the steps I used in estimating the parameters via Kalman filter, and section A.1.3 outlines the optimization technique.

Mostly, the results are in agreement. Notably, the large positive diagonal elements of the above matrices connote a high level of persistent dynamics prevalent in each latent factor. Simultaneously, the very low values of the off-diagonal elements convey weak dynamics or relationships across the factors. The variance-covariance matrix of the innovations from the three methods are:

$$\Sigma_\eta^{2-step} = \begin{bmatrix} 0.0611 & -0.0620 & 0.0266 \\ -0.0620 & 0.1055 & -0.0481 \\ 0.0266 & -0.0481 & 0.5082 \end{bmatrix}, \Sigma_\eta^{2-stepPC} = \begin{bmatrix} 0.4422 & -0.1035 & -0.0585 \\ -0.1035 & 0.1145 & 0.0172 \\ -0.0585 & 0.0172 & 0.0295 \end{bmatrix}$$

$$\Sigma_\eta^{1-step} = \begin{bmatrix} 0.0666 & -0.0643 & 0.0081 \\ -0.0643 & 0.1058 & -0.0286 \\ 0.0081 & -0.0286 & 0.5505 \end{bmatrix}$$

In contrast with the variances represented by the diagonal elements, the covariances between the factors in the off-diagonal elements are very low, further corroborating the aforementioned claim of low cross-factor dynamics. Correspondingly, the lower Cholesky factor of $\Sigma_\eta^{1-step}$ or the lower triangular matrix $Z$ is:

$$Z = \begin{bmatrix} 0.2580 & 0 & 0 \\ -0.2493 & 0.2090 & 0 \\ 0.0313 & -0.0994 & 0.7346 \end{bmatrix}$$

Finally, I estimated the means of the level, slope, and curvature factors calculated in the three approaches in table 3. The average values are very close to each other in both the two-step Diebold-Li, and the one-step Kalman filter methods, albeit the mean of curvature is smaller in the two-step Diebold-Li approach ($-2.4449$) as opposed to that of the one-step approach ($-0.6851$). As the $PCs$ are demeaned, the means of the first three $PCs$ are close to 0.

| Factors | Two-Step Diebold-Li | One-Step Kalman Filter | Two-Step PC |
|---|---|---|---|
| $L_t$ | 5.5365 | 5.6064 | $0.6145 \times 10^{-13}$ |
| $S_t$ | $-2.5536$ | $-1.6641$ | $-0.3810 \times 10^{-13}$ |
| $C_t$ | $-2.4449$ | $-0.6851$ | $-0.1254 \times 10^{-13}$ |

Table 3. Averages of the latent factors from the one-step Kalman and two-step Diebold-Li and PC methods

The Diebold-Li model's unobserved factors are variables primarily of interest in projecting the ex-ante path of the yield curve. While maximum likelihood estimates the unknown parameters, filtering, smoothing and forecasting measure the unobserved states. I analyzed the states inferred from each approach. The coefficients from the $OLS$ regression in the first step are the latent states (factors). Then, the Kalman smoother extracts the optimal latent factors for all $t = 1, 2, ..., T$.

However, the state-space system estimated the demeaned yields as formulated in the Diebold-Li model, instead of the original yields. Consequently, I demeaned the original yields before smoothing. As the actual states (level, slope, and curvature factors) are the variables of interest than their corresponding demeaned values, I readjusted the demeaned states after smoothing by adding the estimated mean, $\mu$, to the factors. Figure 4 displays the latent factors derived from the two-step, and one-step approaches.

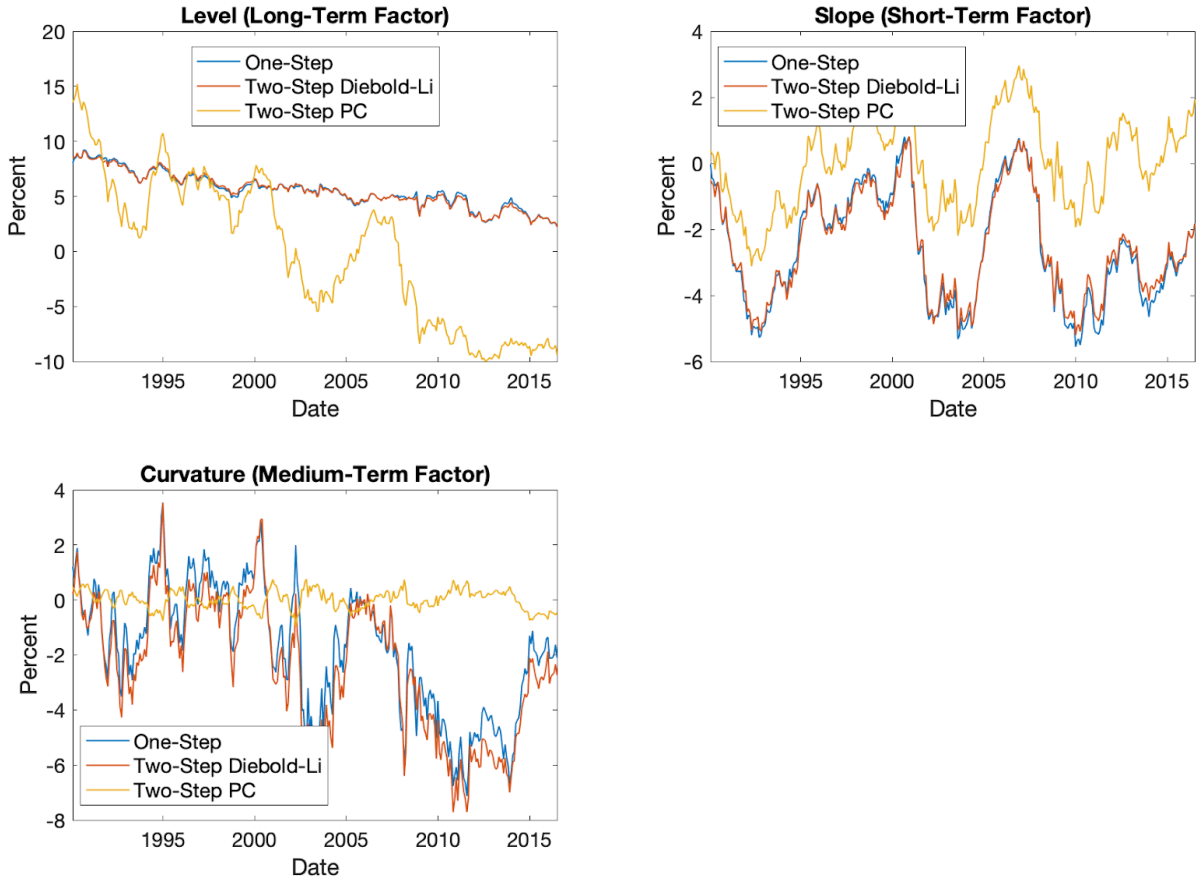

*Fig 4. Empirical latent factors - level, slope, and curvature, estimated from the one-step Kalman filter and two-step methods.*

From the separate graphs of all three factors, the level factor shows considerable persistence as the trajectory indicates a long-term downward trend. Alternatively, the slope and curvature factors are more volatile as they fluctuate from trough to crests, taking in both negative and positive values. Figure 5 shows the concave-shaped curvature loadings given the estimated and constant $\lambda$. Linked with the curvature factor, the estimated decay rate parameter $\lambda$ is 0.047. Unlike the constant decay rate applied in the two-step approach where $\lambda$ is 0.0598, the decay rate parameter estimated from the one-step method is lower. This maximizes the loading on the curvature factor almost a decade later in exactly 39.5 months. We can attribute the interpretation of curvature to the hump-shape (concave looking) curves as a function of the time to maturity for both values of $\lambda$.

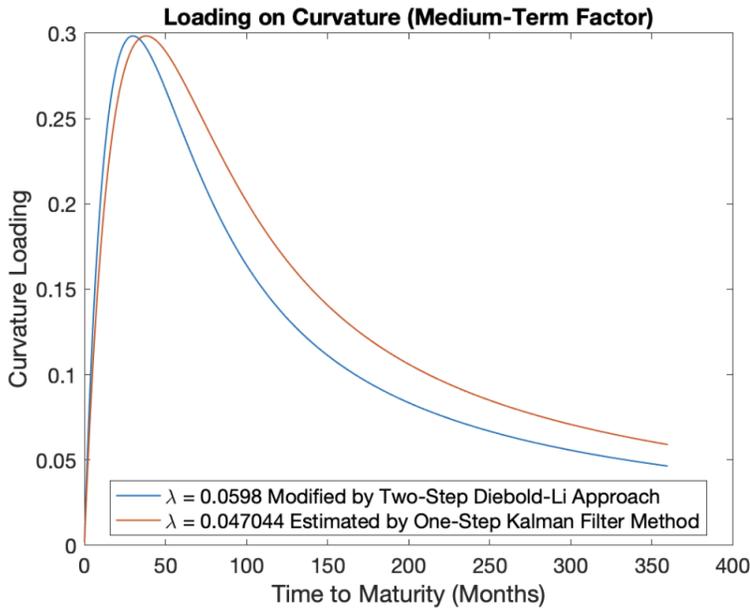

*Fig 5. Empirical medium-term factor and a measure of curvature*

Albeit the differences in the path of the curvature's loading from both the methods, which are prominent after the loadings reach the apex, the factors derived from both methods are reasonably similar. A caveat of the two-step approach is that it does not take into account the uncertainty associated with extracting the latent factors in the first step. Now, figures $6-8$ display the individual forecasts for 56 months ahead alongside the 95, and 80 percent forecast intervals ($FI$) from the three frequentist methods.

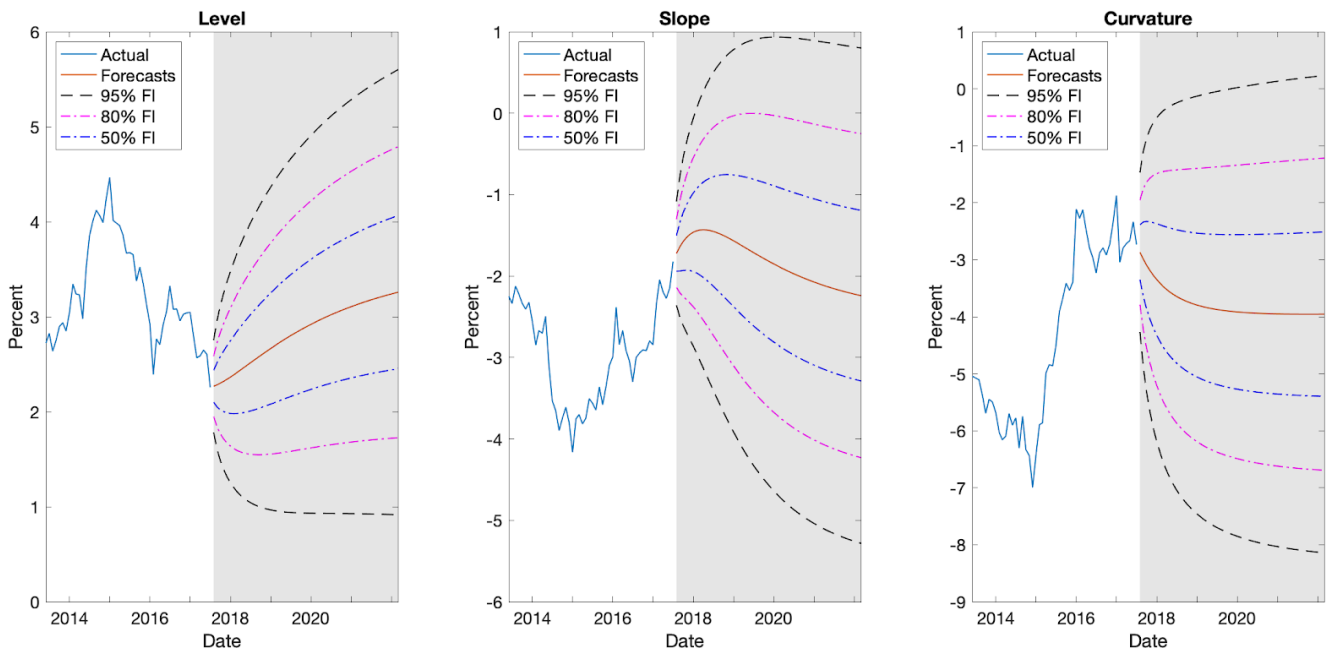

*Fig 6. 56-month ahead path forecasts and their respective 95, 80, and 50 percent forecast intervals of the level, slope, and curvature factors from VAR(1) of the two-step Diebold-Li approach*

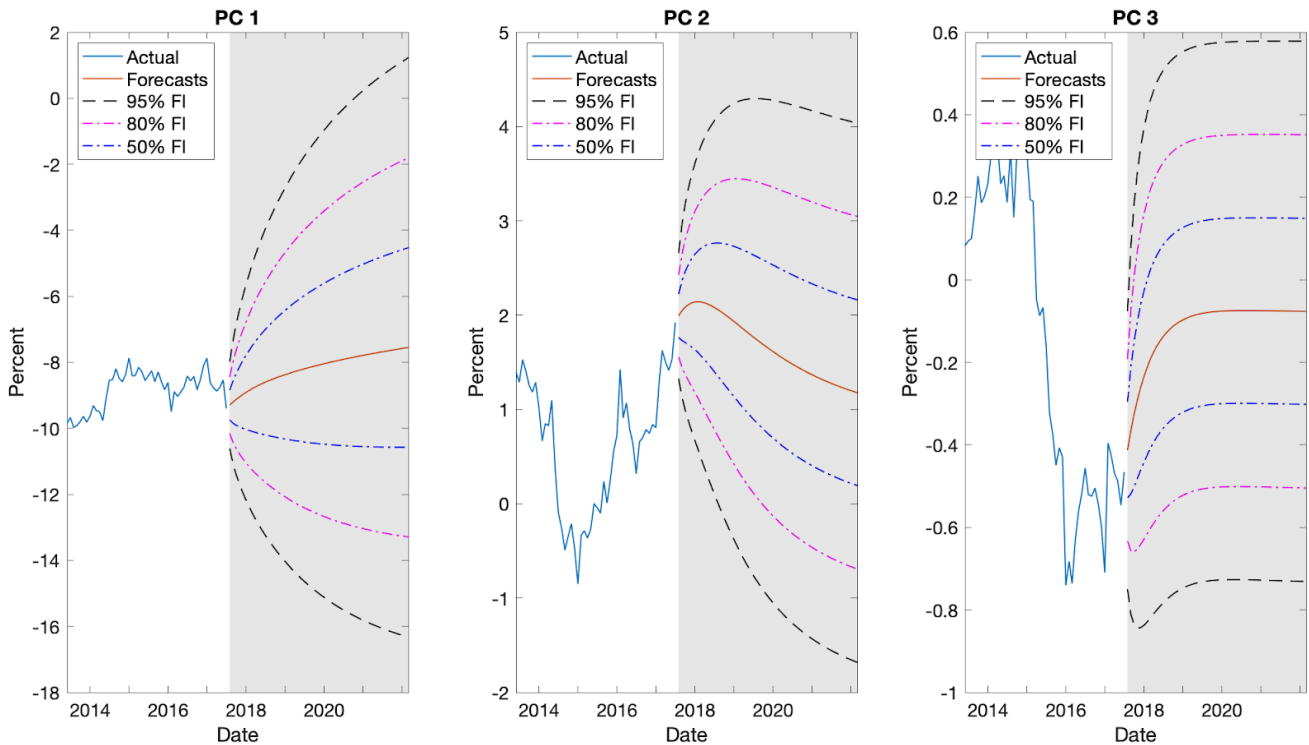

Fig 7. 56-month ahead path forecasts and the 95, 80, and 50 percent forecast intervals of the first three PCs from VAR(1) of the two-step principal components approach

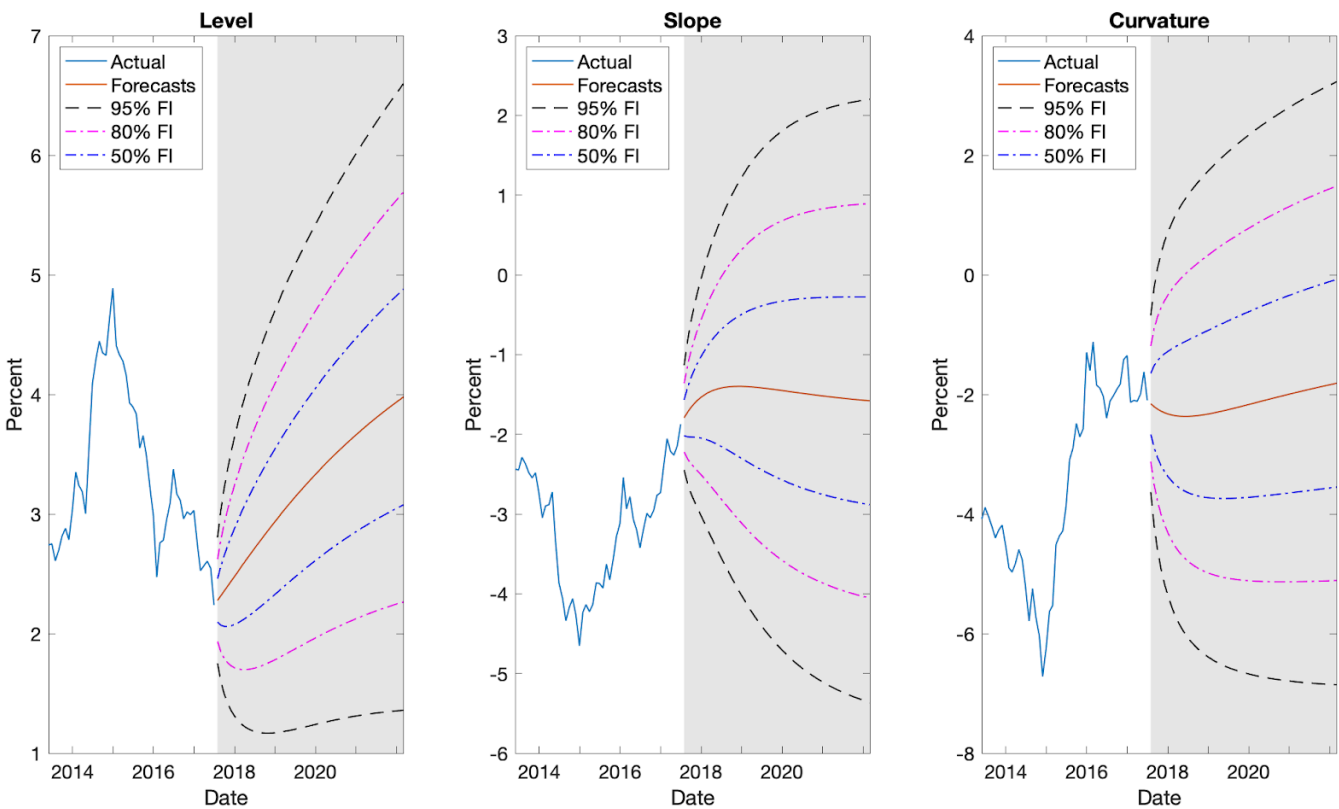

Fig 8. 56-month ahead path forecasts of the latent states from the state transition matrix in the one-step Kalman filter approach and their respective 95, 80, and 50 percent forecast intervals

Finally, table 4 contrasts the standard deviations and means of the residuals of the measurement equation in the one-step approach and those from the first step in the two-step approaches and expresses the results in basis points (bps). Also, I forecasted the paths of the Treasury yields of all 10 maturities in the test set (pseudo- out-of-sample forecasts) using the latent factors estimated from all three frequentist methods. Columns $5-7$ in table 4 shows the mean squared forecast errors ($MSFE$) for all maturities with a few noteworthy observations. Firstly, the $MSFEs$ from the one-step Kalman filter approach are consistently higher for all maturities relative to those from the two-step approaches. The two-step Diebold-Li model beats the two-step $PC$ model for maturities $3, 6, 12, 60,$ and $360$ months due to its superior forecasting performance. Additionally, both the two-step approaches most accurately forecast the longest-term maturity compared to other maturities. This starkly contrasts with yields maturing in 1 to 5 years where errors are higher.

| Maturity (months) | One-Step Kalman Filter | Two-Step Diebold-Li | Two-Step PC | $MSFE$ from One-Step Kalman Filter | $MSFE$ from Two-Step Diebold-Li | $MSFE$ from Two-Step PC |
|---|---|---|---|---|---|---|
| 3 | −6.9024 (12.4976) | −6.4737 (8.1651) | 0.0000 (7.4427) | 1.1449 | 0.9040 | 0.9293 |
| 6 | 0.0000 (0.0000) | 2.8726 (4.4083) | 0.0000 (4.0903) | 1.1457 | 0.8727 | 0.8809 |
| 12 | −0.4479 (8.8588) | 4.4158 (10.2091) | 0.0000 (7.8747) | 1.2163 | 1.0574 | 1.0680 |
| 24 | 3.2340 (5.3119) | 6.3460 (6.0414) | 0.0000 (5.8387) | 1.3613 | 1.1928 | 1.1762 |
| 36 | 0.0000 (0.0000) | −0.5570 (3.5231) | 0.0000 (3.4649) | 1.4932 | 1.0333 | 1.0238 |
| 60 | −0.3473 (6.9423) | −5.8002 (7.6496) | 0.0000 (5.7447) | 1.6587 | 1.1248 | 1.1350 |
| 84 | 1.9235 (6.4895) | −4.6757 (8.7601) | 0.0000 (6.8503) | 1.6604 | 0.9453 | 0.9448 |
| 120 | −1.6243 (3.8546) | −7.2047 (6.6086) | 0.0000 (5.9139) | 1.6812 | 0.8341 | 0.8310 |
| 240 | 10.8117 (18.0057) | 8.8807 (10.5507) | 0.0000 (10.0965) | 1.4457 | 0.7922 | 0.7897 |
| 360 | 2.6889 (15.7415) | 2.1962 (10.0682) | 0.0000 (8.1376) | 1.3358 | 0.7425 | 0.7472 |

*Table 4. Means and standard deviations (in parenthesis) of the residuals from the measurement equations from the three methods for yields of all 10 maturities; MSFE of the test set yields from the three frequentist methods in the last three column*

Though the one-step approach fits better for maturities ranging from 6 months to 10 years, the deviation is extremely high for the shortest (3 month) and long term maturities (20 and 30 years). These extreme values in standard deviation push the average value of the standard deviation of the residuals from one-step approach larger (7.7282), than that from the two-step Diebold-Li approach (7.5984), and the two-step $PC$ approach (6.5454). So, the yield curve fitted with the constant $\lambda$ from the two-step approaches generates more accurate results, and hence, is preferable to the one-step approach.

## 5.2. Bayesian Methods

A caveat of frequentist $VARs$ is that they are heavily parameterized as the total number of parameters to estimate rise proportionally with the number of lags and increase geometrically with the number of variables. Applying such richly parameterized models on small datasets (such as this monthly Treasury yields data) can cause noise to influence the estimates instead of the signal, overfitting the data. To counteract this drawback, I imposed restrictions or a-priori information to curtail the dimension of the parameter space. Bayesian methods shrink the coefficients, albeit not necessarily to 0. As a selection device, it selects regressors based on the magnitude of valuable information that each regressor contains. If the information is valuable, the likelihood of a model will significantly rise. By imposing informative prior information, the model accounts for the uncertainty about the model's parameters via a probability distribution for the vector of parameters. This lessens the risk of overfitting and the signal drives the estimates of the parameters, as opposed to the noise.

Hence, I estimated Bayesian $VARs$ on the latent factors by choosing prior distributions of the parameters and calculating the posterior distribution of the estimates to forecast the yields. Fitting the model on the train set, I separately evaluated each model's performance on the test set using the mean squared forecast error as the loss function. To obtain the posterior parameters, I used a recursive Monte Carlo method for numerical integration, known as the Gibbs sampler, in the independent normal − Wishart, and $SSVS$ priors. In frequent methods, maximizing all the parameters in large scale models using a likelihood function can be computationally inefficient. This contrasts with the Bayesian models wherein we sequentially draw parameters from their conditional posterior distributions, which entails using fewer parameters to approximate the marginal posterior distributions. Sections 5.2.5, and 5.2.6 present the results.

The $VAR(13)$ of latent factors after extending $VAR(1)$ from equation (7) is

$$f_t = a_0 + A_1 f_{t-1} + \ldots + A_{13} f_{t-13} + \eta_t \tag{9}$$

where $f_t$ is a $(3 \times 1)$ vector of latent factors. Let $k$ be the number of variables, so $k = 3$, creating $(3 \times 3)$ matrices $A_1, \ldots, A_{13}$ of coefficients. Thus, there are three equations, and each equation has $r = 1 + kp = 40$ regressors, where $p = 13$ is the lag length. So, there are a total of $kr = 120$ $VAR$ coefficients. We can also stack $VAR(13)$ in (9) over the time period $t = 1, 2, \ldots, T$ in two ways.

In the first representation, $f_t = \Omega_t A + \eta_t$,

where $\Omega_t = I_3 \otimes [1, f_{t-1}', ..., f_{t-13}']$ is a $(3 \times 120)$ matrix;

$A = vec([a_0, A_1, ..., A_{13}]')$ concatenates the coefficients into a $(120 \times 1)$ vector.

Then, I stacked $f = [f_1', ..., f_T']'$ to get $f = \Omega A + \eta$, (10)

where $\Omega = [\Omega_1', ..., \Omega_T']'$ is a $(Tk \times kr)$ or a $(3T \times 120)$ matrix of latent factors as regressors; $f$ has dimension $(3T \times 1)$, and $\eta \sim N(0, I_T \otimes \Sigma)$.

In the second representation, I stacked the latent factors into a $(T \times 3)$ matrix $F$ so that its $i^{th}$ row is $f_t'$

Let $G$ be a $(T \times 40)$ matrix of regressors, where the $t^{th}$ row is $\omega_t' = [1, f_{t-1}', ..., f_{t-13}']$

Let $\Phi = [a_0, A_1, ..., A_{13}]'$ be a $(40 \times 3)$ matrix of VAR coefficients, then we can recast $VAR(1)$ as:

$F = G\Phi + U$ (11)

where $U$ is a $(T \times 3)$ dimensional matrix of innovations wherein $\eta_t'$ is the $t^{th}$ row.

Equation (10) is an analogous representation of equation (11) because: $f = vec(F')$, $A = vec(\Phi)$, $\eta = vec(U')$

Hence, $vec(U) \sim N(0, \Sigma \otimes I_T)$

### 5.2.1 Minnesota Prior

Using the level, slope, and curvature estimates from the first step in the two-step Diebold-Li approach, I applied the Minnesota prior to the first representation of $VAR$ given in equation (10). This presupposes the prior belief that the latent variables follow $AR(1)$ processes. So, the prior for the coefficients in $A$ is $A \sim N(A_{mp}, \Sigma_{mp})$. Rather than estimating the variance covariance matrix of the errors in the $AR(1)$ processes $-\Sigma$, I assumed a fixed estimate, $\hat{\Sigma}$, and thereby only calibrated the coefficient matrix $A$. Accordingly, I first estimated the three above $AR(1)$ equations separately. Then, I equated $\hat{\Sigma} = diag(\hat{\sigma}_{11}^2, \hat{\sigma}_{22}^2, \hat{\sigma}_{33}^2)$, where $\hat{\sigma}_{ii}^2$ is the $OLS$ estimate of the variance of the residuals the $i^{th}$ random walk equation, also the $ii^{th}$ element of $\hat{\Sigma}$. The downside is that it ignores the uncertainty about the parameters, yielding suboptimal density forecasts as we don't integrate out the unknown parameters. Nonetheless, the upside is that this shortens the computation process as we don't have to use Markov Chain Monte Carlo to calibrate the posterior parameters, and only have to estimate the matrix $A$. The Minnesota prior prespecifies empirically relevant values of the prior mean $A_{mp}$ and prior variance $\Sigma_{mp}$ where

$$A_{mp} = \begin{cases} 0.95, & \text{for } a_{ii} \text{ i.e. the coefficient associated with the first lag} \\ 0, & \text{for } a_{ij}, \text{ where } i \neq j \end{cases}$$

The prior mean of $A_{mp}$ expresses the belief that the latent factors are very persistent, as also corroborated from the autocorrelation plots in figure 2. Thus, modeling the paths of the latent factors with $AR(1)$ processes might be suitable. This model slightly deviates from the Minnesota prior proposed by Litterman (1986), who had created a strict random walk with draft process by shrinking the coefficients towards 1, instead of 0.95. The Minnesota prior assumes a diagonal prior variance-covariance matrix $\Sigma_{mp}$ such that the three hyperparameters (which control the prior's tightness) – $d_1$, $d_2$ and $d_3$ determine the diagonal elements of $\Sigma_{mp}$. [4] The hyperparameters to construct the prior covariance are set to $d_1 = 0.001$, $d_2 = 0.001$, and $d_3 = 100$. Given the first representation $f = \Omega A + \eta$, and the normal prior $A \sim N(A_{mp}, \Sigma_{mp})$, estimating the model with linear regression results in the posterior distribution: $A | f \sim N(\hat{A}, \Sigma_A^{-1})$, where $\hat{A}$ is the posterior mean,

$$\hat{A} = \Sigma_A^{-1}(\Sigma_{mp}^{-1} A_{mp} + \Omega'(I_T \otimes \hat{\Sigma}^{-1})f), \quad \Sigma_A = \Sigma_{mp}^{-1} + \Omega'(I_T \otimes \hat{\Sigma}^{-1})\Omega,$$

and I substituted $\Sigma$ with its estimate, $\hat{\Sigma}$; $\Sigma_A$ is a $(kr \times kr)$ or $(120 \times 120)$ precision matrix, and its inverse $\Sigma_A^{-1}$ is the variance-covariance matrix.

### 5.2.2 Natural Conjugate Prior

Unlike the Minnesota prior which fixes $\Sigma$ as a known parameter, the natural conjugate prior estimates both the unknown parameters: the $VAR$ coefficients in $A$, and the error variance-covariance matrix, $\Sigma$. Given the likelihood function in the equation (A.2) in the Appendix, the $VAR$ coefficients is a $(40 \times 3)$ matrix $\Phi$. The natural conjugate prior specifies a joint distribution for $(vec(\Phi), \Sigma)$, wherein I considered a normal inverse – Wishart prior on $(\Phi, \Sigma)$:
$(vec(\Phi), \Sigma) \sim NW^{-1}(s_{h0}, S_{c0})$, where $s_{h0}$ and $S_{c0}$ are the prior values of the shape parameter (or degrees of freedom) $s_h$, and scale matrix $S_c$, respectively. The unconditional prior distribution of $\Sigma$, and the conditional normal inverse – Wishart prior distribution of $vec(\Phi)$ are:

$$\Sigma \sim W^{-1}(s_{h0}, S_{c0}), \quad (vec(\Phi) | \Sigma) \sim N(vec(\Phi_0), \Sigma \otimes \Sigma_\Phi)$$

I subjectively defined the initial values of the informative natural conjugate hyperparameters as:
$vec(\Phi_0) = 0_{(rk \times 1)} = 0_{(120 \times 1)}, \quad s_{h0} = k + kr = 123, \quad S_{c0} = diag(\hat{\sigma}_{11}^2, \hat{\sigma}_{22}^2, \hat{\sigma}_{33}^2), \quad d_1 = 0.03^2, \quad d_2 = 20^2$

This creates a $(3 \times 3)$ diagonal scale matrix $S_{c0}$, where the diagonal entries $\sigma_{kk}$ are $OLS$ estimates of the variances of the residuals from $AR(1)$ equations of the latent factors as done with the Minnesota prior. $d_1$, and $d_2$ are the hyperparameters linked to the diagonal matrix $\Sigma_\Phi$, where $\Sigma_\Phi$ shrinks the coefficients. Finally, I arbitrarily set the shape parameter to 123, as a larger value of $s_{h0}$ diminishes the prior variance of $\Sigma$, making the prior comparatively informative. This is because higher values of degrees of freedom more tightly centers the inverse – Wishart distribution around its prior mean. [5]

---

[4] Section A.2.2 in the appendix displays the structure of the random walk equations, the error variance covariance matrix and elucidates the role of these hyperparameters.
[5] I analytically computed the posterior predictions from the posterior distribution described in section A.2.3.

### 5.2.3 Independent Normal and Inverse – Wishart Prior

The downside of conjugate priors is that the $VAR$ coefficients' prior variances have restrictive forms as we can analytically find the marginal posterior distributions. Therefore, I resorted to a more flexible prior on the joint distribution of $(A, \Sigma)$, known as the independent normal and inverse – Wishart prior. This assumes that $A$ and $\Sigma$ are orthogonal to each other, i.e. $P(A, \Sigma) = P(A) P(\Sigma)$. Using the first formulation of the $VAR$ in equation (10), the form of the prior distribution is:

$$A \sim N(A_{(0)}, V_A), \quad \Sigma \sim W^{-1}(s_{h0}, S_{c0})$$

I set the initial values of the hyperparameters to:

$$A_{(0)} = 0_{(120 \times 1)}, \quad s_{h0} = 123, \quad S_{c0} = diag(\hat{\sigma}_{11}^2, \hat{\sigma}_{22}^2, \hat{\sigma}_{33}^2), \quad d_1 = 0.03^2, \quad d_2 = 0.04^2, \quad d_3 = 20^2$$

As the joint posterior distribution $P(A, \Sigma | f)$ is convoluted, we cannot analytically derive the marginal posterior distributions. So, I approximated them by sequentially drawing samples of each random variable from their univariate conditional posterior distributions, one variable at a time. Using a Gibbs sampler created by Koop and Korobilis (2010), I derived the conditional normal distribution $P(A | f, \Sigma)$, and inverse – Wishart distribution $P(\Sigma | f, A)$. Let the vector of parameters be $\Gamma = \{\Sigma, A\}$. After initializing the algorithm with random values, the simulated samples in the initial iterations may not always be representative of the target posterior distribution. Because the target joint posterior distribution is the stationary distribution, I ran the algorithm for a large number of iterations to ensure that the values in the Markov chain converge to the target posterior distribution. Burn-in samples allow the Markov chain to reach its equilibrium or stationary distribution if the starting point is not suitable (as typically samples from the initial iterations are not from the target posterior distribution). Discarding the initial samples or draws (that are not stationary) precludes oversampling regions with low probabilities. With an initial value of $A$ known as $A_{(0)}$, I started the Gibbs sampler and obtained $\Sigma_{(0)}$ by creating a random variable from the conditional distribution $P(\Sigma | A_{(0)}, f)$. Thereafter, the Gibbs sampler generates a new random variable for $A : A_{(1)}$ by using the observed $\Sigma_{(0)}$. That is, the sampler draws from the conditional distribution $P(A | \Sigma_{(0)}, f)$ after observing $\Sigma_{(0)}$, and the process follows:

$\Sigma_{(1)}$ from $P(\Sigma | A_{(0)}, f)$

$A_{(1)}$ from $P(A | \Sigma_{(1)}, f)$

$\Sigma_{(2)}$ from $P(\Sigma | A_{(1)}, f)$

$A_{(2)}$ from $P(A | \Sigma_{(2)}, f)$

.
.
.

$\Sigma_{(s)}$ from $P(\Sigma | A_{(s-1)}, f)$

$A_{(s)}$ from $P(A | \Sigma_{(s-1)}, f)$

The vectors $\Gamma_{(s)} = \{\Sigma_{(s)}, A_{(s)}\}$ create a Markov chain. Repeating the process $s = 500,000$ times creates a Gibbs sequence of that length. One cycle of the Gibbs sampler completes after iterating over the two univariate distributions. Out of the $500,000$ samples, I stored the last $100,000$ sample points wherein each point refers to a vector of the two parameters. In the process, I discarded the first $400,000$ samples as "burn-in" samples to eliminate the impact of initial sampling estimates. Then, the Markov chain from the sampler converges to the equilibrium or stationary distribution which doesn't depend on the starting values. The empirical distribution of the last $100,000$ draws of each parameter in $\Gamma$ approximates the marginal posterior distributions of the parameters in $\Gamma$. Finally, these retained draws help to compute the $h-$ step ahead forecasts, and impulse response functions of the latent factors.

### 5.2.4 Stochastic Search Variable Selection (SSVS)

Rather than preselecting restrictions in overparameterized models, I employed restrictions that the data supports by building a hierarchical model. This hierarchical model evaluates each restriction as a submodel and utilizes priors to categorize uncertainty around the model's parameters. Requiring minimal input of prior values, the $SSVS$ prior is a more automatic technique of shrinking the coefficients. Although the previous Bayesian methods focus on entirely shrinking the matrices of coefficients, and the residual standard error, this method emphasizes on each element of those matrices. Unlike the $BVARs$ with the previous priors, I applied this prior on $BVARs$ devoid of intercepts: $F_0 = G_0 \Phi_0 + U_0$. As I restricted all the coefficients of $A_0$, I constructed hierarchical priors and established hyperparameters to control these priors. To select the prior hyperparameters, I applied the "default semi-automatic approach" described in George et. al (2008) and Koop and Korobilis (2009). To substantiate, I divided the elements of $A_0$ into two groups: the elements in the first group are strongly shrunk to $0$ as opposed to those in the second group. Thereby, the first group performs the "variable selection" aspect of the procedure by setting the coefficients close to 0, and then only selects those variables in the second group that are unchanged. For that, the $MCMC$ sampler stochastically partitions in each iteration. Let the indicator variable, $\gamma_i \in \{0,1\}$, so that $\gamma = [\gamma_1, \gamma_2, ..., \gamma_{39}]'$. Thereafter, assuming that all coefficients of $A_0$ are independent,[6] each element $A_{0i}$ has a two-part mixture distribution:

$$(A_{0i} | \gamma_i) \sim (1 - \gamma_i) N(0, \sigma_{0i}^2) + \gamma_i N(0, \sigma_{1i}^2), \tag{12}$$

If $\gamma_i = 0$, then we draw $A_{0i}$ from the first normal distribution; thus $A_{0i}$ is starkly shrunk towards 0. Alternatively, if $\gamma_i = 1$, then we draw $A_{0i}$ from the second normal distribution and $A_{0i}$ is comparatively uninformative. The linear combination in equation (21) determines each $\gamma$ 's effect on the prior of $A_{0i}$. Moreover, the prior for the restricted elements in $A_0: (A_0 | \gamma) \sim N_{117}(0, \Sigma_\gamma M \Sigma_\gamma')$, where $M$ is a pre-chosen correlation matrix and $\Sigma_\gamma$ is a diagonal matrix and each diagonal element is defined by: $(1 - \gamma_i) \sigma_{0i}^2 + \gamma_i \sigma_{1i}^2$, $i = 1, ..., 117$.

---

[6] The assumption that the elements of the coefficient matrix are independent of each other may be questionable, particularly since we know that the latent factors are highly autocorrelated. Therefore, we expect larger diagonal entries, and much smaller off-diagonal entries of the coefficient matrix. However, I assumed this to simplify calculations.

In other words,

$$[\Sigma_\gamma]_{i,i} = \begin{cases} \sigma_{0i}, \text{if } \gamma_i = 0 \\ \sigma_{1i} \text{ if } \gamma_i = 1 \end{cases}$$

This apriori's the hyperparameters $\sigma_{0i}^2 \to 0$, and $\sigma_{1i}^2 \to \infty$. In the former case, the prior variance of $A_{0i}$ is very low, which restricts $A_{0i}$'s posterior means to shrink towards 0. In the latter case, $A_{0i}$ is unrestricted, and the likelihood determines the posterior mean. Before choosing the value of the indicator variable $\gamma_i$, I assumed that that the elements of $\gamma$ are independent of one another and sampled each $\gamma_i$ from the Bernoulli density. Consequently, the prior distribution of $\gamma_i$ is $\gamma_i \sim Bernoulli(p_i), i \neq j$, where $P(\gamma_i = 1) = p_i$, and $P(\gamma_i = 0) = 1 - p_i \ \forall \ i = 1, ..., 39$. $p_i$ mirrors the prior belief that the coefficients $A_{0i}$ is unrestricted.

Moreover, I set $\sigma_{0i} = c_0 \sqrt{\widehat{Var}(A_{0i})}$, and $\sigma_{1i} = c_1 \sqrt{\widehat{Var}(A_{0i})}$, wherein $\widehat{Var}(A_{0i})$ is the estimated variance of the coefficient $A_{0i}$ in the unrestricted $VAR$ model. I set $c_0 = 0.01$, $c_1 = 20$, $s_{h0} = k + kr = 123$, and $S_{c0} = I_3$. To preclude the possibility of restricting an element $A_{0i}$ by chance, which yields an uninformative prior when $p_i = 0.5$, I fixed $p_i = 0.2$. To ensure that the elements $A_{0i}$ are apriori independent, I equated the correlation matrix to the identity matrix: $M = I$.

After specifying the priors, I sampled from the posterior distributions. First, I fitted the model with the $SSVS$ prior on only the $A_0$ coefficients, with an $inverse - Wishart$ prior for $\Sigma_0$. Then, I applied the $SSVS$ prior on both $A_0$, and the error variance-covariance $\Sigma_0$ matrices in a separate model. The Gibbs sampler entails normal, $inverse - Wishart$, and Bernoulli distributions to formulate posterior inferences from the model. Therefore, I presupposed an $inverse - Wishart$ prior for $\Sigma_0 : \Sigma_0 \sim W^{-1}(s_{h0}, S_{c0})$.

### 5.2.5 Results of BVARs With Latent Factors

After obtaining the draws of $\Gamma_{(s)} \forall \ s = 1, 2, ..., 500,000$, I calibrated the mean squared forecast errors, log predictive likelihoods, and the distribution of posterior predictive latent factors: predicative means and standard deviations from each of the priors. Table 5 shows the results of the out-of-sampling predictive distribution of the latent variables i.e. the point forecasts for the month of July 2016, which is the first observation in the test set. These are derived from the predictive density $P(f_{T+1}|f_1, f_2, ..., f_T)$, where $T$ is June 2016. In imposing the diffuse or the uninformative prior only, I fit $VAR(9)$ as $p = 9$ is the lag number which maximizes Akaike information criteria $(AIC)$.

| Latent Factors | Diffuse | Minnesota | Natural Conjugate | Independent Normal Inverse Wishart | Partial $SSVS$ prior | Full $SSVS$ prior | Actual Values $f_{T+1}$ |
|---|---|---|---|---|---|---|---|
| $L_{T+1}$ | 2.7482 (0.2320) | 2.6284 (0.2286) | 4.3528 (0.7200) | 2.5978 (0.2047) | 2.5926 (0.2008) | 2.5943 (0.2153) | 1.9817 |
| $S_{T+1}$ | −2.2649 (0.2969) | −2.1045 (0.2856) | −2.1185 (0.4051) | −2.0644 (0.2490) | −2.0367 (0.2799) | −2.0477 (0.2977) | −1.6292 |
| $C_{T+1}$ | −3.2053 (0.6652) | −2.3819 (0.6415) | −2.5644 (0.7827) | −2.3323 (0.5640) | −2.4254 (0.5914) | −2.3258 (0.6706) | −3.8587 |
| Log Predictive Likelihood | −0.1215 | 0.4571 | −0.0697 | −0.1263 | −0.3607 | −0.5176 | |

Table 5. Log predictive likelihood, and predictive means and standard deviations (in the parenthesis) from BVARs with each of the six priors

From table 6, the $BVAR$ with the partial $SSVS$ prior generates minimum mean squared forecast error of level (0.3737), slope (0.18945), and the $BVAR$ with the natural conjugate prior minimizes the error for the curvature factor (1.6751).

| Latent Factors | Diffuse | Minnesota | Natural Conjugate | Independent Normal Inverse Wishart | Partial $SSVS$ prior | Full $SSVS$ prior |
|---|---|---|---|---|---|---|
| $L_{T+1}$ | 0.57936 | 0.41822 | 0.38687 | 0.37964 | 0.3737 | 0.37528 |
| $S_{T+1}$ | 0.40415 | 0.22589 | 0.23949 | 0.18945 | 0.16607 | 0.17519 |
| $C_{T+1}$ | 1.42682 | 2.1807 | 1.6751 | 2.3297 | 2.0544 | 2.3495 |

Table 6. Mean square forecast error of the latent factors

Next, I graphed the impulse response functions ($IRFs$) that delineate how the latent factors in $f_t$ react through time when faced with a one-time innovation. To construct these nonlinear functions of the $BVAR$ coefficients, and error variance-covariance matrices, I used $MCMC$ simulations. In calculating the impulse responses, I triangularized the $BVARs$ by decomposing $\Sigma_\eta : \Sigma_\eta = ZZ'$, where $Z$ is the lower Cholesky factor. So, the first variable $L_t$ is not sensitive to a contemporaneous innovations to the other two variables succeeding in the system in $f_t$, namely $S_t$, and $C_t$. Alternatively, contemporaneous innovations to latent factors preceding $C_t$ impact $C_t$. Hence, the recursive causal order of the variables in the orthogonal impulse response graphs are pertinent as the order influences the estimates. The red-colored lines depict the posterior median of the latent factors through time and the blue and yellow colored bands are the $5^{th}$, and $95^{th}$ percentiles, respectively. The zig-zags in $IRFs$ from the independent normal inverse − Wishart prior might signal that the models have overfit the train set due to a large number of parameters. In this case, those parameters try to capture the idiosyncrasies in the train set. These juxtapose with the very smooth $IRFs$ from the Minnesota and natural conjugate prior, and the less smooth from those of the diffuse prior. The drawback is that we cannot interpret the $IRFs$ in figure $9-15$ as innovations to the term structure factors arise due to disturbances in exogenous macro shocks, and not directly to the factors themselves.

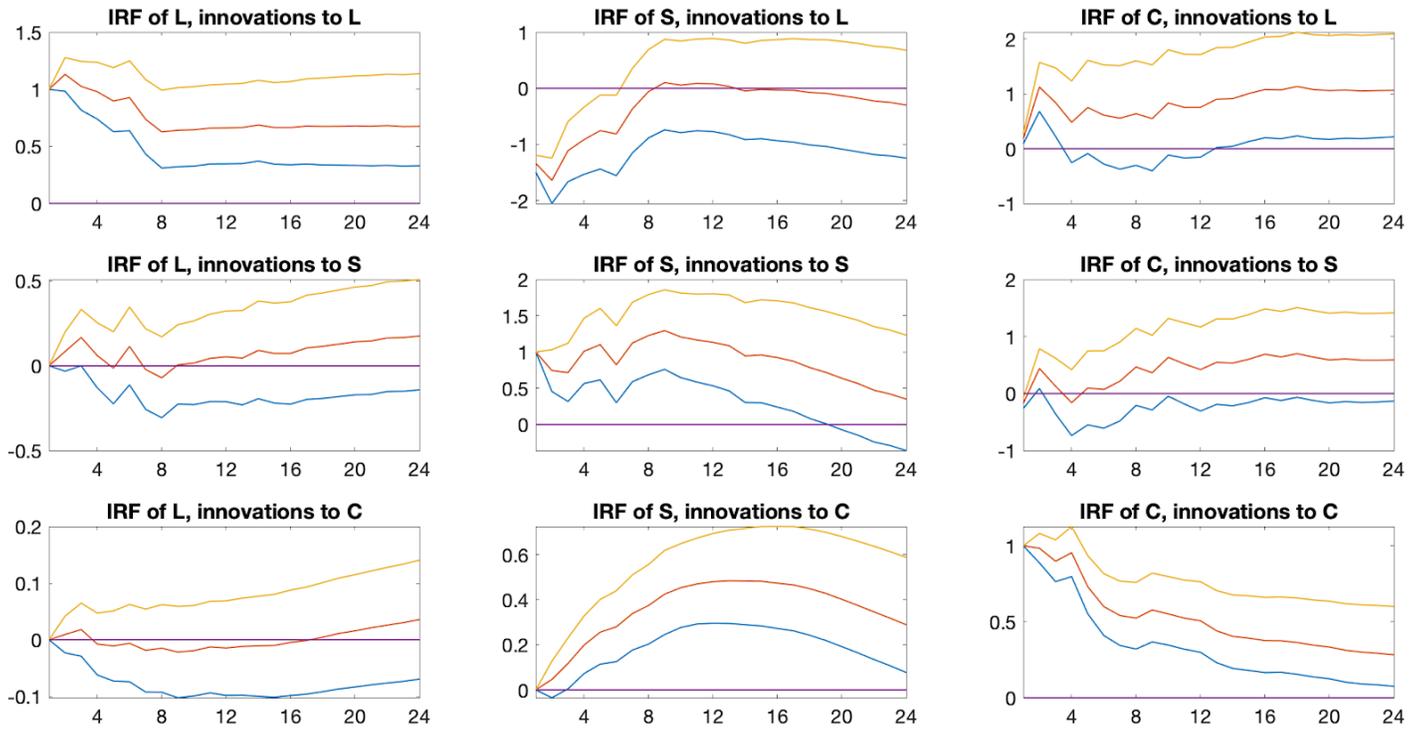

*Fig 9. IRFs of the latent factors under the diffuse prior*

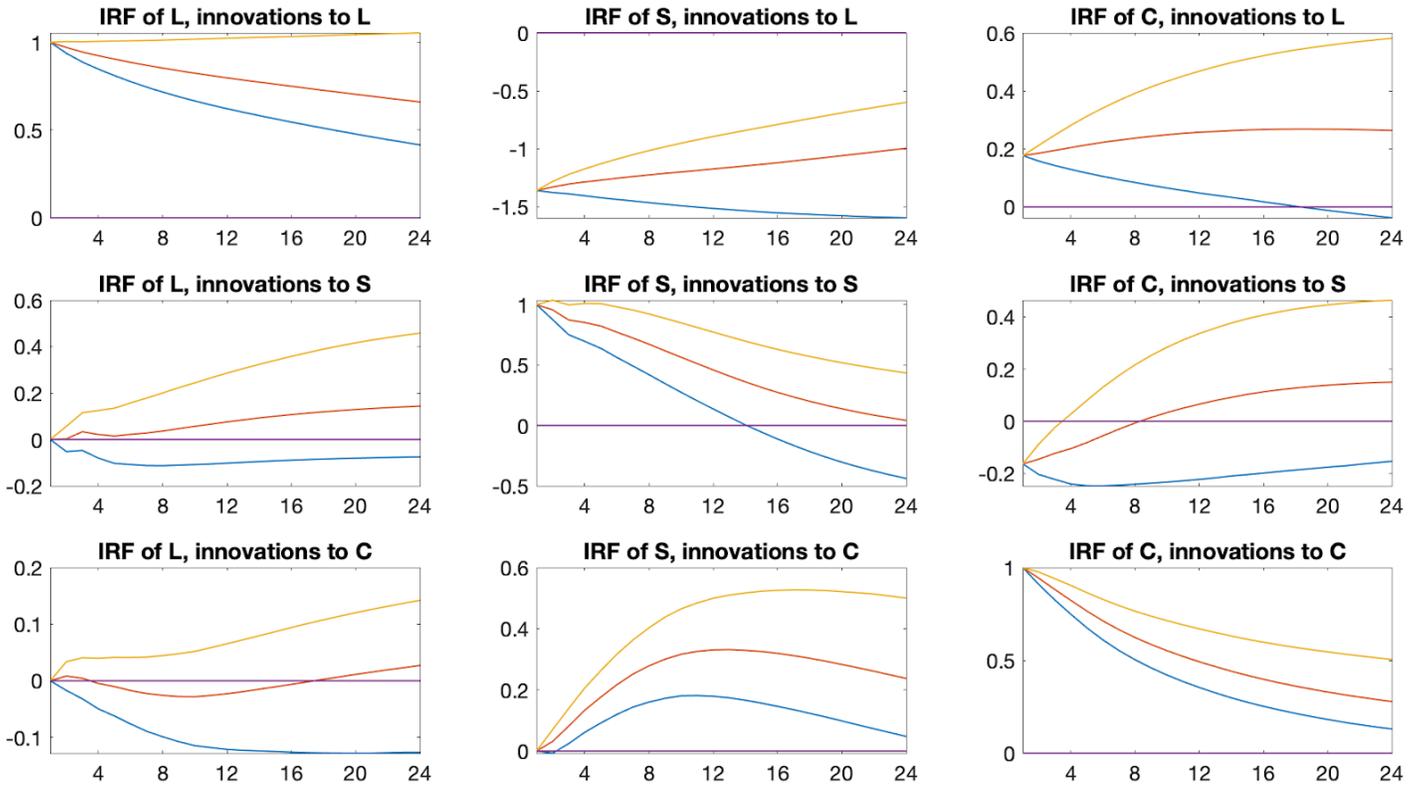

Fig 10. IRFs of the latent factors under the Minnesota prior

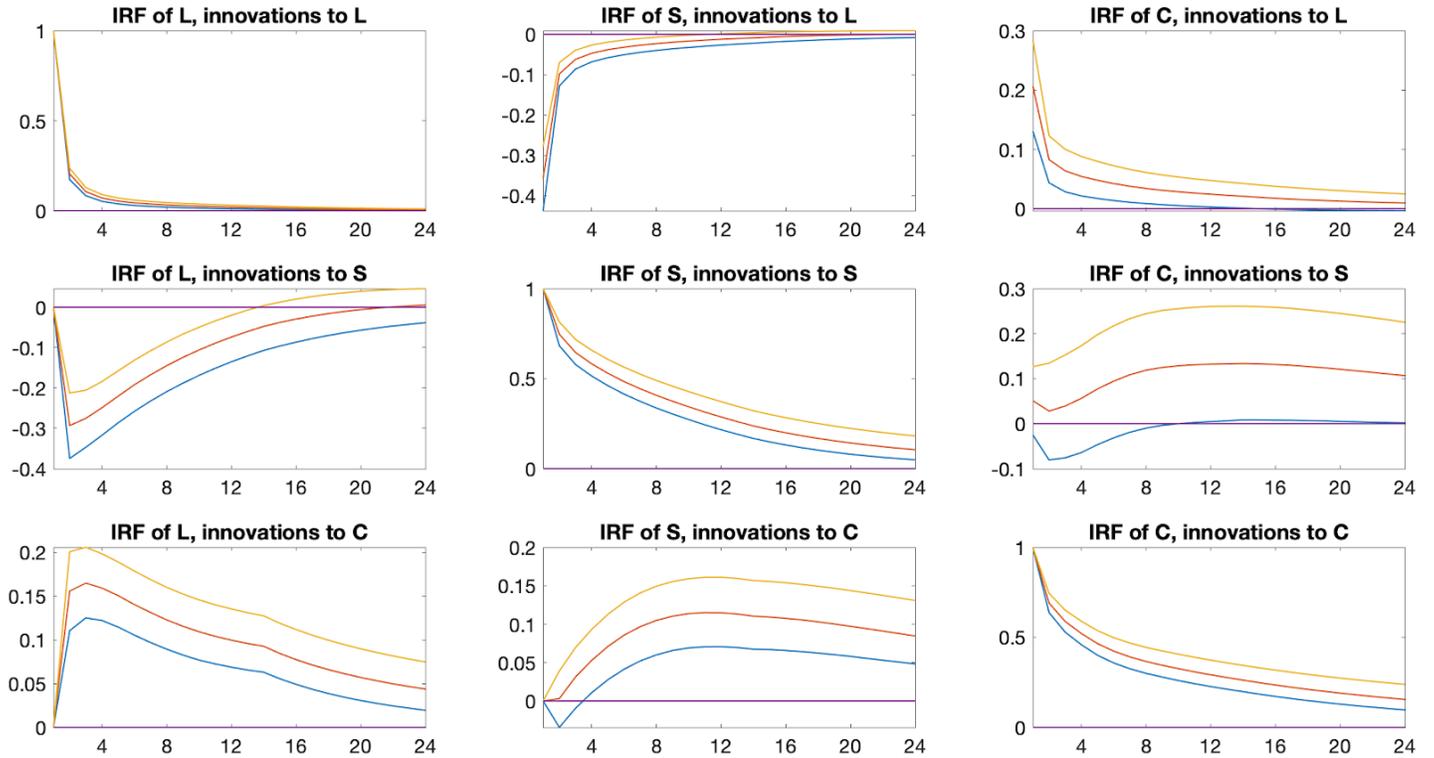

Fig 11. IRFs of the latent factors under the natural conjugate: normal Wishart prior

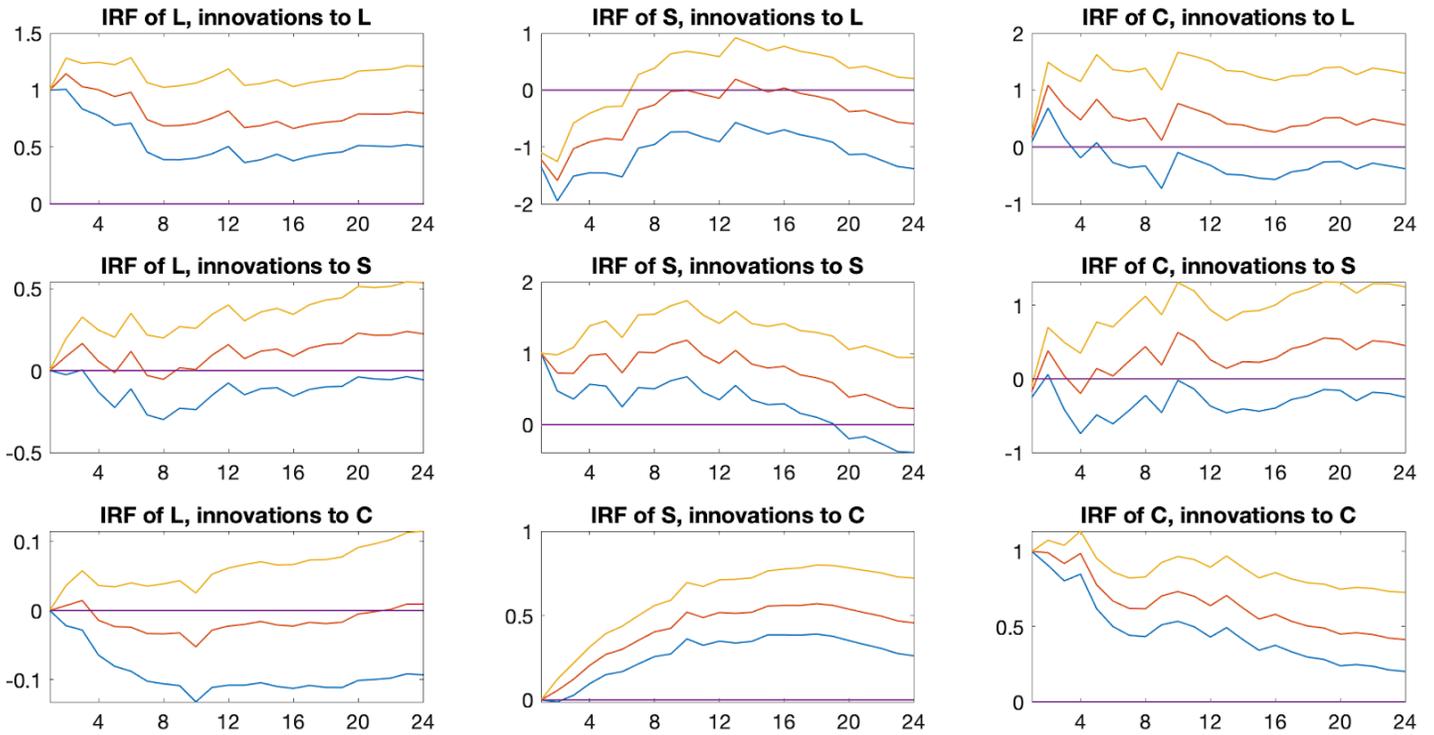

Fig 12. IRFs of the latent factors under the independent normal inverse – Wishart prior

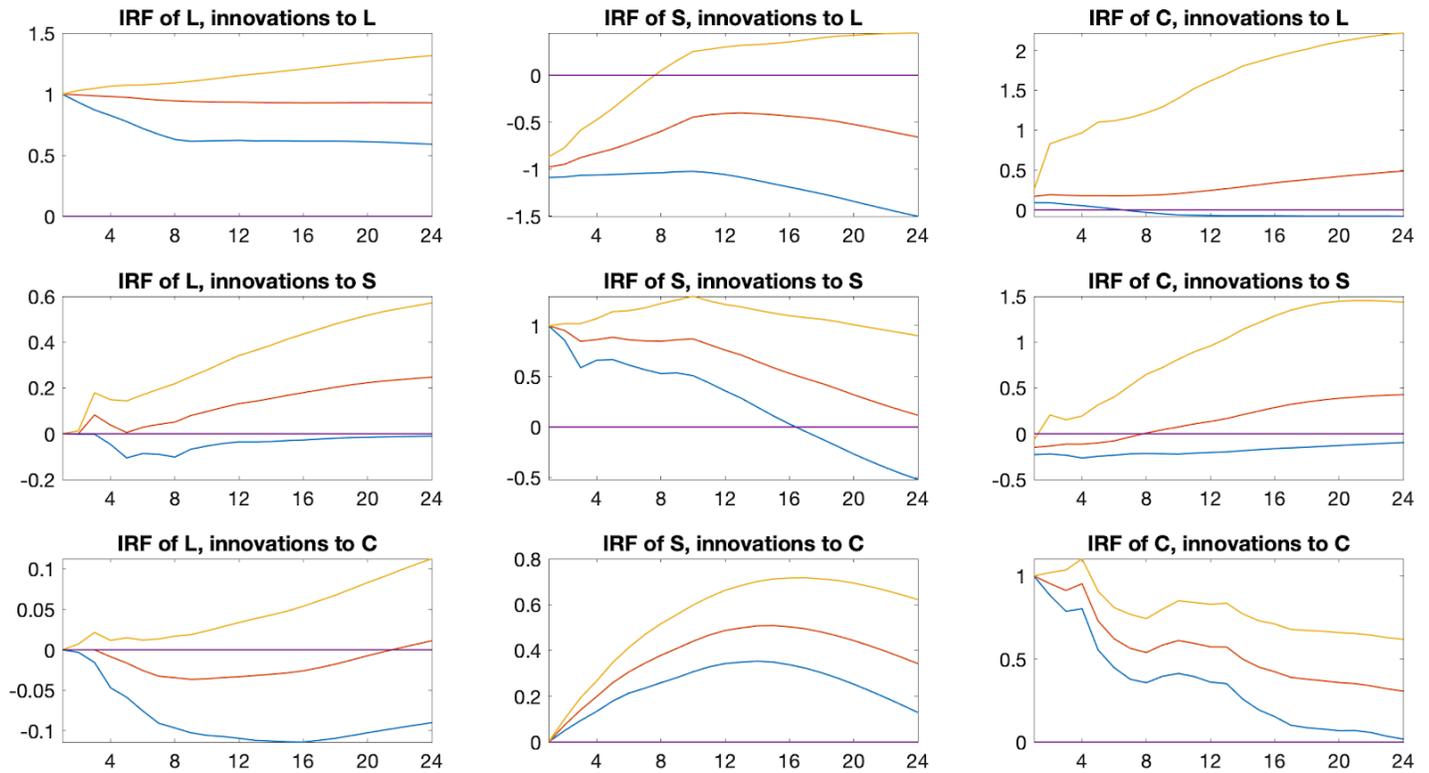

Fig 13 . IRFs of the latent factors under the SSVS prior on VAR coefficients, and inverse − Wishart prior on the error covariance matrix.

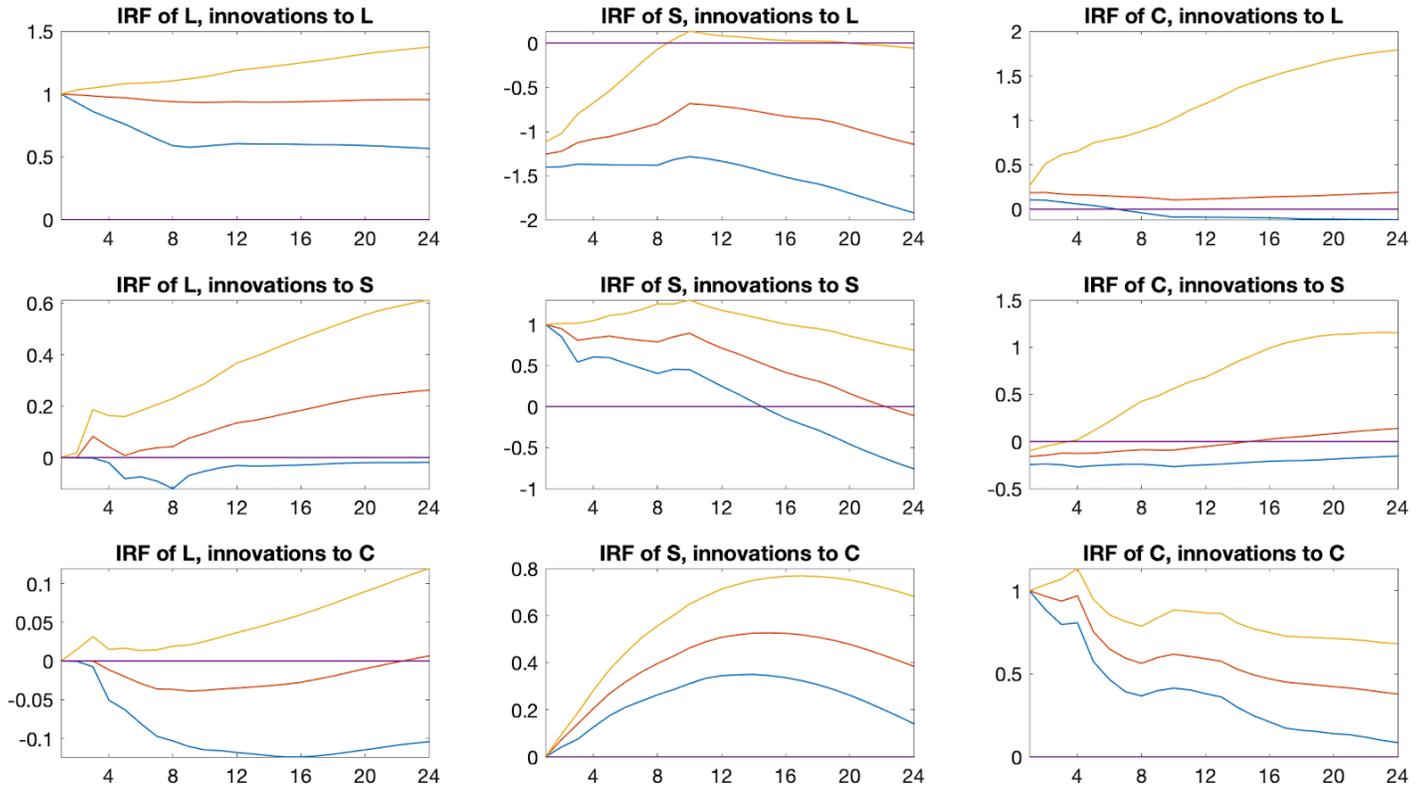

Fig 14. IRFs of the latent factors under the full SSVS prior (on both VAR coefficients and error covariance matrices)

Tables $7-14$ depict the $h-$step ahead forecast distributions and $MSFE$ for Treasury yields of all maturities. The out-of-sample forecast horizons are $h = 6, 12, 18, 24, 30, 36, 42, 48,$ and $54$ months. Overall, the predictive mean gradually rises as the term to maturity grows. Furthermore, the predictive standard deviation is broadly smaller for yields with long-term than short and medium-term maturities. For the 6-month horizon, the $BVARs$ with natural conjugate prior outperforms those from other priors as it generates the lowest $MSFE$ for most maturities. Mainly, the $BVARs$ with Minnesota prior yield the lowest $MSFE$ for all other time horizons. However, for later forecast horizons: $36, 42,$ and $54$ months ahead horizon, the vector autoregression with the independent normal inverse $-$ Wishart prior produces results that are somewhat comparable to those of the Minnesota prior, although these priors generate negative yields for short and medium term maturities not seen in shorter forecast horizons. Lastly, the forecast error is generally lower for shorter than for longer horizons as the degree of uncertainty rises when $h$ increases. However, the results from table $11$ indicate an exception to this case $-$ forecast errors for most maturities in all priors are commonly smaller for the $30$ month as opposed to the $24$ month forecast horizon.[7]

---

[7] I encountered a computational issue while forecasting yields of a few maturities for horizons greater than 6 months using the natural conjugate prior. So, I have not added the forecast distribution and MSFE for this prior in tables 9 to 16.

| Maturity | Minnesota | Natural Conjugate | Independent Normal Inverse Wishart | Partial $SSVS$ |
|---|---|---|---|---|
| 3 | 0.7173 (0.5176) | 1.7393 (0.6847) | 0.7668 (0.4163) | 0.7150 (0.3984) |
| 6 | 0.7996 (0.5591) | 1.8536 (0.7137) | 0.8358 (0.4561) | 0.8032 (0.4386) |
| 12 | 0.8837 (0.6293) | 1.9200 (0.7416) | 0.9695 (0.5137) | 0.9093 (0.4899) |
| 24 | 0.9697 (0.6739) | 2.0561 (0.7645) | 1.3023 (0.5306) | 0.7773 (0.5244) |
| 36 | 1.0936 (0.6488) | 2.2140 (0.7558) | 0.7668 (0.4163) | 1.1419 (0.5239) |
| 60 | 1.4134 (0.6671) | 2.5407 (0.7260) | 1.6874 (0.5423) | 1.4682 (0.5119) |
| 84 | 1.7094 (0.6209) | 2.8505 (0.6942) | 1.7158 (0.5014) | 1.6886 (0.4782) |
| 120 | 1.8817 (0.5595) | 3.0994 (0.6624) | 2.1261 (0.4530) | 1.9171 (0.4407) |
| 240 | 2.2552 (0.4721) | 3.6156 (0.6341) | 2.3147 (0.3811) | 2.1744 (0.3615) |
| 360 | 2.7047 (0.4634) | 3.7577 (0.6063) | 2.3025 (0.3746) | 2.3140 (0.3754) |

*Table 7. Predictive distribution of the 6-month ahead point forecasts of the Treasury yields*

| Maturity | Minnesota | Natural Conjugate | Independent Normal Inverse Wishart | Partial $SSVS$ |
|---|---|---|---|---|
| 3 | 0.25251 | 0.16058 | 0.23889 | 0.24008 |
| 6 | 0.37065 | 0.39739 | 0.40142 | 0.35061 |
| 12 | 0.50466 | 0.20503 | 0.50986 | 0.50507 |
| 24 | 0.66471 | 0.22363 | 0.75965 | 0.73288 |
| 36 | 0.77216 | 0.31164 | 0.83796 | 0.82136 |
| 60 | 0.88323 | 0.32991 | 0.86156 | 0.87474 |
| 84 | 0.79533 | 0.24585 | 0.75565 | 0.76745 |
| 120 | 0.56078 | 0.34713 | 0.54716 | 0.53057 |
| 240 | 0.31088 | 0.15523 | 0.26244 | 0.64196 |
| 360 | 0.27572 | 0.27572 | 0.45285 | 0.52188 |

Table 8. MSFE of the 6-month ahead point forecasts of the Treasury yields

| | $h = 12$ | | | $h = 18$ | | |
|---|---|---|---|---|---|---|
| Maturity | Minnesota | Independent Normal Inverse Wishart | Partial $SSVS$ | Minnesota | Independent Normal Inverse Wishart | Partial $SSVS$ |
| 3 | 0.6610 (0.8991) | 1.5184 (0.7485) | 0.8137 (0.6906) | 0.6007 (1.1598) | 2.6862 (0.9778) | 2.0341 (0.8788) |
| 6 | 0.7445 (0.9121) | 1.7097 (0.7425) | 1.3921 (0.7237) | 0.7018 (1.1467) | 2.7155 (0.9279) | 2.3254 (0.8889) |
| 12 | 0.8214 (0.9536) | 1.8137 (0.7996) | 1.1127 (0.7272) | 0.8117 (1.1377) | 2.8771 (0.9809) | 1.7388 (0.8673) |
| 24 | 0.9242 (0.9112) | 1.5527 (0.9997) | 1.0584 (0.7101) | 0.9520 (1.0395) | 2.26 (0.9216) | 1.7111 (0.8327) |
| 36 | 1.0623 (0.8167) | 2.3044 (0.6791) | 1.6084 (0.6964) | 1.1406 (0.8888) | 2.879 (0.7498) | 2.0963 (0.7357) |
| 60 | 1.4233 (0.8033) | 2.1795 (0.6687) | 1.3907 (0.6168) | 1.5044 (0.8716) | 3.0163 (0.7321) | 2.4286 (0.6653) |
| 84 | 1.7448 (0.7068) | 2.1228 (0.5839) | 1.4034 (0.5438) | 1.6669 (0.6982) | 2.5940 (0.5893) | 2.1207 (0.5646) |
| 120 | 1.9429 (0.6327) | 2.6065 (0.5193) | 2.2330 (0.4853) | 2.0748 (0.6102) | 3.3258 (0.4981) | 3.2654 (0.4745) |
| 240 | 2.3127 (0.5339) | 2.6891 (0.4374) | 2.4695 (0.8815) | 2.3922 (0.5484) | 3.1261 (0.4463) | 3.2755 (0.4330) |
| 360 | 2.7904 (0.5104) | 2.6056 (0.4285) | 3.0018 (0.4054) | 2.8386 (0.4807) | 3.1459 (0.4229) | 3.6494 (0.3683) |

Table 9. Predictive distribution of the 12-month ahead (columns 2 – 4), and 18-month ahead (columns 5 – 7) point forecasts of the Treasury yields

| | h = 12 | | | h = 18 | | |
|---|---|---|---|---|---|---|
| Maturity | Minnesota | Independent Normal Inverse Wishart | Partial $SSVS$ | Minnesota | Independent Normal Inverse Wishart | Partial $SSVS$ |
| 3 | 0.36117 | 2.217 | 0.56814 | 0.29239 | 6.8972 | 3.897 |
| 6 | 0.45496 | 2.6887 | 1.7479 | 0.39918 | 6.9988 | 5.0869 |
| 12 | 0.5204 | 2.9369 | 1.0255 | 0.50657 | 7.7125 | 2.6857 |
| 24 | 0.66306 | 2.0815 | 0.89938 | 0.7093 | 4.6223 | 2.5635 |
| 36 | 0.76095 | 4.4706 | 2.012 | 0.90368 | 7.231 | 3.6341 |
| 60 | 0.94726 | 2.991 | 0.88483 | 1.1117 | 6.5857 | 3.9149 |
| 84 | 0.91172 | 1.7763 | 0.37622 | 1.1596 | 3.2544 | 1.7707 |
| 120 | 0.69366 | 2.2394 | 1.2611 | 0.93084 | 4.9096 | 4.6458 |
| 240 | 0.40026 | 1.0182 | 0.62335 | 0.50724 | 2.0911 | 2.5455 |
| 360 | 0.84705 | 0.54111 | 1.281 | 0.93819 | 1.628 | 3.1661 |

Table 10. MSFE of the 12-month ahead (columns 2 – 4), and 18-month ahead (columns 5 – 7) forecasts of the Treasury yields of all the maturities

| | h = 24 | | | h = 30 | | |
|---|---|---|---|---|---|---|
| Maturity | Minnesota | Independent Normal Inverse Wishart | Partial $SSVS$ | Minnesota | Independent Normal Inverse Wishart | Partial $SSVS$ |
| 3 | 0.5568 (1.2277) | 2.3422 (1.0999) | 1.4655 (0.9562) | 0.3231 (1.2086) | 0.7629 (1.2258) | 0.3100 (0.8973) |
| 6 | 0.6715 (1.1734) | 2.0337 (0.9942) | 1.6105 (0.9182) | 0.4166 (1.0681) | 0.9664 (1.0219) | 0.3255 (0.8243) |
| 12 | 1.3158 (0.8898) | 2.2844 (1.0814) | 1.3051 (0.8896) | 0.5107 (0.1575) | 0.5423 (1.2406) | 0.5157 (0.8770) |
| 24 | 0.9202 (1.0380) | 2.399 (1.0317) | 1.1102 (0.8231) | 0.6099 (1.0148) | 1.2116 (1.2221) | 0.5543 (0.8066) |
| 36 | 1.1348 (0.8850) | 3.3216 (0.4100) | 1.3088 (0.7054) | 0.7861 (0.8557) | 1.2379 (0.9309) | 0.5220 (0.6631) |
| 60 | 1.4949 (0.8933) | 2.8699 (0.8135) | 2.0204 (0.6773) | 1.2110 (0.8797) | 1.1227 (0.9371) | 0.5089 (0.6644) |
| 84 | 1.8943 | 3.1545 | 2.7857 | 1.5777 | 1.7667 | 1.8815 |
|     | (0.6841) | (0.5213) | (0.4680) | (0.6807) | (0.7708) | (0.5290) |
| 120 | 2.0869   | 3.1557   | 2.7711   | 1.7863   | 1.8127   | 2.0946   |
|     | (0.6025) | (0.5234) | (0.4704) | (0.5918) | (0.6095) | (0.4641) |
| 240 | 2.3766   | 3.1708   | 3.1174   | 2.1216   | 2.7032   | 2.4762   |
|     | (0.5107) | (0.4346) | (0.4062) | (0.4813) | (0.4445) | (0.3752) |
| 360 | 2.8329   | 3.3240   | 3.5877   | 2.6245   | 3.1561   | 3.2172   |
|     | (0.4347) | (0.4092) | (0.3371) | (0.4618) | (0.4416) | (0.3480) |

Table 11. Predictive distribution of the 24-month ahead (columns 2 – 4), and 30-month ahead (columns 5 – 7) point forecasts of the Treasury yields

| | $h = 24$ | | | $h = 30$ | | |
|---|---|---|---|---|---|---|
| Maturity | Minnesota | Independent Normal Inverse Wishart | Partial $SSVS$ | Minnesota | Independent Normal Inverse Wishart | Partial $SSVS$ |
| 3   | 0.24681 | 5.2085 | 1.9754 | 0.0692  | 0.49405 | 0.06256  |
| 6   | 0.36184 | 3.856  | 2.3733 | 0.12016 | 0.80362 | 0.065293 |
| 12  | 1.4781  | 4.7715 | 1.4522 | 0.16869 | 0.19502 | 0.38148  |
| 24  | 0.65647 | 5.2394 | 1.0004 | 0.24989 | 1.2134  | 1.2134   |
| 36  | 0.89259 | 2.1073 | 1.2498 | 0.19737 | 0.3553  | 1.0981   |
| 60  | 1.0914  | 5.8559 | 2.4662 | 0.3553  | 1.0981  | 0.11024  |
| 84  | 1.2195  | 4.1798 | 2.808  | 0.57917 | 0.45254 | 0.003465 |
| 120 | 0.95439 | 4.1848 | 2.7591 | 0.62041 | 0.95395 | 1.1913   |
| 240 | 0.48521 | 2.2224 | 2.0662 | 0.45741 | 0.49374 | 0.96947  |
| 360 | 0.92715 | 2.114  | 2.9505 | 0.56923 | 1.6539  | 1.8149   |

Table 12. MSFE of the 24-month ahead (columns 2 – 4), and 30-month ahead (columns 5 – 7) forecasts of the Treasury yields of all the maturities

|         | h = 36 | | | h = 42 | | |
|---------|--------|---|---|--------|---|---|
| Maturity | Minnesota | Independent Normal Inverse Wishart | Partial SSVS | Minnesota | Independent Normal Inverse Wishart | Partial SSVS |
| 3   | − 0.353 (1.1403)   | − 0.4332 (1.0936) | − 1.6787 (0.8472) | − 0.8578 (1.0945) | − 0.8415 (1.0168) | − 1.8520 (0.8089) |
| 6   | − 0.3935 (0.9097)  | 0.2260 (0.8222)   | − 1.0618 (0.7227) | − 0.9365 (0.8892) | 0.0674 (0.8054)   | − 1.2663 (0.6860) |
| 12  | − 0.1676 (1.0686)  | − 0.3368 (1.1309) | − 1.2979 (0.8212) | − 0.6229 (1.0439) | − 0.6021 (1.0716) | − 1.8529 (0.7916) |
| 24  | − 0.0658 (0.9426)  | 0.3348 (1.1400)   | − 1.6664 (0.7272) | − 0.4703 (0.9579) | − 0.2728 (1.1507) | − 1.1262 (0.7472) |
| 36  | 0.0692 (0.7555)    | 1.8887 (0.8684)   | − 0.6837 (0.6253) | − 0.4035 (0.7336) | 1.1707 (0.8700)   | − 0.6636 (0.6002) |
| 60  | 0.6786 (0.7834)    | 0.1079 (0.8440)   | − 0.7382 (0.5943) | 0.2993 (0.7721)   | − 0.1626 (0.7969) | − 0.6075 (0.5874) |
| 84  | 1.0651 (0.6321)    | 1.1303 (0.7293)   | 0.7242 (0.4857)   | 0.7276 (0.6688)   | 0.6609 (0.7224)   | − 0.4371 (0.5041) |
| 120 | 1.3460 (0.5426)    | 2.1537 (0.5850)   | 1.1781 (0.4175)   | 1.0529 (0.5569)   | 1.6214 (0.5837)   | 0.8650 (0.4357)   |
| 240 | 1.7312 (0.4604)    | 2.4754 (0.4126)   | 1.8584 (0.3535)   | 1.4823 (0.4934)   | 1.8693 (0.4266)   | 1.6011 (0.3704)   |
| 360 | 2.3697 (0.4371)    | 3.2342 (0.3872)   | 2.6601 (0.3335)   | 2.1941 (0.4585)   | 2.65581 (0.3886)  | 2.1041 (0.3479)   |

Table 13. Predictive distribution of the 24-month ahead (columns 2 – 4), and 30-month ahead (columns 5 – 7) point forecasts of the Treasury yields

|         | h = 36 | | | h = 42 | | |
|---------|--------|---|---|--------|---|---|
| Maturity | Minnesota | Independent Normal Inverse Wishart | Partial SSVS | Minnesota | Independent Normal Inverse Wishart | Partial SSVS |
| 3  | 0.17053  | 0.24326  | 3.0232 | 0.84233 | 0.81278 | 3.6556 |
| 6  | 0.21483  | 0.02435  | 1.281  | 1.013   | 0.00699 | 1.7858 |
| 12 | 0.071609 | 0.19079  | 1.9541 | 0.52257 | 0.49296 | 3.8137 |
| 24 | 0.030896 | 0.050554 | 3.1556 | 0.33678 | 0.14655 | 1.5282 |
| 36 | 0.014583 | 2.8855 | 0.7634 | 0.35224 | 0.96186 | 0.7286 |
| 60 | 0.052276 | 0.11704 | 1.4117 | 0.02271 | 0.37526 | 1.1184 |
| 84 | 0.075673 | 0.11584 | 0.0043323 | 0.0038971 | 0.016659 | 1.5057 |
| 120 | 0.055697 | 1.0894 | 0.0046425 | 0.0032612 | 0.26149 | 0.060025 |
| 240 | 0.0026235 | 0.63272 | 0.031835 | 0.039094 | 0.035852 | 0.062196 |
| 360 | 0.24974 | 1.861 | 0.62426 | 0.10506 | 0.61751 | 0.054793 |

Table 14. MSFE of the 36-month ahead (columns 2 – 4), and 42-month ahead (columns 5 – 7) forecasts of the Treasury yields of all the maturities

| | $h = 48$ | | | $h = 54$ | | |
|---|---|---|---|---|---|---|
| Maturity | Minnesota | Independent Normal Inverse Wishart | Partial $SSVS$ | Minnesota | Independent Normal Inverse Wishart | Partial $SSVS$ |
| 3 | − 1.0522 (1.0950) | − 0.9697 (1.0666) | − 1.4198 (0.8538) | − 1.0492 (1.1076) | − 1.5710 (1.0225) | − 2.1235 (0.8571) |
| 6 | − 1.0758 (0.9677) | − 0.4000 (0.9343) | − 0.6946 (0.7644) | − 0.9913 (1.0608) | − 1.4748 (0.9969) | − 1.4392 (0.8710) |
| 12 | − 0.8126 (1.0428) | − 0.6981 (1.0820) | − 0.6875 (0.8109) | − 0.8428 (1.0514) | − 1.4250 (1.0170) | − 0.6391 (0.8294) |
| 24 | − 0.6972 (0.9441) | − 0.4458 (1.1420) | − 0.4712 (0.7131) | − 0.8070 (0.9425) | − 1.2631 (1.0383) | − 0.8792 (0.7671) |
| 36 | − 0.6320 (0.7437) | − 0.2673 (0.8731) | − 0.9870 (0.6089) | − 0.7269 (0.7863) | − 1.5703 (0.8546) | − 0.7624 (0.6128) |
| 60 | 0.0678 (0.7258) | 0.1211 (0.7493) | 0.6074 (0.5556) | − 0.0625 (0.7251) | 0.4053 (0.7084) | 0.5162 (0.5352) |
| 84 | 0.5366 (0.6358) | 1.1591 (0.6897) | 1.3654 (0.4879) | 0.4498 (0.6326) | 1.5684 (0.6138) | 1.2504 (0.4796) |
| 120 | 0.9197 (0.4913) | 1.6999 (0.5370) | 1.8046 (0.3892) | 0.8907 (0.4819) | 1.3624 (0.4866) | 1.4357 (0.3633) |
| 240 | 1.4108 (0.4726) | 2.8949 (0.4431) | 2.2602 (0.3606) | 1.4120 (0.4773) | 2.9845 (0.4563) | 2.3645 (0.3706) |
| 360 | 2.1659 (0.4433) | 2.2780 (0.4008) | 2.7183 (0.3356) | 2.1758 (0.4488) | 2.5337 (0.4142) | 2.9469 (0.3728) |

Table 15. Predictive distribution of the 48-month ahead (columns 2 – 4), and 54-month ahead (columns 5 – 7) point forecasts of the Treasury yields

|  | h = 48 | | | h = 54 | | |
|---|---|---|---|---|---|---|
| Maturity | Minnesota | Independent Normal Inverse Wishart | Partial $SSVS$ | Minnesota | Independent Normal Inverse Wishart | Partial $SSVS$ |
| 3 | 1.237 | 1.0604 | 2.1898 | 1.2304 | 2.6602 | 4.7679 |
| 6 | 1.3129 | 0.22094 | 0.58465 | 1.1263 | 2.3864 | 2.2776 |
| 12 | 0.83288 | 0.63698 | 0.62022 | 0.88885 | 2.3255 | 0.54626 |
| 24 | 0.65151 | 0.30887 | 0.33775 | 0.84088 | 1.8854 | 0.96607 |
| 36 | 0.67563 | 0.20909 | 1.3853 | 0.84074 | 3.0985 | 0.90714 |
| 60 | 0.14608 | 0.1082 | 0.024762 | 0.26265 | 0.002 | 0.004379 |
| 84 | 0.064222 | 0.13621 | 0.33109 | 0.11577 | 0.60587 | 0.21201 |
| 120 | 0.036213 | 0.34803 | 0.48241 | 0.048083 | 0.063723 | 0.1061 |
| 240 | 0.072443 | 1.476 | 0.33659 | 0.071841 | 1.7016 | 0.4686 |
| 360 | 0.087539 | 0.16644 | 0.71953 | 0.093515 | 0.44048 | 1.1598 |

*Table 16. MSFE of the 48-month ahead (columns 2 – 4), and 54-month ahead (columns 5 – 7) forecasts of the Treasury yields of all the maturities*

### 5.2.6 Augmented Bayesian VARs: Yields with Macroeconomic Variables

Finally, I extended the previous Bayesian $VAR$ models by incorporating four macroeconomic variables: a variable that measures economic activity: year-to-year percent change in total capacity utilization $-\%\Delta tcu$, inflation rate measured via year-to-year percent change in the personal consumption expenditures $-\%\Delta pce$, unemployment rate $-unrate$, and a monetary policy instrument: effective federal funds rate $-effr$. According to Dijk et. al (2013), long-run inflation expectations propel the level factor's path. As the actual or ex-post inflation rate influences the ex-ante inflation rate, I incorporated the inflation rate as a variable in the dataset. Similarly, the expectations hypothesis of the term structure of interest rates propounds that we can determine the long and short term interest rates as the current and expected value of the future short term interest rates. Hence, we can link the whole yield curve (decomposed into its factors) to the present and future ex-ante policy rates − federal funds rate. The vector of endogenous[8] variables is ordered as:

$f_t = [L_t, S_t, C_t, \%\Delta tcu_t, \%\Delta pce_t, unrate_t, effr_t]'$

---

[8] By "endogenous", I am not referring to the endogeneity assumption in the Gauss Markov Theorem. Instead, I mean that all the coefficients of the endogenous variables are determined by the VAR system.

Since changes in monetary policy affect inflation and unemployment rates with least one month of lag, I ordered $effr_t$ last in $f_t$. Furthermore, $A_1,...,A_{13}$, and the variance of residuals, $\Sigma_\eta$ from equation (9) each have $(7 \times 7)$ dimension; and there are a total of 644 coefficients including the intercepts. Accordingly, the dimensions of other matrices and vectors in the two representations of $BVARs$ expand as well. Thus, the $BVARs$ consider a bidirectional linkage between the unobserved factors and macro variables. With the exception of the diffuse prior, which I applied on $BVAR(4)$, each of the other informative priors are based on $BVAR(13)$. As before, table 15 displays the predictive distribution of the one-step-ahead forecasts of the latent factors and the four macroeconomic variables; and table 16 presents the mean squared forecast errors for all variables from the five prior distributions.

|  | Diffuse | Minnesota | Natural Conjugate | Independent Normal Inverse Wishart | Partial $SSVS$ | Full $SSVS$ | Actual values $f_{T+1}$ |
|---|---|---|---|---|---|---|---|
| $L_{T+1}$ | 3.1209 (0.2279) | 2.5990 (0.2227) | 3.5862 (0.4059) | 2.4924 (0.1378) | 2.4857 (0.1331) | 2.5189 (0.1949) | 1.9817 |
| $S_{T+1}$ | −2.9117 (0.2960) | −2.1255 (0.2799) | −2.2168 (0.2832) | −2.0513 (0.1678) | −1.9788 (0.1850) | −1.9709 (0.2606) | −1.6292 |
| $C_{T+1}$ | −2.8227 (0.6580) | −2.3910 (0.6237) | −2.5977 (0.5366) | −2.4317 (0.3730) | −2.4917 (0.3704) | −2.4465 (0.6557) | −3.8587 |
| $\%\Delta tcu_{T+1}$ | −2.8626 (0.7242) | −3.8436 (0.5992) | −3.5514 (0.7979) | −3.9853 (0.3635) | −4.2646 (0.3582) | −4.1263 (0.6455) | −5.3968 |
| $\%\Delta pce_{T+1}$ | 0.2757 (0.2370) | 0.6570 (0.1800) | 0.8086 (0.3076) | 0.6468 (0.1093) | 0.6439 (0.1218) | 0.6028 (0.1918) | 1.1162 |
| $unrate_{T+1}$ | 5.3392 (0.1342) | 5.0190 (0.1255) | 5.4449 (0.3862) | 4.9345 (0.0809) | 4.9925 (0.0824) | 4.9790 (0.1304) | 6.7 |
| $effr_{T+1}$ | 0.1570 (0.3456) | 0.3597 (0.2837) | 1.2832 (0.4477) | 0.2361 (0.1718) | 0.3583 (0.1677) | 0.4179 (0.2857) | 0.09 |
| Log Predictive Likelihood | −1.6138 | −0.7608 | −6.1235 | −9.1313 | −6.9611 | −3.0456 |  |

Table 15. Log predictive likelihood, and predictive means and standard deviations (in the parenthesis) from BVARs with each of the five priors

|  | Diffuse | Minnesota | Natural Conjugate | Independent Normal Inverse Wishart | Partial $SSVS$ | Full $SSVS$ |
|---|---|---|---|---|---|---|
| $L_{T+1}$ | 1.2979 | 0.38107 | 2.5744 | 0.26081 | 0.25406 | 0.28858 |
| $S_{T+1}$ | 1.645 | 0.24634 | 0.34525 | 0.17817 | 0.12224 | 0.11679 |
| $C_{T+1}$ | 1.0732 | 2.154 | 1.5901 | 2.0361 | 1.8686 | 1.9941 |
| $\%\Delta tcu_{T+1}$ | 6.4218 | 2.4125 | 3.4053 | 1.9921 | 1.2819 | 1.6141 |
| $\%\Delta pce_{T+1}$ | 0.70641 | 0.2109 | 0.094621 | 0.22034 | 0.22314 | 0.26357 |
| $unrate_{T+1}$ | 1.8518 | 2.8259 | 1.5753 | 3.1169 | 2.9155 | 2.9618 |
| $effr_{T+1}$ | 0.004484 | 0.07272 | 1.4237 | 0.021332 | 0.071988 | 0.10754 |

*Table 16. Mean square forecast error of the latent factors and macroeconomic variables*

The results from table 8 suggest that the $BVAR$ with partial $SSVS$ prior minimizes $MSFE$ for $L_{T+1}$ (0.25406), and $\%\Delta tcu_{T+1}$ (1.2819). Alternatively, the $BVAR$ with the full $SSVS$, and diffuse priors yields the lowest mean squared error for $S_{T+1}$ (0.11679), and $C_{T+1}$ (1.0732), respectively.

Figures $15-19$ depict the impulse responses to examine the dynamics of the system of augmented $BVARs$ with macro variables.[9] The specific order of observed and unobserved variables in $f_t$ imposes a recursive causal framework wherein contemporaneous $effr_t$ depends on other contemporaneous variables in $f_t$, but not vice-versa. As before, the zig-zags become very conspicuous in the independent normal inverse $-$ Wishart prior, predominantly due to overfitting. However, the informative priors imposed in the Minnesota, natural conjugate, and to some extent the $SSVS$ priors appear to have smoothed out the $IRFs$. Although we cannot interpret the $IRFs$ in the previous section, we can consider the innovations to the effective federal funds rate as monetary policy shocks as $effr_t$ is ordered last in the set of variables. From all four priors, $effr_t$ hikes by approximately 0.8 in response to innovations to $S_t$ (in Diffuse prior), and by 0.4 (in Minnesota prior), implying that the monetary policy instrument and the slope factor are well associated with each other. This could be because the Federal Reserve (Fed) may hike the federal funds rate in response to changes observed in the Treasury yields market, and vice versa.

We can somewhat interpret the $IRFs$ of $effr_t$ to shocks to macro variables (graphs in the fourth row). Economic theory suggests that we would expect the Fed to hike the federal funds rate when $\%\Delta tcu_t$ rises, as it anticipates the economy to heat when more finished goods are generated. The $IRFs$ from all except from the $SSVS$ priors in figures $19-20$, support this theory as the federal funds rate's median response is fairly flat in the latter two graphs. In fact, the federal funds rate fell in the partial $SSVS$ prior in figure 19. Likewise, the median response of the federal funds rate to a unit shock to inflation, and unemployment rate defies the standard Taylor rule, except in the case of natural conjugate prior: a unit shock to inflation rate sparks inflationary pressure, causing the Fed to tighten the monetary policy in the initial periods before cutting the rates; a unit rise to unemployment rate paves way for a looser monetary policy. However, in other priors, the federal funds rate drops when the inflation rate spikes. Similarly, the federal funds rate is flat throughout the time horizon (Minnesota prior), flat upto 8 months before falling (full $SSVS$ prior), flat in later months (independent normal inverse − Wishart prior), and shows similar behavior in other priors.

---

[9] The following pages contain the interpretable responses to innovations to latent factors, and the IRFs of all the variables to shock to *effr;* resulting in 12 instead of 49 graphs.

Along the same veins, with the exception of the independent normal inverse − Wishart prior, $unrate$, $\%\Delta pce$, and $\%\Delta tcu$ in other priors don't respond typically to an upward shock to the federal funds rate, as justified by economic theory − a contractionary monetary policy escalates economic headwinds wherein capacity utilization shrinks, inflationary spiral slows, and unemployment rate grows. Nonetheless, the median response of $\%\Delta tcu$, and $\%\Delta pce$ is flat (diffuse, and Minnesota priors), and flat for one year before falling (full $SSVS$ prior). To some extent the $IRFs$ in figure 18 (graphs in the second row) are compatible with economic theory: a downward trend of $\%\Delta tcu$ for after 6 months, deflation sets in, and $unrate$ gradually gains momentum but only after 8 months. Lastly, except in figures 17 − 18, the term structure factors scarcely move in response to federal funds rate shocks. In figure 17, $L_t$ spikes when faced with a unit shock to the federal funds rate. Since the 10-year Treasury yield is an empirical measure of the level factor as they are highly correlated with each other, a rise in $effr_t$ tantamounts to a rise in the long term Treasury yield, having a contractionary effect in the economy. Additionally, $S_t$ plunges, which indicates that its empirical measure : $3mo − 10y$ or the yield spread falls presumably because the 10-year Treasury yield rises. Largely, the $IRFs$ illustrate that the macro variables don't comport with the general economic theory, underscoring the limitations of using the recursive order scheme as a way to identify monetary policy.

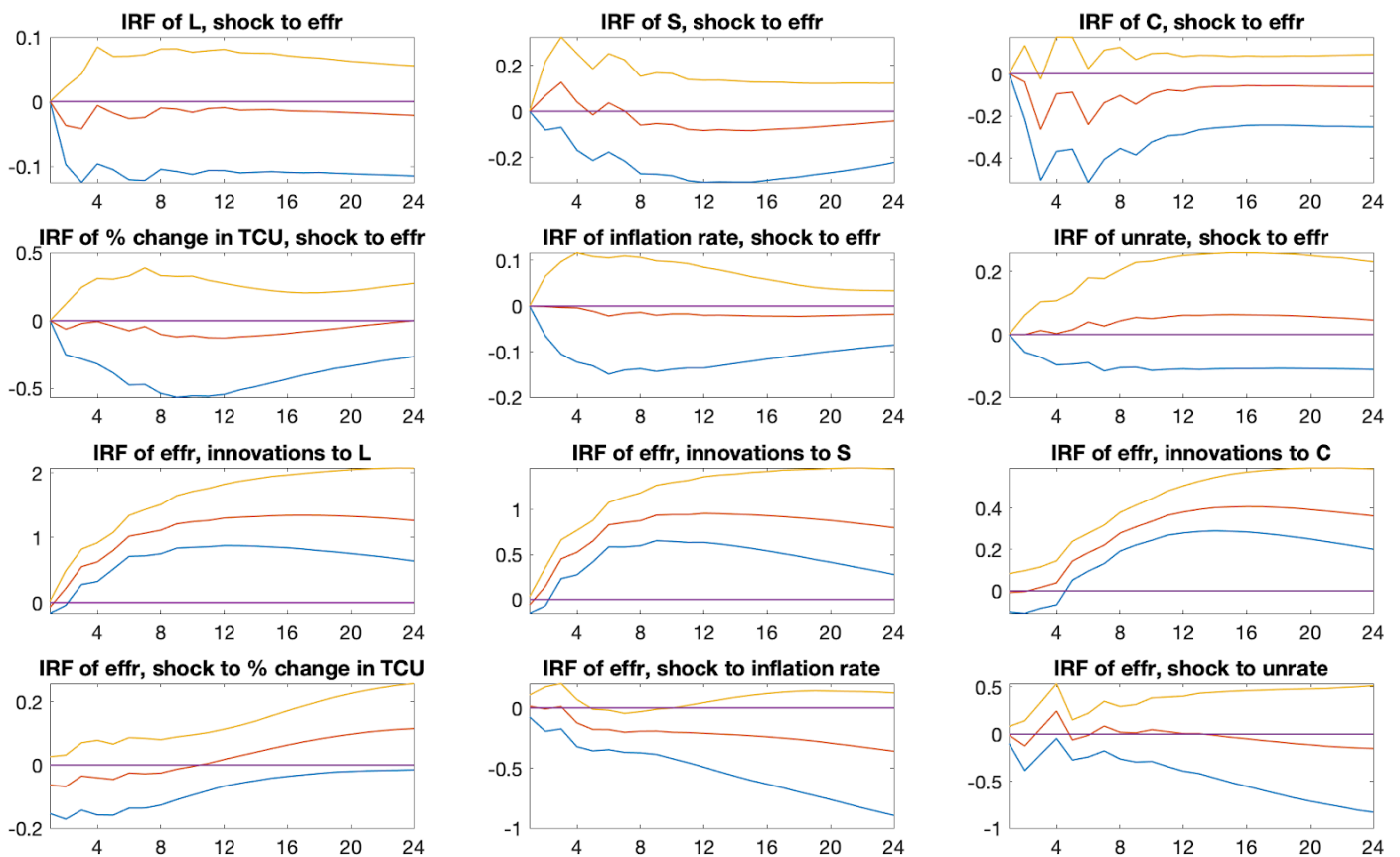

*Fig 15. IRFs of the variables under the Diffuse Prior*

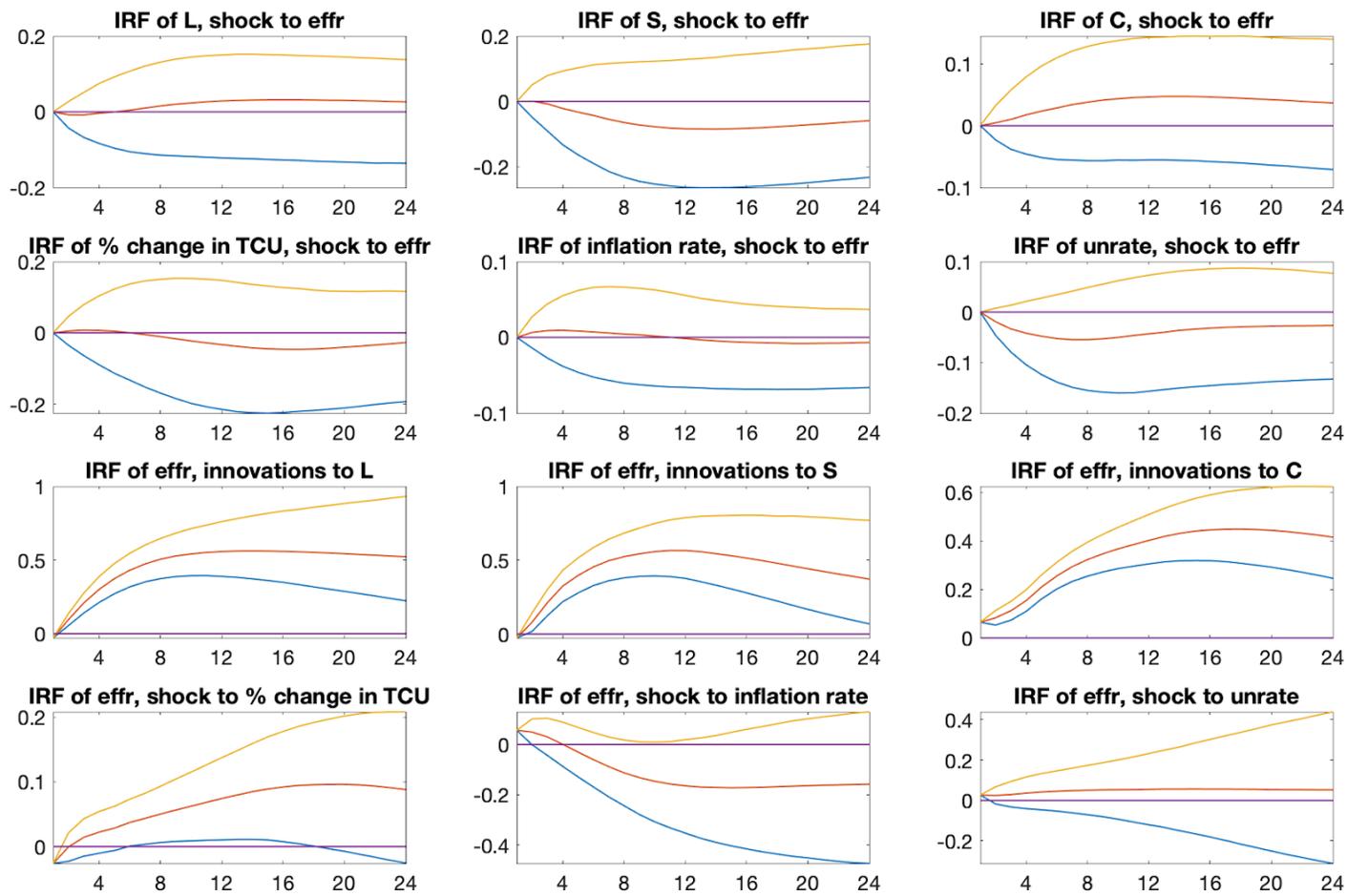

*Fig 16. IRFs of variables to shock to effr; and the IRFs of effr to shock to latent and macro variables under the Minnesota Prior*

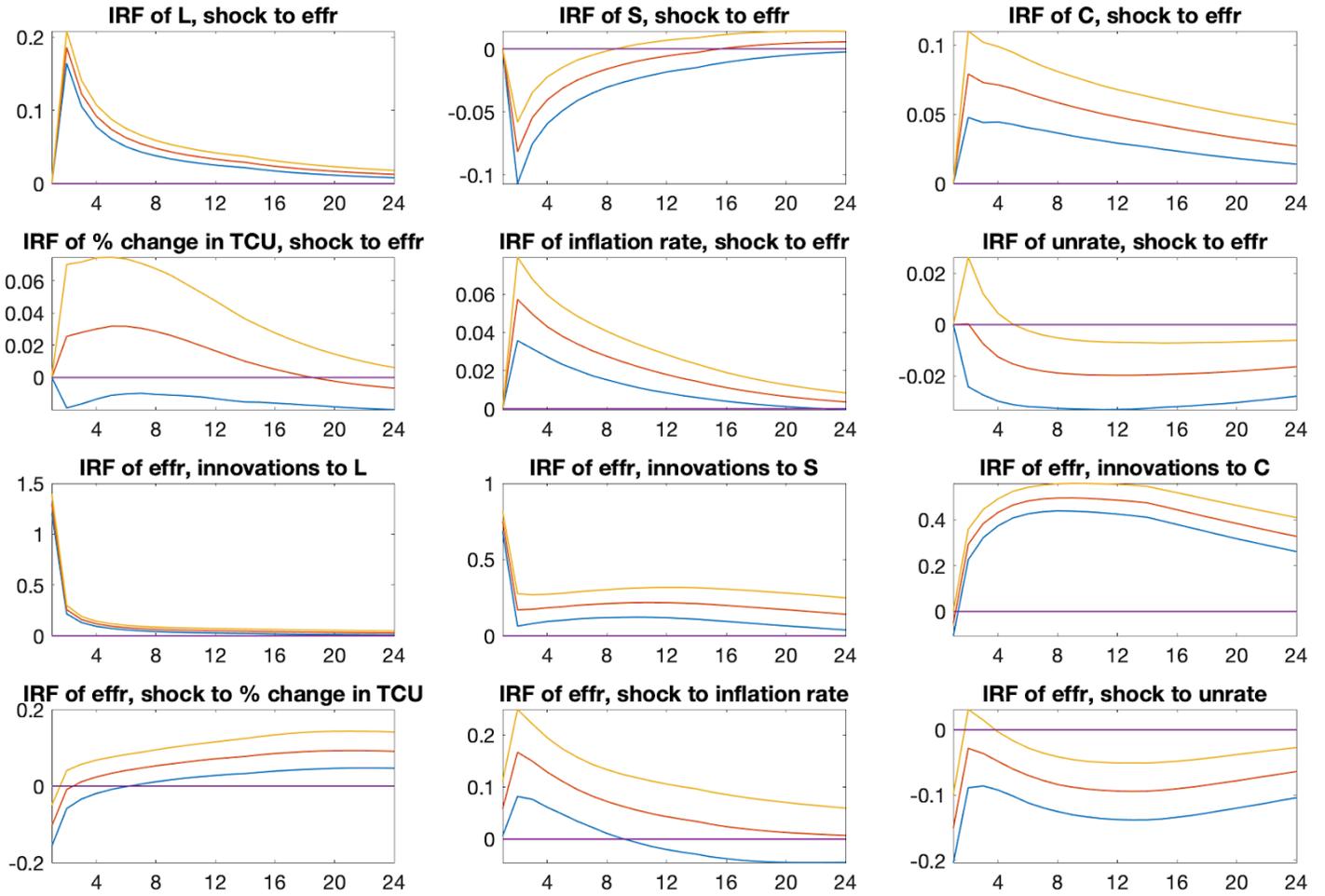

Fig 17. IRFs of variables to shock to effr; and the IRFs of effr to shock to latent and macro variables under the natural conjugate: normal-Wishart prior

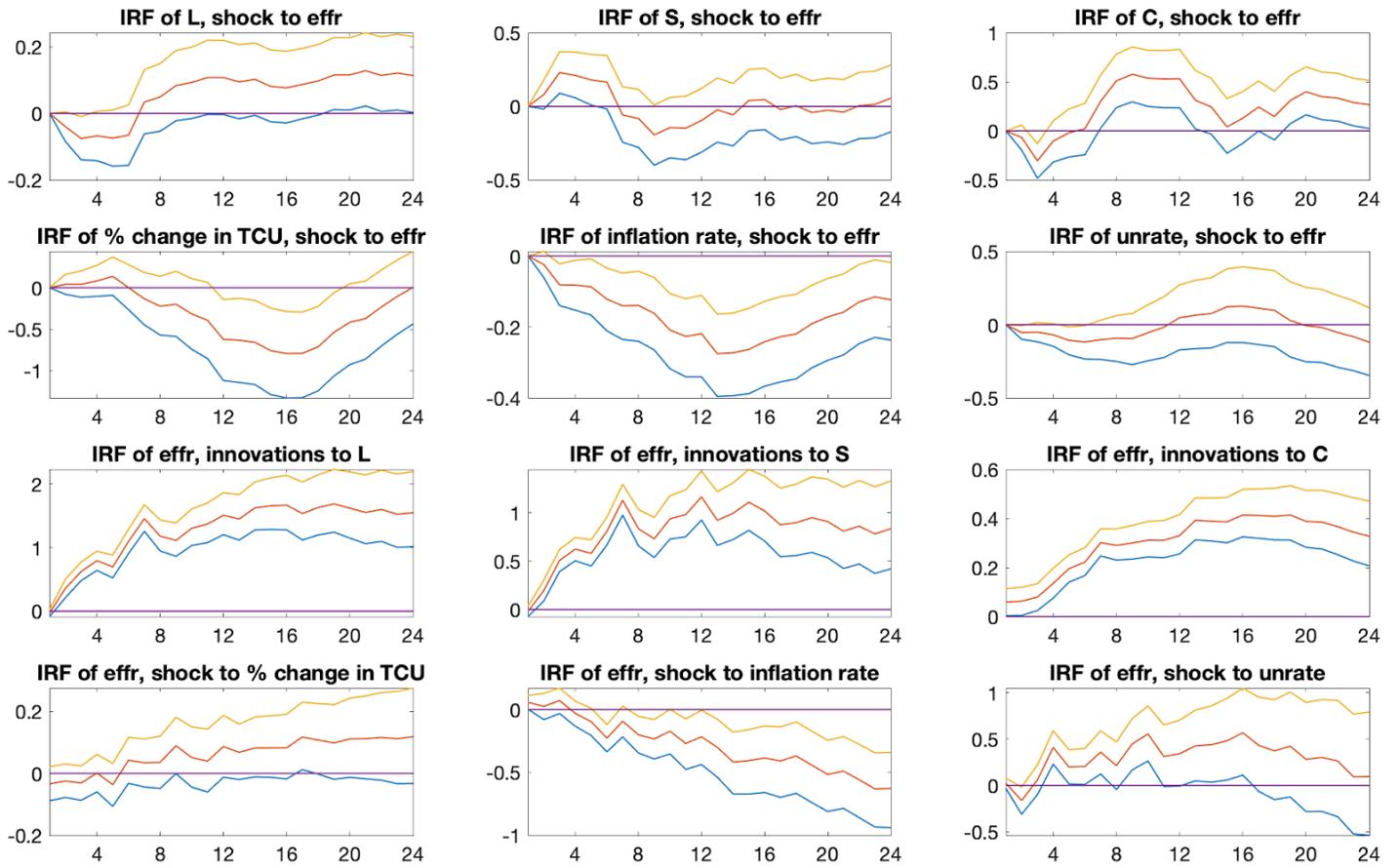

Fig 18. IRFs of variables to shock to effr; and the IRFs of effr to shock to latent and macro variables under the Independent Normal Wishart Prior

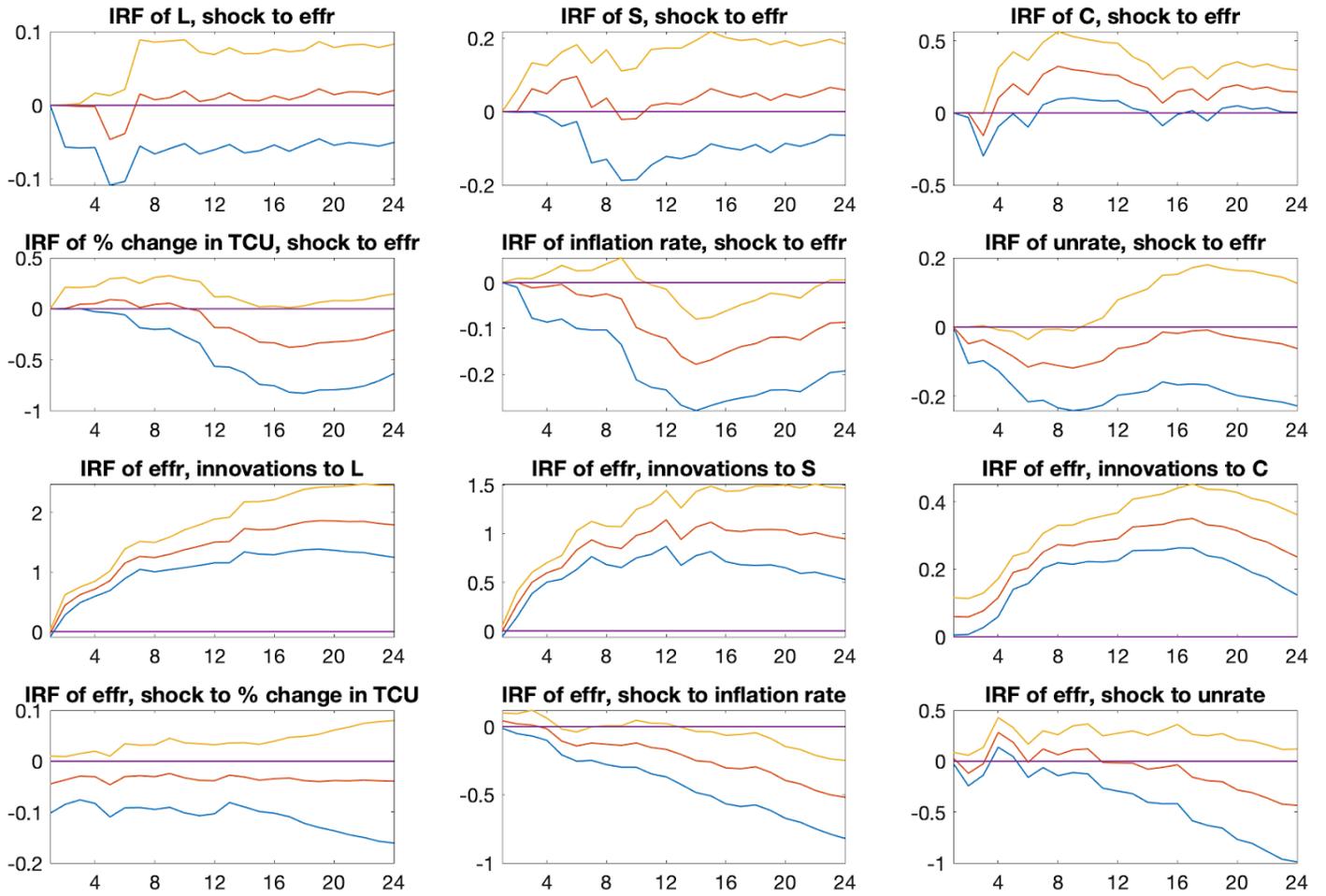

Fig 19. IRFs of variables to shock to effr; and the IRFs of effr to shock to latent and macro variables under the SSVS prior on VAR coefficients, and inverse − Wishart prior on the error covariance matrix

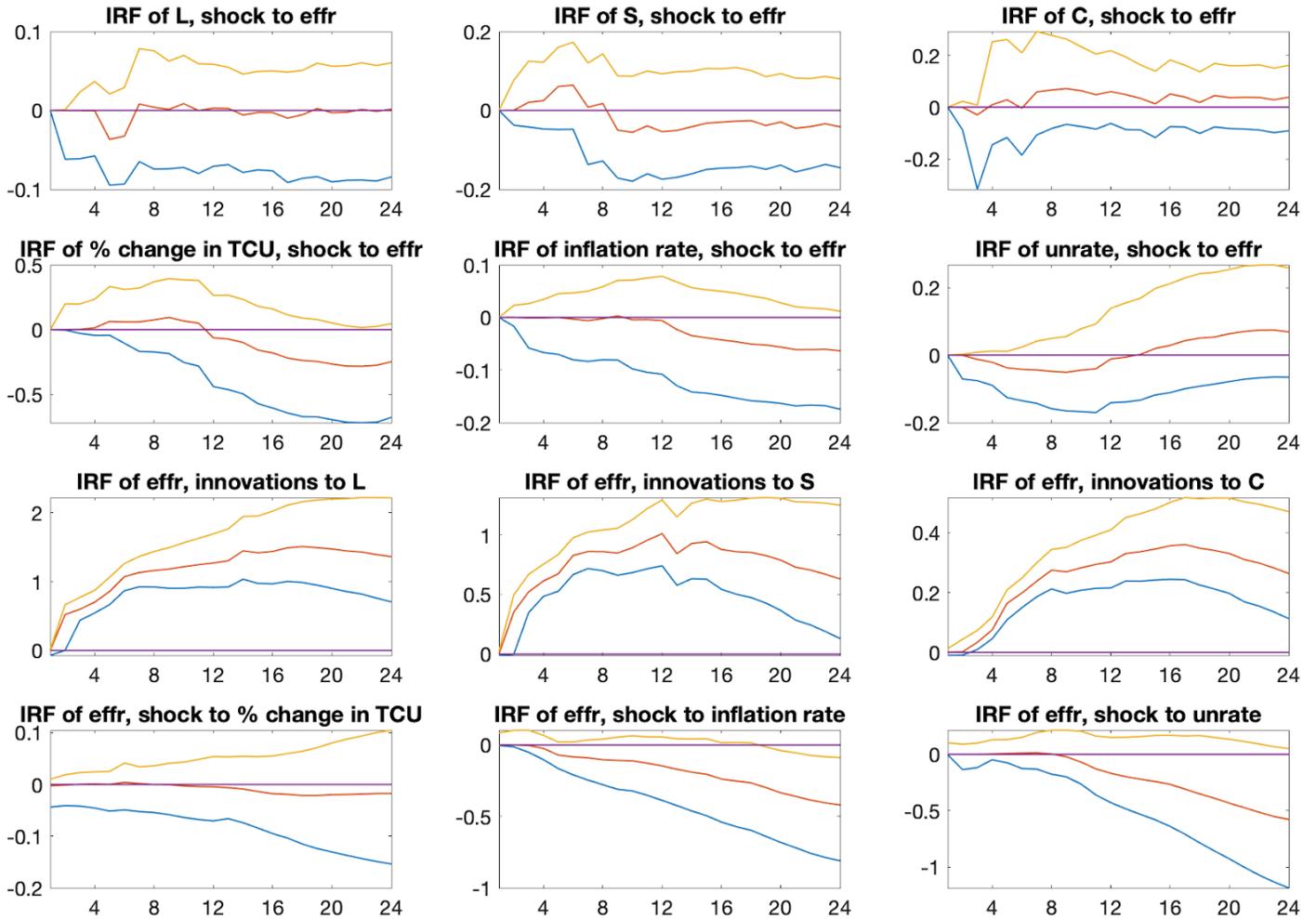

*Fig 20. IRFs of variables to shock to effr; and the IRFs of effr to shock to latent and macro variables under the full SSVS prior (on both VAR coefficients and error covariance matrices)*

### 5.2.7 BVAR with Dummy Observations and Sign Restrictions

So far in section 5.2 on Bayesian methods, I utilized the frequentist estimates of the latent states obtained in the two-step Diebold-Li model as data and applied various priors in estimating the models using $BVARs$. As this is not a complete Bayesian approach, now I estimated a fully Bayesian model without using the frequentist estimates, namely the time-invariant $BVAR$ with dummy or pseudo observations. After incorporating five macro variables, there are fifteen variables in the dataset. The overall data consists of the following variables:

$Y_t = [tres3mo_t, tres6mo_t, tres12mo_t, tres24mo_t, tres36mo_t, tres60mo_t, tres84mo_t, tres120mo_t, tres240mo_t, tres360m, \%\Delta tcu_t, \%\Delta pce_t, unrate_t, ted_t, effr_t]$

I modeled a $VAR(2)$ of the form:

$$Y_t = \alpha + A_1 Y_{t-1} + A_2 Y_{t-2} + e_t, \quad e_t \sim WN(0, \Sigma_e) \tag{13}$$

where, $\alpha = [\alpha_1, ..., \alpha_{15}]$ is a vector of constants, while $A_1$, and $A_2$ are autoregressive matrices. Combining all the parameters and data, I rewrote the $VAR(2)$ as:

$$Y = X\Theta + E \tag{14}$$

where, $Y = [Y_1, ..., Y_T]'$, and $X = [X_1, ..., X_T]'$, such that $X_t = [Y_{t-1}', Y_{t-2}', 1]'$, $E = [e_1, ..., e_T]'$, and the matrix of coefficients is $\Theta = [A_1, A_2, \alpha]'$. Following the procedure of Banbura et. al. (2007), I casted the Minnesota prior in the form of a normal inverse-Wishart prior of the coefficients:[10]

$$vec(\Theta) \mid \Sigma_e \sim N(vec(\Theta_0), \Sigma_e \otimes \Sigma_0), \qquad \Sigma_e \sim W^{-1}(s_{h0}, S_{c0}) \tag{15}$$

To enforce the prior in (15), I first added dummy observations $Y_{d1}$, and $X_{d1}$ to the system of equations in (14). Equivalently, this establishes a normal inverse-Wishart prior:

$$\Theta_0 = (X_d' X_d)^{-1} X_d' Y_d, \quad \Sigma_0 = (X_d' X_d)^{-1}, \quad S_{c0} = (Y_d - X_d \Theta_0)'(Y_d - X_d \Theta_0), \quad s_{h0} = T - 31$$

Moreover, since the variables are non-stationary, I constructed dummies to indicate the prior beliefs that the coefficients on the first and second lags of the dependent variables sum to 1. These dummy observations are:

$$Y_{d2} = \frac{\text{diag}(m_1 \mu_1 \ldots m_{15} \mu_{15})}{\theta}, \quad X_{d2} = \left[ \frac{(1,2) \otimes \text{diag}(m_1 \mu_1 \ldots m_{15} \mu_{15})}{\theta} \quad 0_{(15 \times 1)} \right]$$

$\mu_i$ is a sample average of variable $i$ in $Y_t$. $\theta$ regulates the magnitude of shrinkage wherein larger values of $\theta$ shrinks the parameters less. Also, I fixed a relatively loose prior: $\theta = 12f$ to represent the uncertain prior beliefs about the coefficients. After appending the dummy observations with the actual data, the appended data comprises of $Y_a$, and $X_a$ of size $T_a$. Expanding the $VAR(2)$ in (14) with this augmented dataset yields $Y_a = X_a \Theta + E_a$ where $E_a = [E', E_d']'$. Now, the conditional posterior distributions of the parameters are

$$vec(\Theta) \mid \Sigma_e, Y \sim N(vec(\Theta_a), \Sigma_e \otimes (X_a' X_a)^{-1}), \qquad \Sigma_e \mid Y \sim W^{-1}(\Sigma_a, T_d + T),$$

where $\Theta_a = (X_a' X_a)^{-1} X_a' Y_a, \quad \Sigma_a = (Y_a - X_a \Theta_a)'(Y_a - X_a \Theta_a)$.

I obtained the posterior estimates after the Gibbs sampler for a total of $100,000$ times, and retained only the last $10,000$ estimates of the Markov chain. Employing this time-invariant $BVAR$ with dummies is computationally efficient as to draw the variance-covariance matrix associated with $P(vec(\Theta)|\Sigma, Y)$, we only have to invert a matrix of dimension 31. A shortcoming of recursive ordering identification scheme implemented in previous sections is that several macro variables responded contrary to the paths suggested by economic theory. To overcome it, I resorted to restricting the signs of the impulse responses. Using the saved draws of the posterior parameters, I computed the $IRFs$ of the US Treasury yields and macro variables when the economy is hit by an exogenous upward monetary policy shock or a hike to the federal funds rate. Identifying the monetary policy using sign restrictions, I fitted a structural $VAR$, and assumed that with a rise in the federal funds, we generally see a rise in Treasury yields, $TED$ spread, and the unemployment rate, but the inflation rate and total capacity utilization slump – conforming with the standard economic theory. Furthermore, the Fed tightens the monetary policy in response to higher inflation expectations, which indirectly raises the interest rates on the Treasury securities. Although the cost of borrowing for short, medium, and long term horizons escalate, according to Edelberg and Marshall (1996), the spike is typically higher for short-term maturities, than for medium and long-term maturities. Relatedly, with a contractionary monetary policy, we notice a widening $TED$ spread – the gap between the three-month Treasury bill rate and the $LIBOR$ rate, a reflection of heightened credit risk.

---

[10] Section A.2.7 describes the prior expectation and variance associated with the Minnesota prior, hyperparameters, dummy observations linked to the natural conjugate prior, and how I appended the dummy with the actual observations.

Given each retained draw of the Gibbs sampler, from the Cholesky decomposition of $\Sigma_e$, the structural impact matrix is $\Omega$ as the $SVAR$ is exactly identified. Rather than directly sampling $\Omega$, this indirectly estimates the $SVARs$ by using the Gibbs draw of $\Sigma_e$, resulting in the posterior distribution of $\Omega$. To calculate the $(15 \times 15)$ matrix $\Omega$ after burning the initial draws, firstly, for every retained Gibbs draw I calculated the $(15 \times 15)$ matrix $M$, and I drew its $ij$ elements from the standard normal distribution: $M_{ij} \sim N(0,1)$. Secondly, I obtained the matrix $Q$ from the $QR$ decomposition of the square matrix $M$ such that $Q$ is orthonormal: $Q'Q = I$. Thirdly, I calculated the Cholesky factor of the latest draw of $\Sigma_e : \Sigma_e = \Omega_0' \Omega_0$. Finally, the relevant candidate matrix $\Omega$ is $\Omega = Q \Omega_0$. As $Q'Q = I$, thereby $\Omega' \Omega = \Sigma_e$. Whilst calibrating $Q \Omega_0$ alters $\Omega_0$, it preserves the property that $\Sigma_e = \Omega_0' \Omega_0$. The form of the candidate matrix $\Omega$ is

$$\Omega = \begin{bmatrix} \omega_{1,1} & \omega_{1,2} & \cdots & \omega_{1,15} \\ \omega_{2,1} & \cdots & & \vdots \\ \vdots & & \ddots & \\ \omega_{15,1} & \cdots & & \omega_{15,15} \end{bmatrix}$$

Here, the last row of the matrix represents the response of variables due to a shock in $effr_t$. The sign restrictions constrain the coefficients of the matrix $\Omega$. To corroborate, $[\omega_{15,1}, \omega_{15,2}, ..., \omega_{15,10}] > 0$ implies that the US Treasury yields respond positively or rise when $effr_t$ rises. Similarly, $[\omega_{15,13}, \omega_{15,14}, \omega_{15,15}] > 0$, indicating that $unrate_t$, $ted_t$, and $effr_t$, respectively, also increase due to a contractionary monetary policy. These contrast with $[\omega_{15,12}, \omega_{15,13}] < 0$ wherein $\%\Delta tcu_t$, and $\%\Delta pce_t$, respectively, shrink. After matrix $M$ fulfilled the sign restrictions in $\Omega$, I used $\Omega$ to measure and trace the impulse responses of the variables for up to 24 months since the time of shock in figure 20.

These $IRFs$ illustrate the median responses and the 68% error bands for a horizon of 24 months. The error bands are the $16^{th}$, and $84^{th}$ quantiles of retained draws of the impulse responses.

Overall, these variables don't change much over time. Whilst the Treasury yields and $TED$ rates move upward by a slight margin at the time of impact, the effect of the shock on the federal funds rate dissipates. I also constructed the impulse responses without imposing any identification assumptions on the signs of responses of the Treasury yields and the $TED$ rates to check if the responses of these variables are substantive, or simply hinged on assumptions. However, I maintained the same sign restrictions on the responses of $effr$, $\Delta tcu$, $\Delta pce$, and $unrate$. Figure 21 depicts these $IRFs$ with very wide intervals around the median responses, implying hugely dispersed posterior distributions. Additionally, the interest rates in the Treasury yields of several maturities appear to fall, rather than rise, which contradicts the responses in figure 20 wherein I restricted all the signs, defying the "conventional wisdom." An exception to this is the response of the $TED$ spread that rises on impact before following a downward trajectory after a few months.

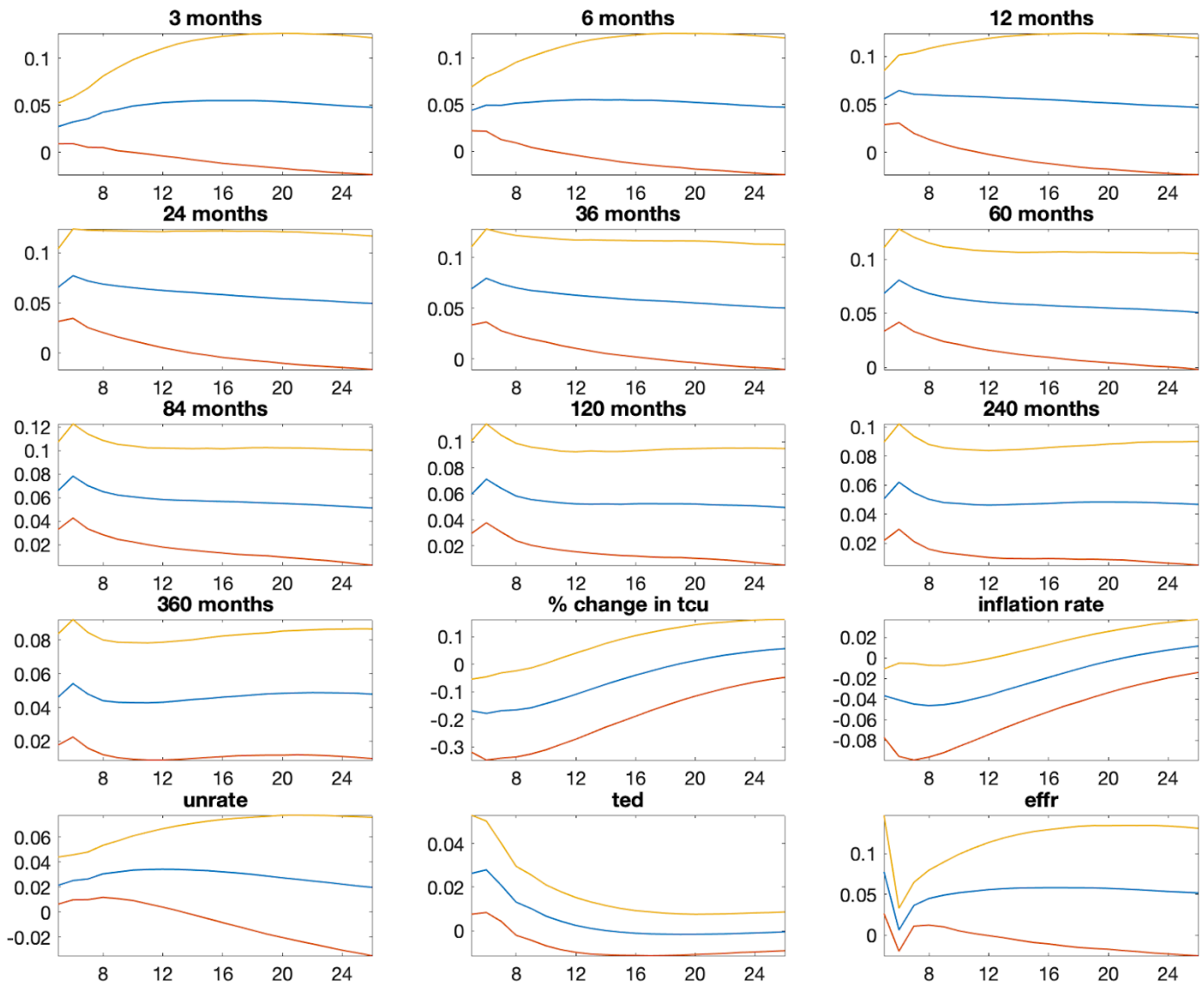

Fig 20. IRFs with sign restrictions on all variables due to an upward shock on the federal funds rate

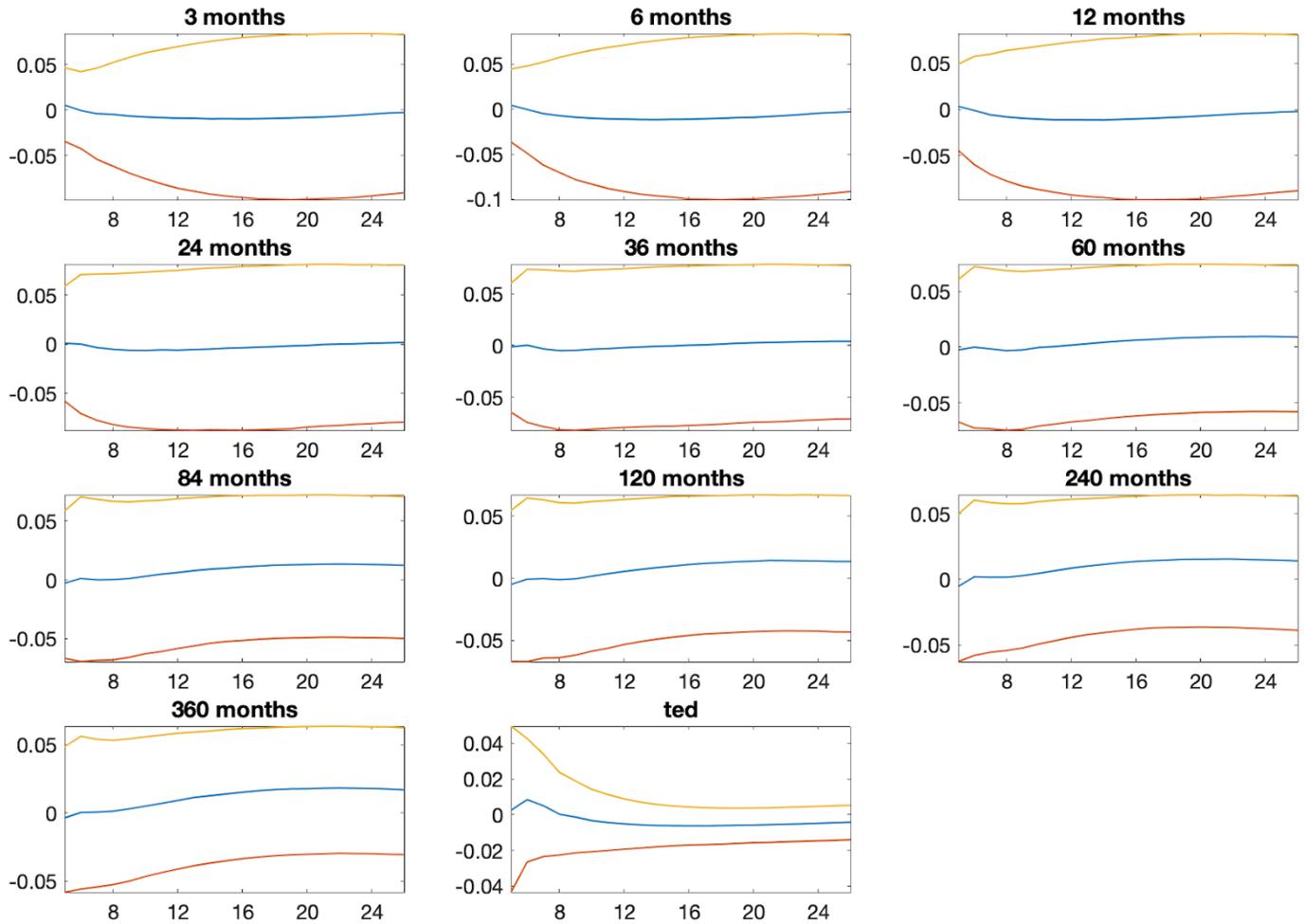

Fig 21. IRFs with sign restrictions only on federal funds rate, capacity utilization, unemployment rate, and inflation rate

## 6. Conclusion

In this research paper, I examined the various frequentist and Bayesian methods to forecast the US Treasury yields by first decomposing the aggregate of the yields into its latent factors – level, slope, and curvature. Previously, extensive literature has confirmed that the high-dimensional data on bond yields follows a factor structure. Therefore, instead of fitting an unrestricted $VAR$ that wastes degrees of freedom, I fitted restricted $VARs$ on the latent factors in the train set. Besides using the dynamic Nelson Siegal model to acquire the three latent variables, I obtained the first three principal components from the data of the Treasury yields and additional macroeconomic variables and then modeled the two-step principal components approach. Furthermore, I compared the $VAR$ coefficients, residuals, 56-month path forecasts of the latent factors, and various forecast intervals obtained from the three frequentist methods – two-step Diebold-Li, two-step $PC$, and the one-step Kalman filter approaches in the test set. Recognizing the drawbacks of the frequentist methods, I fitted Bayesian $VARs$ on the latent factors with various priors – uninformative or diffuse, Minnesota, natural conjugate, independent normal inverse – Wishart, and $SSVS$ priors. Moreover, I augmented these models by adding macroeconomic variables such as the inflation rate, unemployment rate, percent change in $TCU$, and the effective federal funds rate.

Lastly, I plotted the impulse response functions to gauge the impact of innovations to latent variables on other variables; contrasted the mean squared forecast errors to check which priors on $BVARs$ forecast accurately; and quantified the predictive means and standard deviations from all priors. Finally, I constructed a fully Bayesian $VAR$ with dummy observations, and constructed sign-restricted impulse response functions.

The empirical results can be summarized as follows: firstly, the one-step Kalman filter approach better fits the yields ranging from 6 months to 10 years of maturity. Alternatively, when comparing the standard deviations, on average the residuals from the two-step $PC$ approach are the least volatile followed by that of the two-step Diebold-Li and one-step approaches, respectively. Secondly, applying the $SSVS$ priors in both the augmented $BVARs$, and those with the latent factors only most precisely forecast the variables one-month ahead. Lastly, I contrasted the $MSFEs$ for all the Treasury yields with forecast horizons ranging from $6-54$ months, and found that in general the the $BVARs$ Minnesota prior minimizes the mean squared forecast error for all but the six month forecast horizon. Overall, the forecasts from the Minnesota prior are the most accurate compared to those from the two-step Diebold-Li approach because the former generates a lower $MSFE$ in general. A limitation is that I imposed the priors on the $BVARs$ with the frequentist estimates of the latent factors, instead of the Bayesian estimates of those factors. It will be worthwhile to construct fully Bayesian $TVP-VAR$ by using a Bayesian analog of the Kalman filter approach. This entails estimating the unobserved factors and the observed Treasury yields by using the Carter and Kohn algorithm.

# 7. Appendix

## A.1 Frequentist Methods

### A.1.1 Estimation of the Reduced-Form Frequentist VAR

In equation (10), $f = \Omega A + \eta$, I assumed that the errors $\eta$ are exogenous, so $E[\Omega' \eta] = 0$, and the error from equation (10): $\eta = f - \Omega A$.

Then, the exogeneity assumption becomes, $E[\Omega'(f - \Omega A)] = 0$

$E[\Omega'f - \Omega'\Omega A] = 0 \Rightarrow A = (\Omega'\Omega)^{-1}\Omega'f$

Recursing equation (7) 56-months ahead results in $f_{t+56} = \sum_{k=0}^{55} A_1^k a_0 + A_1^{55} f_t + \sum_{k=0}^{55} A_1^k \eta_{t+56-k}$

It's conditional mean is the 56-months forward forecast: $E[f_{t+56} | f_t] = \sum_{k=0}^{55} A_1^k a_0 + A_1^{56} f_t$

At time $T$, if the $h-$step forecast of a Treasury yield of maturity $\tau$ is $\hat{y}_{T+h}(\tau_m)$, and actual yield is $y_{t+h}(\tau_m)$, then the mean square forecast error is $MSFE = E[(\hat{y}_{T+h}(\tau_m) - y_{m,T+h})(\hat{y}_{T+h}(\tau_m) - y_{m,T+h})']$

## A.1.2 Steps to Compute the Kalman Filter

A generic representation of the linear state-space system is:

$X_t = AX_{t-1} + Zu_t$, $Y_t = BX_t + Wv_t$, where $u_t \sim WN(0, \Sigma_u)$, $v_t \sim WN(0, \Sigma_v)$

Equations (5) and (8) are equivalent with the generic system above when $\eta_t = Zu_t$, $\varepsilon_t = Wv_t$, $Y_{1,t} = Y_t$, and $\Lambda = B$. As the errors $u_t$ and $v_t$ are white noise process, I further assumed that their variance-covariance matrices $\Sigma_u$ and $\Sigma_v$ are identity matrices (unit variances), so we can write: $\Sigma_u = I_3$ and $\Sigma_v = I_{10}$.

Thus, $Var(\eta_t) = Var(Zu_t) \Rightarrow \Sigma_\eta = Z \Sigma_u Z' \Rightarrow \Sigma_\eta = Z I_3 Z' \Rightarrow \Sigma_\eta = Z Z'$

Likewise, $\Sigma_\varepsilon = WW^T$

In the state-space system, the state or latent variables are unknown and I estimated them using the signal or information provided from the known parameters in the measurement equation.

Let $Y_1^{t-1} = \{y_{1,t-1}, y_{1,t-2}, ..., y_{1,1}\}$ be the historical values of the yields adjusted for intercepts. Kalman filter recursively estimates the latent factors and their variances at each month $t$ using information on the yields-adjusted for intercepts upto that month $- X_{t/t}$, and $\Sigma_{X, t/t}$, using the state-space parameters $- A$, $\Sigma_\eta$, $\Lambda$, and $\Sigma_\varepsilon$.

$X_{t|t} = E[X_t | Y_1^t] = E[X_t | y_{1,1}, y_{1,2}, ..., y_{1,t-1}, y_{1,t}]$, $X_{t|t-1} = E[X_t | Y_1^{t-1}]$,

$\Sigma_{X, t|t} = Var[X_t | Y_1^t]$, $\Sigma_{X, t|t-1} = Var[X_t | Y_1^{t-1}]$

The Kalman filter comprises the following steps:

Firstly, I began the Kalman filter by affixing the initial values of the demeaned latent variables $X_{t-1|t-1} = X_{0/0}$, and the variance $\Sigma_{X, t-1|t-1} = \Sigma_{X, 0/0}$

Secondly, I predicted the demeaned state and its variance at time $t$ given the historical values upto time $t-1$:
$X_{t|t-1} = AX_{t-1|t-1}$, and $\Sigma_{X, t|t-1} = A\Sigma_{X, t-1|t-1}A' + \Sigma_\eta$ given $X_{t-1|t-1}$, and $\Sigma_{X, t-1|t-1}$

Thirdly, the fitted intercept-adjusted yields at that time period are $Y_{1, t|t-1} = \Lambda X_{t|t-1}$

Fourthly, the forecast error is $E_{t|t-1} = Y_{1,t} - E[Y_{1,t} | Y_1^{t-1}] = Y_{1,t} - \Lambda X_{t|t-1}$, and its variance is

$\Sigma_{E, t|t-1} = \Lambda \Sigma_{X, t|t-1} \Lambda' + \Sigma_\varepsilon$

Finally, I updated the beliefs about the latent factors, which is the current state $X_t$ at each month $t$ each time the system receives a new signal in the form of $Y_{1,t}$. Based on the information in the forecast error, this updates the demeaned states:
$X_{t|t} = X_{t|t-1} + K E_{t|t-1}$, and the variance $\Sigma_{X, t|t} = [1 - K\Lambda] \Sigma_{X, t|t-1}$ given $X_{t|t-1}$, and $\Sigma_{X, t|t-1}$.

Here, $K = \Sigma_{X, t/t-1} \Lambda' \Sigma_{E, t/t-1}^{-1}$ is the Kalman gain, which is the weight associated with the forecast error. As a function of the state-space parameters, the Kalman filter computes the likelihood function and updates the beliefs of the state equation by calculating $X_{T|T}$, and $\Sigma_{X, T|T}$. Unlike the filtering recursions, which forecast the demeaned yields based on information up to the most recent point in time, the Kalman smoothing recursion calibrates $X_{t|T}$, and $\Sigma_{X, t|T}$ using all the values $t < T$.

### A.1.3 Optimization Technique to Evaluate the Maximum Likelihood Estimates by the Kalman Filter

To maximize the likelihood over the 29 free parameters, I used a numerical optimizer – Quasi-Newton algorithm. It evaluates the likelihood function a maximum of 30,000 times, after each iteration, it updates the parameter values. The lower bound on the step size is $10^{-9}$. Used as a relative bound, this is the stopping criteria which implies that the iterations end when $|x^* - x^{j+1}| < 10^{-9} |x^* - x^j|^2$, where $j$ is the iteration number and $x^*$ is the true solution. With each passing iteration, the error $|x^* - x^{j+1}|$ rapidly diminishes, and after a total of 275 iterations, the convergence criteria is met. The set of graphs in Figure $A1$ showcase the function values at every iteration for all the variables.

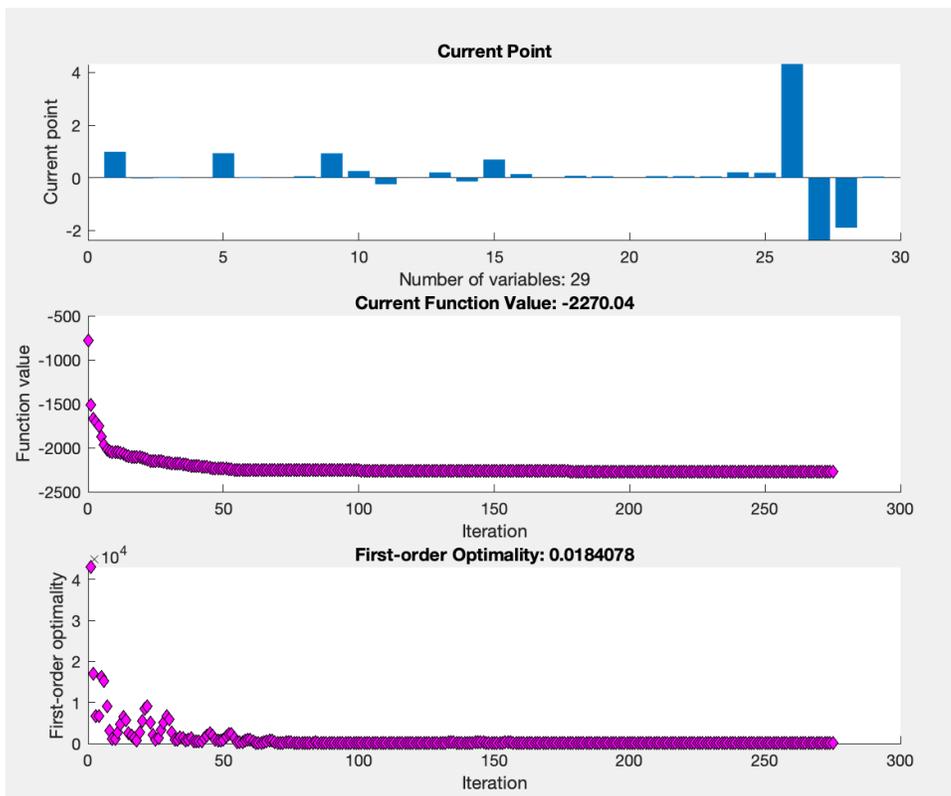

Fig A1. Values of the optimization function and the first-order optimality condition at each iteration for the 29 parameters

The exponential decline indicates that the solver significantly decreases the objective value, and finds the minimum value in the remaining iterations. The first-order optimality condition measures the gap between the current value of $x_j$ and the true value $x^*$. It is 0 at the optimal level indicating that the function has reached the minimum value i.e. the gradient is 0. The output of the optimization results in the estimated values of the 29 parameters.

## A.2 Bayesian Methods

### A.2.1 Likelihood Functions

From equation (10), $(f \mid A, \Sigma) \sim N(\Omega A, I_T \otimes \Sigma)$. So, the likelihood function is:

$$P(f \mid A, \Sigma) = (2\pi)^{-\frac{3T}{2}} |\Sigma|^{-\frac{T}{2}} e^{-\frac{1}{2}(f-\Omega A)'(I_T \otimes \Sigma^{-1})(f-\Omega A)} \tag{A.1}$$

Equation (A.1) holds as $\left| I_T \otimes \Sigma \right| = |\Sigma|^T$ and $(I_T \otimes \Sigma)^{-1} = I_T \otimes \Sigma^{-1}$

Since the two depictions of $VAR(1)$ are similar, equation (11)'s likelihood will be the same as (A.1). Recasting (A.1) with respect to $F, G, \Phi$ results in $f - \Omega A = \eta = vec(U')$.
Consequently, the quadratic form in (A.1) is the same as:

$$(f - \Omega A)'(I_T \otimes \Sigma^{-1})(f - \Omega A) = vec(U')'(I_T \otimes \Sigma^{-1}) vec(U') = tr(U\Sigma^{-1}U') = tr(\Sigma^{-1}U'U)$$

Based on the above quadratic form and $U = F - G\Phi$, the likelihood derived from equation (11) is:

$$P(F \mid \Phi, \Sigma) = (2\pi)^{-\frac{Tk}{2}} |\Sigma|^{-\frac{T}{2}} e^{-\frac{1}{2}tr[\Sigma^{-1}U'U]}$$

$$= (2\pi)^{-\frac{3T}{2}} |\Sigma|^{-\frac{T}{2}} e^{-\frac{1}{2}tr[\Sigma^{-1}(F-G\Phi)'(F-G\Phi)]} \tag{A.2}$$

### A.2.2 Minnesota Prior

The parameters of the model are the $VAR$ coefficients $A_{mp}$, the variance of the coefficients, $\Sigma_{mp}$, and the variance-covariance matrix of the innovations, $\Sigma$. The $AR(1)$ process of the latent variables is

$$\begin{bmatrix} L_t \\ S_t \\ C_t \end{bmatrix} = \begin{bmatrix} 0 \\ 0 \\ 0 \end{bmatrix} + \begin{bmatrix} 0.95 & 0 & 0 \\ 0 & 0.95 & 0 \\ 0 & 0 & 0.95 \end{bmatrix} \begin{bmatrix} L_{t-1} \\ S_{t-1} \\ C_{t-1} \end{bmatrix} + \begin{bmatrix} 0 & 0 & 0 \\ 0 & 0 & 0 \\ 0 & 0 & 0 \end{bmatrix} \begin{bmatrix} L_{t-2} \\ S_{t-2} \\ C_{t-2} \end{bmatrix} + \ldots + \begin{bmatrix} 0 & 0 & 0 \\ 0 & 0 & 0 \\ 0 & 0 & 0 \end{bmatrix} \begin{bmatrix} L_{t-13} \\ S_{t-13} \\ C_{t-13} \end{bmatrix} + \begin{bmatrix} \eta_L \\ \eta_S \\ \eta_C \end{bmatrix}$$

The elements of $\Sigma_{mp}$ where $l$ is the lag length from $l = 1, \ldots, 13$ are:

$$\Sigma_{i,jj} = \begin{cases} \dfrac{d_1}{l^2}, & \text{when the coefficient is the own lag itself} \\ \dfrac{d_2\,\sigma_{ii}^2}{l^2\sigma_{jj}^2}, & \text{when the coefficients are the lags of different variables } i \neq j \end{cases}$$

Lastly, I restricted the variance for a constant to be $d_3\sigma_{ii}$. These hyperparameters control the prior tightness or the magnitude of uncertainty about the coefficients in $A_{mp}$. For instance, $d_1$ regulates the prior variance of own lags. As $d_1 \to 0$, then $(a_{11}, a_{22}, a_{33}) \to (0.95, 0.95, 0.95)$, which imposes the prior more tightly. $d_2$ controls the tightness of priors on the lags of the independent endogenous variables. As $d_2 \to 0$, then $(a_{12}, ..., a_{ij}, ..., a_{32}) \to 0$. $d_3$ regulates the prior variance or tightness of constants. As $d_3 \to 0$, the constants increasingly shrink to $0$. Finally, $\dfrac{\sigma_{ii}}{\sigma_{jj}}$ accommodates for the varying scales of the latent variables, if present.

### A.2.3 Natural Conjugate Prior

Since $\Sigma$ cannot be negative, the support of $\Sigma$ takes all positive values. Therefore, I assumed the inverse − Wishart prior as the conjugate prior for $\Sigma$. The joint prior distribution is $(vec(\Phi), \Sigma) \sim NW^{-1}(vec(\Phi_0), \Sigma_\Phi, s_{h0}, S_{c0})$, with the hyperparameters: $vec(\Phi_0), \Sigma_\Phi, s_{h0}, S_{c0}$, and its prior joint density function is:

$$P(\Phi, \Sigma) = P(\Sigma)\,P(\Phi\,|\,\Sigma)$$

$$P(\Phi, \Sigma) \propto |\Sigma|^{-\frac{(s_{h0}+44)}{2}} e^{-\frac{1}{2}\{\Sigma^{-1}tr[S_{c0} + (\Phi-\Phi_0)'\Sigma_\Phi^{-1}(\Phi-\Phi_0)]\}} \qquad (A.3)$$

Similar to $\Sigma_{i,jj}$ in the Minnesota prior, the following function defines the prior for $\Sigma_\Phi$:

$$\Sigma_{\Phi,kk} = \begin{cases} \dfrac{d_1}{l^2\sigma_{kk}^2}, & \text{when the coefficient is lag } l \text{ of variable } k \\ d_2 & \text{for constants} \end{cases}$$

As opposed to the Minnesota prior, the kronecker structure in $\Sigma \otimes \Sigma_\Phi$ in the natural conjugate prior inhibits us to include different prior variances for the lags of dependent variables, and those of independent variables in each equation of $VAR(13)$. Additionally, the prior variance is more restrictive than $\Sigma_{i,jj}$ in Minnesota prior. To measure the posterior distribution of the parameters $\Phi$ and $\Sigma$, I multiplied the likelihood in equation (A.2) with the natural conjugate prior in (A.3):

$$P(\Phi,\Sigma\,|\,F) \propto P(\Phi,\Sigma)\,P(F\,|\,\Phi,\Sigma)$$

$$\propto |\Sigma|^{-\frac{(s_{h0}+44+T)}{2}} e^{-\frac{1}{2}\{\Sigma^{-1}tr[S_{c0}+(\Phi-\Phi_0)'\Sigma_\Phi^{-1}(\Phi-\Phi_0)+(F-G\Phi)'(F-G\Phi)]\}} \qquad (A.4)$$

Equation (A.4) comprises of two quadratic terms, which can be reorganized as:

$(\Phi - \Phi_0)' \Sigma_\Phi^{-1}(\Phi - \Phi_0) + (F - G\Phi)'(F - G\Phi)$ as $(\Phi - \hat{\Phi})' V_\Phi (\Phi - \hat{\Phi})$,

where, $\hat{\Phi}$ is a $(r \times k)$ or $(40 \times 3)$ matrix, and $V_\Phi$ is a symmetric $(40 \times 40)$ matrix.

$(\Phi - \Phi_0)' \Sigma_\Phi^{-1}(\Phi - \Phi_0) + (F - G\Phi)'(F - G\Phi)$

$= (\Phi' \Sigma_\Phi^{-1}\Phi - \Phi' \Sigma_\Phi^{-1}\Phi_0 - \Phi_0' \Sigma_\Phi^{-1}\Phi + \Phi_0' \Sigma_\Phi^{-1}\Phi_0) + (F'F - F'G\Phi - \Phi'G'F + \Phi'G'G\Phi)$

$= (\Phi - \hat{\Phi})' V_\Phi (\Phi - \hat{\Phi}) - \hat{\Phi}' V_\Phi \hat{\Phi} + \Phi_0' \Sigma_\Phi^{-1}\Phi_0 + F'F,$  (A.5)

where, $V_\Phi = \Sigma_\Phi^{-1} + G'G, \qquad \hat{\Phi} = V_\Phi^{-1}(\Sigma_\Phi^{-1}\Phi_0 + G'F)$

Plugging in equation (A.5) in (A.4), the joint posterior distribution is

$P(\Phi, \Sigma | F) = |\Sigma|^{-\frac{(s_{h0} + 44 + T)}{2}} e^{-\frac{1}{2}\{\Sigma^{-1} tr[S_{c0} + (\Phi - \hat{\Phi})' V_\Phi (\Phi - \hat{\Phi}) - \hat{\Phi}' V_\Phi \hat{\Phi} + \Phi_0' \Sigma_\Phi^{-1}\Phi_0 + F'F]\}}$

$= |\Sigma|^{-\frac{(s_{h0} + 44 + T)}{2}} e^{-\frac{1}{2}\{\Sigma^{-1} tr[S_c + (\Phi - \hat{\Phi})' V_\Phi (\Phi - \hat{\Phi})]\}},$

where $\hat{S}_c = S_{c0} - \hat{\Phi}' V_\Phi \hat{\Phi} + \Phi_0' \Sigma_\Phi^{-1}\Phi_0 + F'F$

Therefore, $(\Phi, \Sigma | F) \sim NW^{-1}(\hat{\Phi}, V_\Phi^{-1}, s_{h0} + T, \hat{S}_C)$

To make posterior inferences about $\Phi$ and $\Sigma$, I analytically integrated out the relevant parameter to obtain their marginal posterior distributions as

$P(\Phi | F) = \int_0^\infty P(\Phi, \Sigma | F) d\Sigma, \qquad P(\Sigma | F) = \int_0^\infty P(\Phi, \Sigma | F) d\Phi$

The posterior means of $\Phi$ and $\Sigma$ are $\hat{\Phi}$, and $\frac{\hat{S}_c}{s_{h0} + T - 1}$, respectively.

### A.2.4 Independent Normal and Inverse−Wishart Prior

The prior variance $V_A$ follows the exact framework as $\Sigma_{i,jj}$ in Minnesota prior does. So, $V_A = \Sigma_{i,jj}$; $d_1$, $d_2$, and $d_3$ control the extent that the coefficients in $A$ shrink. The unconditional prior densities are:

$P(A) = (2\pi)^{-\frac{kr}{2}} |V_A|^{-\frac{1}{2}} e^{-\frac{1}{2}(A-A_0)' V_A^{-1}(A-A_0)} = (2\pi)^{-60} |V_A|^{-\frac{1}{2}} e^{-\frac{1}{2}(A-A_0)' V_A^{-1}(A-A_0)}$  (A.6)

$P(\Sigma) = \frac{|S_{c0}|^{\frac{s_{h0}}{2}}}{2^{\frac{3s_{h0}}{2}} \Gamma_3(\frac{s_{h0}}{2})} |\Sigma|^{-\frac{s_{h0}+4}{2}} e^{-\frac{1}{2} tr(S_{c0}\Sigma^{-1})}$  (A.7)

Given the likelihood in (A.1) and the prior density of $A$ in (A.6), the conditional posterior distribution of $A$:
$(A | f, \Sigma) \sim N(\hat{A}, \Sigma_A^{-1}),$

where $\Sigma_A = V_A^{-1} + \Omega'(I_T \otimes \Sigma^{-1})\Omega, \quad \hat{A} = \Sigma_A^{-1}(V_A^{-1}A_0 + \Omega'(I_T \otimes \Sigma^{-1})f)$

Next, I derived the conditional posterior distribution $P(\Sigma|f, A)$. Binding the likelihood in (A.1) and $\Sigma$'s prior in equation (A.7) results in the kernel of the inverse − Wishart density function:

$$P(\Sigma|f, A) \propto P(f|A, \Sigma)P(\Sigma)$$

$$\propto |\Sigma|^{-\frac{T+s_{h0}+4}{2}} e^{-\frac{1}{2}[tr(S_{c0}\Sigma^{-1}) + \sum_{t=1}^{T}(f_t - \Omega_t A)(f_t - \Omega_t A)'\Sigma^{-1}]}$$

Consequently, $(\Sigma|f, A) \sim W^{-1}(s_{h0} + T, S_{c0} + \sum_{t=1}^{T}(f_t - \Omega_t A)(f_t - \Omega_t A)')$  (A.8)

Using Gibbs sampler, I calculated the marginal posterior distributions of the parameters. Taking the sample mean and the sample variance of the retained $100,000$ draws in the markov chain outputs the mean and the variance of the marginal distributions of $\Sigma$, and $A$.

### A.2.5 Stochastic Search Variable Selection

Before applying this prior, I fitted a $VAR$ without the intercepts, resulting in 117 coefficients in $A$, instead of 120. Similar to equation (11): $F = G\Phi + U$, where $A = vec(\Phi)$, I fitted $F_0 = G_0\Phi_0 + U_0$

The maximum likelihood ($ML$) estimates of $\Phi_0$ is $\hat{\Phi}_{0,ML} = (G_0'G_0)^{-1}G_0'F_0$, and the sum of squares of the $ML$ estimated residuals is $SSE(\hat{\Phi}_{0,ML}) = (F_0 - G_0\Phi_0)'(F_0 - G_0\Phi_0)$.

Correspondingly, the $ML$ estimate of the error variance-covariance matrix $\hat{\Sigma}_0$ is $\hat{\Sigma}_{0,ML} = \frac{SSE(\hat{\Phi}_{0,ML})}{T}$

Furthermore, $A_o = vec(\Phi_0)$, where $A_0$ is a $(117 \times 1)$ vector. Then, I found the identifying restrictions of this form of $VAR$, wherein I restricted the elements of $\Sigma_0$, and the nine coefficients of the latent factors in $A_0$. By Cholesky factorization, I decomposed $\Sigma_0$ into a $(3 \times 3)$ upper triangular matrix $U: \Sigma_0 = U'U$.

Let $\xi_j = [u_{1j}, ..., u_{j-1,j}]' \ \forall j = 2, ..., k \Rightarrow j = 2, 3$. This implies that $\xi_j = [u_{12}, u_{13}, u_{23}]'$ − the off-diagonal elements of $U$. Furthermore, I stored those off-diagonal elements as $\xi = [\xi_2', \xi_3']'$, and, let $u = [u_{11}, u_{22}, u_{33}]'$ be the diagonal elements of $U$.

I constructed hierarchical priors for $(A_0, \xi, u)$. Next, I selected the priors on the off-diagonal entries of the upper Cholesky factor of $\Sigma_0: \xi_j = [u_{12}, u_{13}, u_{23}]'$. Let $\delta_j = [\delta_{1j}, ..., \delta_{j-1,j}]'$ be a vector of $0 - 1$ variables, and let $H_j = diag(h_{ij}, ..., h_{j-1,j})$, where:

$$h_{ij} = \begin{cases} \tau_{0,ij}, \text{ if } \delta_{ij} = 0 \\ \tau_{1,ij} \text{ if } \delta_{ij} = 1 \end{cases}$$

As before, I fixed the hyperparameters to $\tau_{0,ij} = 0.01$, and $\tau_{1,ij} = 10 \ \forall i = 1, ..., 117$. These hyperparameter values control the prior in equation (12).

The prior for $\xi_j \mid \delta_j \sim N_{j-1}(0, H_j M_j H_j) \; \forall \; j = 2, 3$.

With this prior, the distribution of each element of $\xi_j$:

$$(\xi_{ij} \mid \delta_{ij}) \sim (1 - \delta_{ij}) N(0, \tau_{0,ij}^2) + \delta_{ij} N(0, \tau_{1,ij}^2) \; \forall \; i = 1, ..., j - 1$$

Now, I assumed that the elements of $\delta = [\delta_{12}, \delta_{13}, \delta_{23}]'$ are independent Bernoulli random variables where $q_{ij} \in (0, 1)$ such that $P(\delta_{ij} = 1) = q_{ij}$, and $P(\delta_{ij} = 0) = 1 - q_{ij}$, $i = 1, 2, 3$; $j = 1, 2$. Thus, $q_{ij}$ reflects the prior probability that $u_{ij}$ is unrestricted, and I set $q_{ij} = 0.2$. Finally, in determining the priors on the diagonal elements $u = [u_{11}, u_{22}, u_{33}]$, I assumed that $u_{ii}^2 \sim Gamma(a_i, b_i)$, where $(a_i, b_i) > 0$ are constants. The prior density for $u_{ii} \; \forall i = 1, 2, 3$ is:

$$[u_{ii}] = \frac{2 b_i^{a_i}}{\Gamma(a_i)} u_{ii}^{2(a_i - 1)} e^{(-b_i u_{ii}^2)} \; \forall \; u_{ii} > 0$$

### A.2.6 Calculation of Impulse Responses

If $B$ is a backshift or a lag operator, first I rewrote $VAR(13)$ equivalently into its companion form to get $VAR(1)$ as follows: $F_t = a_0 + A_{11} F_{t-1} + \eta_t$

Then I inverted the above $VAR(1)$ to an $MA(\infty)$ form:

$F_t = a_0 + A_{11} B F_t + \eta_t \Rightarrow (I - A_{11} B) F_t = a_0 + \eta_t$

Hence, $F_t = (I - A_{11} B)^{-1} (a_0 + \eta_t)$

Equivalently, $F_t = (a_0 + \eta_t) + A_{11} B (a_0 + \eta_t) + A_{11}^2 B^2 (a_0 + \eta_t) + ...$

$F_t = (I + A_{11} B + A_{11}^2 B^2 + ...) a_0 + (I + A_{11} B + A_{11}^2 B^2 + ...) \eta_t$

Therefore, $F_t = (I - A_{11} B)^{-1} a_0 + \eta_t + A_{11} \eta_{t-1} + A_{11}^2 \eta_{t-2} + ...,$ \hfill (A.9)

where $\eta_t$ in $VAR(1)$ are the reduced-form innovations such that $E[\eta_t] = 0$, and $E[\eta_t \eta_t'] = \Sigma_\eta$

The innovations are mutually correlated since $\Sigma_\eta$ is symmetric, positive-definite with non-zero off diagonal elements. Henceforth, I hypothesized that the reduced-form innovations $\eta_t$ are a linear transformation of the orthogonal impulses or structural shocks $\varepsilon_t : \eta_t = Z \varepsilon_t$, and the variance is normalized to be the identity matrix: $\Sigma_\varepsilon = I_k$. $\varepsilon_t$ is a $(k \times 1)$ vector containing three structural shocks wherein $\varepsilon_{i,t}$ refers to the shock from variable $i$ contained in the vector of variables $F_t$. Additionally, I imposed an identifying assumption on $Z$ to help interpret the structural shocks. Therefore, $Z$ is a lower triangular matrix obtained from the Cholesky decomposition of $\Sigma_\eta : \Sigma_\eta = Z Z'$ as done in the one-step Kalman filter approach. The Cholesky decomposition orthogonalizes the innovations and produces orthogonal impulse responses. So, only the first structural shock $- \varepsilon_{L,t}$ influences $L_t$, and the first two structural shocks $- \varepsilon_{L,t}, \varepsilon_{S,t}$, affect $S_t$.

Substituting $\eta_t = Z \varepsilon_t$ in equation (A.9) yields:

$$F_t = (I - A_{11}B)^{-1}a_0 + Z\varepsilon_t + A_{11}Z\varepsilon_{t-1} + A_{11}{}^2 Z\varepsilon_{t-2} + ... \tag{A.10}$$

Now, I recursed (A.10) $h-$ steps forward:

$$F_{t+h} = (I - A_{11}B)^{-1}a_0 + Z\varepsilon_{t+h} + A_{11}Z\varepsilon_{t+h-1} + A_{11}{}^2 Z\varepsilon_{t+h-2} + ... + A_{11}{}^h Z\varepsilon_t + ...$$

Taking the partial derivatives of $F_{i,t+h}$ with respect to $\varepsilon_{j,t}$ results in the response of the variable $i$ at $t+h$ months ahead due to a structural shock $\varepsilon_{j,t}$ from another variable $j \neq i$:

$$\frac{dF_{i,t+h}}{d\varepsilon'_{j,t}} = [A_{11}{}^h Z]_{(i,j)}$$

Therefore, the impulse response function is a plot of $\frac{dF_{i,t+h}}{d\varepsilon'_{j,t}}$ $\forall\, h = 1, 2, ..., 24,$ where the time horizon is 24 months.

### A.2.7 BVAR with Dummy Observations

In the reduced-form $VAR$, the means and variances of the Minnesota prior:

$$E[A_{q,ij}] = \begin{cases} m_i, & q = 1, \text{ element } i = j \\ 0, & \text{otherwise} \end{cases}, \quad V[A_{q,ij}] = \begin{cases} \dfrac{f^2}{q^2}, & i = j \\ \dfrac{f^2 \sigma_i^2}{q^2 \sigma_j^2}, & \text{otherwise} \end{cases}$$

The hyperparameter $f = 0.95$ controls the degree of prior tightness around the white noise or random walk model. $\sigma_i$ is standard deviation of the errors obtained after estimating the $AR$ coefficients by $OLS$. $m_i \forall i = 1, ..., 15$ are the prior means of the coefficients associated with the dependent variables' first lag. However, I presumed that the residuals in a structural $VAR$ are correlated, therefore, the residual variance-covariance matrix cannot be diagonal and fixed. The dummy observations linked to the inverse-Wishart prior are:

$$Y_{d1} = \begin{bmatrix} \dfrac{\text{diag}(m_1\sigma_1 ... m_{15}\sigma_{15})}{f} \\ 0_{(15\times 15)} \\ \hdashline \text{diag}(\sigma_1 ... \sigma_{15}) \\ \hdashline 0_{(1\times 15)} \end{bmatrix}, \quad X_{d1} = \begin{bmatrix} \dfrac{D \otimes \text{diag}(\sigma_1 ... \sigma_{15})}{f} & 0_{(30\times 1)} \\ \hdashline 0_{(15\times 30)} & 0_{(15\times 1)} \\ \hdashline 0_{(1\times 30)} & c \end{bmatrix}$$

where, $D = diag(1,2)$, and another hyperparameter is $c = 0.95$ that controls the prior variance of the constants. The dummy observations in the first and the second block apply the prior beliefs on the $AR$ coefficients, and the variance-covariance matrix, respectively. Finally, the dummies on the third block demonstrate that the constants have a diffuse prior.

Combining the dummy observations for the sum-of-coefficients, and the natural conjugate priors results in:

$Y_d = [Y_{d1}, Y_{d2}]; X_d = [X_{d1}, X_{d2}]$

The augmented dataset is $Y_a = [Y', Y_d']'; X_a = [X', X_d']$ and $T_a = T + T_d$